\documentclass[12pt]{article}
\usepackage{catmac}
\usepackage{amssymb}
\usepackage{bbm}
\newtheorem{Theorem}{Theorem}[section]
\newtheorem{Proposition}[Theorem]{Proposition}
\newtheorem{Lemma}[Theorem]{Lemma}
\newtheorem{Corollary}[Theorem]{Corollary}
\newtheorem{Definition}[Theorem]{Definition}
\newcommand{\CC}{{\mathbb{C}}}
\newcommand{\II}{{\mathbbm{1}}}

\newcommand{\RR}{{\mathbb{R}}}\newcommand{\ZZ}{{\mathbb{Z}}}
\newcommand{\ab}[2]{#1 \,\!^{[\,#2\,]}}

\newcommand{\bfa}{{\bf a}}
\newcommand{\bfk}{{\bf k}}
\newcommand{\bfm}{{\bf m}}
\newcommand{\BSUJ}{\rmB\SUJ}
\newcommand{\BUJ}{\rmB\rmU J}
\newcommand{\Bun}[2]{{\mbox{\rm Bun}(#1,#2)}}
\newcommand{\calJ}{{L}}
\newcommand{\con}{{\cal A}}

\newcommand{\CP}{\CC{\rm P}}
\newcommand{\CW}{{\it CW}}
\newcommand{\diag}{{\mbox{\rm diag}}}
\newcommand{\diff}{{\rm d}}
\newcommand{\End}{{\mbox{\rm End}}}
\newcommand{\Ext}{\mbox{\rm Ext}}
\newcommand{\gau}{{\cal G}}
\newcommand{\gref}[1]{{\rm (\ref{#1})}}
\newcommand{\Hom}{{\mbox{\rm Hom}}}
\newcommand{\Howe}{{\mbox{\rm Howe}}}
\newcommand{\id}{{\rm id}}
\newcommand{\im}{{\rm im}\,}
\newcommand{\ka}[1]{f_{#1}}
\newcommand{\lens}[2]{{{\rm L}_{#1}^{#2}}}
\newcommand{\mod}{{\mbox{\rm mod}}}
\newcommand{\onepoint}{\ast}
\newcommand{\orb}{{\cal M}}
\newcommand{\OT}{{\rm OT}}

\newcommand{\pr}{{\rm pr}}
\newcommand{\prU}[2]{\pr^{\rmU #1}_{#2}}
\newcommand{\qed}{\hspace*{0.1cm} \hfill \rule[0.3mm]{2mm}{2mm}}
\newcommand{\Red}[2]{{\mbox{\rm Red}(#1,#2)}}
\newcommand{\rmB}{{\rm B}}
\newcommand{\rmC}{{\rm C}}
\newcommand{\rmE}{{\rm E}}
\newcommand{\rmeven}{{\rm even}}
\newcommand{\rmGL}{{\rm GL}}
\newcommand{\rmgl}{{\rm gl}}
\newcommand{\rmi}{{\rm i}}
\newcommand{\rmK}{{\rm K}}
\newcommand{\rmL}{{\rm L}}

\newcommand{\rmP}{{\rm P}}
\newcommand{\rmS}{{\rm S}}
\newcommand{\rmSO}{{\rm SO}}
\newcommand{\rmSU}{{\rm SU}}
\newcommand{\rmU}{{\rm U}}
\newcommand{\rref}[1]{{\rm \ref{#1}}}
\newcommand{\sphere}[1]{{{\rm S}^{#1}}}
\newcommand{\SUJ}{{\rmSU J}}
\newcommand{\torus}[1]{{{\rm T}^{#1}}}
\newcommand{\Tr}{{\rm Tr}}
\newcommand{\twbfm}{\widetilde{\bfm}}
\newcommand{\tweta}{\tilde{\eta}}
\newcommand{\twgamma}{{\tilde{\gamma}}}
\newcommand{\twH}{\tilde{H}}
\newcommand{\twL}{\tilde{L}}

\newcommand{\twm}{\widetilde{m}}
\newcommand{\twPhi}{\tilde{\Phi}}
\newcommand{\twQ}{\tilde{Q}}
\newcommand{\twxi}{\tilde{\xi}}
\newcommand{\UJ}{{\rmU J}}
\newcommand{\ddeckmatrix}[4]{\left(\begin{array}{ccc}   
#1&\cdots&#2\\ \vdots&\ddots&\vdots\\#3&\cdots&#4   
\end{array}\right)}
\newcommand{\vveckmatrix}[9]{\left(\begin{array}{cccc}  
#1&#2&\cdots&#3\\#4&#5&\cdots&#6\\          
\vdots&\vdots&\ddots&\vdots\\#7&#8&\cdots&#9
\end{array}\right)}
\begin{document}

\begin{titlepage}
\begin{center}
{\LARGE{\bf Classification of Gauge Orbit Types\\[0.3cm]
for $\rmSU n$-Gauge Theories}}

\vspace{2.5cm}

{\large {\bf G. Rudolph, M. Schmidt and I.P. Volobuev$^1$}}

\vskip 0.5 cm

Institute for Theoretical Physics\\
University of Leipzig\\
Augustusplatz 10\\
04109 Leipzig \& Germany
\\[0.2cm]
E-mail: matthias.schmidt@itp.uni-leipzig.de
\\
Fax:  +49-341-97/32548
\\[0.5cm]
$^1$Nuclear Physics Institute, Moscow State University\\
119899 Moscow \& Russia

\end{center}

\vspace{2.5 cm}

\begin{abstract}
\noindent 
A method for determining the
orbit types of the action of the group of gauge transformations on
the space of connections for gauge theories with gauge group
$\rmSU n$ in space-time dimension $d\leq 4$ is presented. 
The method is based on the 1:1-correspondence between orbit types 
and holonomy-induced Howe subbundles of the underlying principal
$\rmSU n$-bundle. It is shown that the orbit types are labelled by
certain cohomology elements of space-time satisfying two relations.
Thus, for every principal $\rmSU n$-bundle the corresponding 
stratification of the gauge orbit space can be determined explicitly. 
As an application, a criterion characterizing kinematical nodes for 
physical states in $2+1$-dimensional Chern-Simons theory proposed by 
Asorey {\it et al.} is discussed.
\end{abstract}

\vspace{1 cm}
\end{titlepage}

\tableofcontents
\thispagestyle{empty}
\setcounter{page}{0}
\newpage
\section{Introduction}
\label{Sintro}
One of the basic principles of modern theoretical physics is the
principle of local gauge invariance. Its application to the theory of
particle interactions gave rise to the standard model, which proved to
be a success from both theoretical and phenomenological points of view.
The most impressive results of the model were obtained within the
perturbation theory, which works well for high energy processes. On
the other hand, the low energy hadron physics, in particular, the
quark confinement, turns out to be dominated by nonperturbative effects,
for which there is no rigorous theoretical explanation yet.

The  application of geometrical methods to non-abelian gauge
theories revealed their rich geometrical and topological
properties. In particular, it showed that the configuration space of such
theories, which is the space of gauge group orbits in the space of
connections, may have a highly nontrivial structure. In general,
the orbit space possesses not only orbits of the so called principal type,
but also orbits of other types, which may give rise to singularities of 
the configuration space.
This stratified structure of the gauge orbit space is believed to be of
importance for both classical and quantum properties of
non-abelian gauge theories in nonperturbative approach, and it has
been intensively studied in recent years.
Let us discuss some aspects indicating its physical relevance. 

First,
studying the geometry and topology of the generic (principal)
stratum, one gets a deeper understanding of the Gribov-ambiguity
and of anomalies in terms of index theorems. In particular, one
gets anomalies of purely topological type, which can not be seen 
by perturbative quantum field theory. These are
well-known results from the eighties. Moreover, there are partial
results and conjectures concerning the relevance of nongeneric
strata. First of all, nongeneric gauge orbits affect the
classical motion on the orbit space due to boundary conditions
and, in this way, may produce nontrivial contributions to the path
integral. They may also lead to localization of certain quantum
states, as it was suggested by finite-dimensional examples
\cite{EmmrichRoemer}. Further, the gauge field configurations
belonging to nongeneric orbits can possess a magnetic charge, i.e.
they can be considered as a kind of magnetic monopole
configurations, which are responsible for quark confinement. This
picture was found in 3-dimensional gauge systems
\cite{Asorey:Nodes}, and it is conjectured that it can hold for
4-dimensional theories as well \cite{Asorey:99}. Finally, it was
suggested in \cite{Heil:Anom} that non-generic strata may lead to
additional anomalies.

Most of the problems mentioned here are still awaiting a
systematic investigation. In a series of papers we are going to make
a new step in this direction. We give a complete solution to
the problem of determining the strata that are present in the
gauge orbit space for $\rmSU n$ gauge theories in compact Euclidean 
space-time of dimension $d=2,3,4$. Our analysis is based on the results of 
a paper by Kondracki and Rogulski \cite{KoRo}, where it was
shown that the gauge orbit space is a stratified topological space
in the ordinary sense, cf.~\cite{KoRo:Strat} and references therein.
Moreover, these authors found an interesting relation between orbit types
and certain bundle reductions, which we are going to use.  We also refer to
\cite{FSS} for the discussion of a very simple, but instructive
special example (orbit types of $\rmSU 2$-gauge theory on $\sphere{4}$).

The paper is organized as follows. In Section \rref{Sprel} we introduce the
basic notions related to the action of the group of gauge transformations on 
the space of connections, state the definitions of stabilizer and orbit type 
and recall basic results
concerning the stratification structure of the gauge orbit space.
In Section \rref{Sot} we introduce the
notions of Howe subgroup and holonomy-induced Howe subbundle and establish 
the relation between orbit types and holonomy-induced Howe subbundles. 
Section \rref{SHSG} is devoted to the study of the Howe subgroups of 
$\rmSU n$. In Section \rref{SHSB} we
give a classification of the Howe subbundles of $\rmSU n$-bundles for 
space-time dimension $d\leq 4$. In
Section \rref{Sholind} we prove that any Howe subbundle of $\rmSU n$-bundles
is holonomy-induced. In Section \rref{SP} we implement the equivalence 
relation of Howe subbundles due to the action of $\rmSU n$. As an example, in 
Section \rref{SSU2} we determine the orbit types for gauge group $\rmSU 2$. 
Finally, in Section \rref{SNodes} we discuss
an application to Chern-Simons theory in $2+1$ dimensions.

In two subsequent papers we shall investigate the natural partial
ordering on the set of orbit types and the structure of another,
coarser stratification, see \cite{KoSa}, obtained by first
factorizing with respect to the so called pointed gauge group and
then by the structure group.
\section{Gauge Orbit Types and Stratification}
\label{Sprel}
We consider a fixed topological sector of a gauge theory with gauge
group $G$ on a Riemannian
manifold $M$. In the geometrical setting it means that we are given 
a smooth right principal fibre bundle $P$ with structure group $G$ over 
$M$. $G$ is assumed to be a compact connected Lie group and $M$ is 
assumed to be compact, connected, and orientable. 

Denote the sets of connection forms
and gauge transformations of $P$ of Sobolev class $W^k$ (locally square
integrable up to the $k$-th derivative) by $\con^k$ and
$\gau^k$, respectively. Here $k$ is a nonnegative integer. For generalities
on Sobolev spaces of cross sections in fibre bundles, see \cite{Palais}.
Provided $2k>\dim M$, $\con^k$ is an affine Hilbert space and $\gau^{k+1}$ is
a Hilbert Lie group acting smoothly from the right on $\con^k$
\cite{KoRo,MitterViallet,Singer:Gribov}. 
We shall even assume $2k>\dim M+2$. Then, by the Sobolev Embedding Theorem,
connection forms are of class $C^1$ and, therefore, have continuous
curvature. 
If we view elements of $\gau^{k+1}$ as $G$-space morphisms $P\rightarrow G$, 
the action of $g\in\gau^{k+1}$ on $A\in\con^k$ is given by
\begin{equation} \label{Ggauact}
A^{(g)}=g^{-1}Ag + g^{-1}\diff g \,.
\end{equation}
Let $\orb^k$ denote the quotient topological space
$\con^k/\gau^{k+1}$. This space represents the configuration
space of our gauge theory.

For this to make sense, $\orb^k$ should not depend essentially on the
purely technical parameter $k$. Indeed, let $k'>k$. Then one has natural
embeddings $\gau^{k'+1}\hookrightarrow\gau^{k+1}$ and
$\con^{k'}\hookrightarrow\con^k$. As a consequence of the first, the latter
projects to a map $\varphi:\orb^{k'}\rightarrow\orb^k$. Since the image of
$\con^{k'}$ in $\con^k$ is dense, so is $\varphi\left(\orb^{k'}\right)$ in
$\orb^k$. To see that $\varphi$ is injective, let $A_1,A_2\in\con^{k'}$ and
$g\in\gau^{k+1}$ such that $\varphi(A_2)=\varphi(A_1)^{(g)}$. Then
\gref{Ggauact} implies
\begin{equation} \label{Gdiffg}
\diff g=g\varphi(A_2)-\varphi(A_1)g\,.
\end{equation}
Due to $2k'>2k>\dim M$, by the multiplication rule for Sobolev functions,
the rhs.~of \gref{Gdiffg} is of class $W^{k+1}$. Then $g$ is of class
$W^{k+2}$. This can be iterated until
the rhs.~is of class $W^{k'}$. Hence, $g\in\gau^{k'+1}$, so that $A_1$ and
$A_2$ are representatives of the same element of $\orb^{k'}$. This shows
that $\orb^{k'}$ can be identified with a dense subset of $\orb^k$.
Another question is whether the stratification structure of $\orb^k$, which
will be discussed in a moment, depends on $k$. Fortunately, the answer to this question 
is negative, see Theorem \rref{Totgau}.

In general, the orbit space of a smooth Lie group action does not admit a
smooth manifold structure w.r.t.~which the projection is smooth. 
The best one can expect is that it admits a stratification.
For the notion of stratification of a topological space, see
\cite{KoRo:Strat} or \cite[\S 4.4]{KoRo}. For the gauge orbit space
$\orb^k$, a stratification was
constructed in \cite{KoRo}, using a method which is known from compact Lie 
group actions on completely regular spaces \cite{Bredon:CTG}.
In order to explain this, let us recall the notions of stabilizer and
orbit type. The {\it stabilizer}, or isotropy subgroup, of $A\in\con^k$ is
the subgroup
$$
\gau^{k+1}_A=\{ g\in\gau^{k+1} ~|~ A^{(g)}=A \}
$$
of $\gau^{k+1}$. It has the following transformation property: For any 
$A\in\con^k$ and $g\in\gau^{k+1}$, 
$$
\gau^{k+1}_{A^{(g)}}=g^{-1}\gau^{k+1}_A g\,.
$$
Thus, there exists a natural map, called {\it type map}, assigning to each
element of $\orb^k$ the conjugacy class in $\gau^{k+1}$ made up by the
stabilizers of its representatives in $\con^k$. Let
$\OT\left(\con^k,\gau^{k+1}\right)$
denote the image of this map. The elements of
$\OT\left(\con^k,\gau^{k+1}\right)$ are
called {\it orbit types}. The set $\OT\left(\con^k,\gau^{k+1}\right)$ carries a 
natural
partial ordering: $\tau\leq\tau'$ iff
there are representatives $\gau^{k+1}_A$ of $\tau$ and $\gau^{k+1}_{A'}$ of
$\tau'$ such that $\gau^{k+1}_A\supseteq \gau^{k+1}_{A'}$. Note that this
definition is consistent with \cite{Bredon:CTG} but not with
\cite{KoRo} and several other authors who define it just inversely.

As was shown in \cite{KoRo}, the subsets $\orb^k_\tau\subseteq\orb^k$, consisting
of gauge orbits of type $\tau$, can be equipped with a smooth Hilbert
manifold structure and the family
\begin{equation} \label{Gorbdecomp}
\left\{\orb^k_\tau~|~\tau\in\OT\left(\con^k,\gau^{k+1}\right)\right\}
\end{equation}
is a
stratification of $\orb^k$. Accordingly, the manifolds $\orb^k_\tau$ are 
called strata. In particular,
$$
\orb^k=\bigcup_{\tau\in\OT\left(\con^k,\gau^{k+1}\right)}
~~\orb^k_\tau\,,
$$
where for any $\tau\in\OT\left(\con^k,\gau^{k+}\right)$ there holds
\begin{equation} \label{Gdensitythm}
\mbox{\it $\orb^k_\tau$ is open and dense in
$\bigcup_{\tau'\leq\tau}\orb^k_{\tau'}$.}
\end{equation}
Similarly to the case of compact Lie groups, there exists a principal orbit 
type $\tau_0$ obeying $\tau_0\geq\tau$ for all 
$\tau\in\OT\left(\con^k,\gau^{k+1}\right)$. Due to \gref{Gdensitythm}, the
corresponding stratum $\orb^k_{\tau_0}$ is open and dense in $\orb^k$. For
this reason it is called the generic stratum.

The above considerations show that the set $\OT\left(\con^k,\gau^{k+1}\right)$ 
together with its natural partial ordering carries the information about which 
strata occur and how they are patched together. 

To conclude,
let us remark that instead of using Sobolev techniques one can also stick to
smooth connection forms and gauge transformations. Then one obtains
essentially analogous results about the stratification of the corresponding 
gauge orbit space where, roughly speaking, one has to replace
'Hilbert manifold' and 'Hilbert Lie group' by 'tame Fr\'echet manifold' and
'tame Fr\'echet Lie group', see \cite{Abbati:Gau,Abbati:OrbSpa}.
\section{Correspondence between Orbit Types and Bundle Reductions}
\label{Sot}
In this section, let $p_0\in P$ be fixed. For $A\in\con^k$, let $H_A$ and
$P_A$ denote the holonomy group and holonomy subbundle, respectively, of
$A$ based at $p_0$. We assume $2k>\dim M+2$. Then, by the Sobolev Embedding
Theorem, $A$ is of class $C^1$ so that $P_A$ is a subbundle of $P$ of class
$C^2$. In particular, it is a subbundle of class $C^0$. 

For any $g\in\gau^{k+1}$, let $\vartheta_g$ denote the associated vertical 
automorphism of $P$. Then
\begin{equation} \label{Gdefthg}
\vartheta_g(p) = p\cdot g(p)~~~\forall~p\in P\,.
\end{equation}
Let $\rmC_G(H)$ denote the centralizer of $H\subseteq G$ in $G$. We
abbreviate
$\rmC_G^2(H)=\rmC_G\left(\rmC_G(H)\right)$.

Let $A\in\con^k$ and $g\in\gau^{k+1}$. Since the elements of $\gau^{k+1}_A$
map $A$-horizontal paths in $P$ to $A$-horizontal paths they are constant
on $P_A$. Conversely, any gauge transformation which is constant on $P_A$
leaves $A$ invariant. Thus, for any $g\in\gau^{k+1}$ one has
\begin{equation} \label{GconstPA}
g\in\gau^{k+1}_A~~~\Longleftrightarrow~~~g|_{P_A} \mbox{ is constant.}
\end{equation}
This suggests the idea to characterize orbit types by certain classes of
subbundles of $P$. For any subgroup $S\subseteq\gau^{k+1}$ define a subset
$\Theta(S)\subseteq P$ by
\begin{equation} \label{GdefThS}
\Theta(S) = \left\{p\in P ~|~ g(p) = g(p_0)~\forall g\in S\right\}\,.
\end{equation}
\begin{Lemma} \label{LThS}
~\\[0.1cm]
{\rm (a)} For any $A\in\con^k$, $\Theta\left(\gau^{k+1}_A\right)=P_A\cdot
\rmC_G^2\left(H_A\right)$.
\\[0.1cm]
{\rm (b)} Let $A,A'\in\con^k$. Then $\Theta\left(\gau^{k+1}_A\right)
=\Theta\left(\gau^{k+1}_{A'}\right)$ implies $\gau^{k+1}_A=\gau^{k+1}_{A'}$.
\\[0.1cm]
{\rm (c)} Let $g\in\gau^{k+1}$. For any subgroup $S\subseteq\gau^{k+1}$,
$\Theta\left(gSg^{-1}\right) = \vartheta_g\left(\Theta(S)\right)\cdot
g(p_0)^{-1}$.
\end{Lemma}
{\it Remark:} In \cite{KoRo}, $P_A\cdot\rmC_G^2(H_A)$ is called the {\it
evolution bundle} generated by $A$. Thus, in this terminology, (a) states
that $\Theta$ maps the stabilizer of a connection to its evolution bundle.
\\

{\it Proof:}

(a) Let $A\in\con^k$. Recall that $P_A$ has structure group $H_A$. Hence, in
view of \gref{GconstPA}, the equivariance property of gauge transformations 
implies
\begin{equation} \label{GctrHA}
\left\{ g(p_0) ~\left|~ g\in\gau^{k+1}_A \right.\right\}
=
\rmC_G(H_A)\,.
\end{equation}
Thus, by equivariance again,
\begin{equation} \label{GconstPAC}
g\in\gau^{k+1}_A ~~~\Longrightarrow~~~g|_{P_A\cdot\rmC_G^2(H_A)} \mbox{ is
constant.}
\end{equation}
This shows $P_A\cdot\rmC_G^2(H_A)\subseteq \Theta\left(\gau^{k+1}_A\right)$. 
Conversely, let $p\in P$ such that
$g(p)=g(p_0)$ for all $g\in\gau^{k+1}_A$. There exists $a\in G$ such that
$p\cdot a^{-1}\in P_A$. Then \gref{GconstPA} yields
$$
g(p_0) = g\left(p\cdot a^{-1}\right) = ag(p)a^{-1} = ag(p_0)a^{-1}\,,
$$
for all $g\in\gau^{k+1}_A$. Due to \gref{GctrHA}, then $a\in\rmC_G^2(H_A)$.
Hence, $p=(p\cdot a^{-1})\cdot a\in P_A\cdot\rmC_G^2(H_A)$.

(b) Let $A,A'\in\con^k$ be given. For any $g\in\gau^{k+1}$, we have
$$
g\in\gau^{k+1}_A
~~~\Longleftrightarrow~~~
g|_{\Theta\left(\gau^{k+1}_A\right)} \mbox{ is constant.}
$$
Here implication from left to right is due to \gref{GconstPAC} and assertion
(a), the inverse implication follows from
$P_A\subseteq\Theta\left(\gau^{k+1}_A\right)$ and \gref{GconstPA}. Since
a similar characterization holds for $\gau^{k+1}_{A'}$, the assertion
follows.

(c) Let $p\in P$, $h\in S$. Using \gref{Gdefthg} we compute
$$
g(p_0)^{-1}g(p)h(p)g(p)^{-1}g(p_0)
=
h\left(\vartheta_{g^{-1}}(p)\cdot g(p_0)\right)\,.
$$
This allows us to write down the following chain of equivalences:
$$
\begin{array}{rcl}
p\in\Theta\left(gSg^{-1}\right)
& \Longleftrightarrow &
g(p)h(p)g(p)^{-1} = g(p_0)h(p_0)g(p_0)^{-1}
\\
& \Longleftrightarrow &
g(p_0)^{-1}g(p)h(p)g(p)^{-1}g(p_0) = h(p_0)
\\
& \Longleftrightarrow &
h\left(\vartheta_{g^{-1}}(p)\cdot g(p_0)\right) = h(p_0)
\\
& \Longleftrightarrow &
\vartheta_{g^{-1}} (p) \cdot g(p_0)\in\Theta(S)\,.
\end{array}
$$
This proves assertion (c).
\qed
\begin{Definition}
Let $G$ be a Lie group and let $P$ be a principal $G$-bundle.
A subgroup $H\subseteq G$ is called {\em Howe subgroup} iff there exists
a subset $K\subseteq G$ such that $H=\rmC_G(K)$. A reduction
of $P$ to a Howe subgroup of $G$ will be called {\em Howe subbundle}. A
subbundle $Q\subseteq P$ will be called {\em holonomy-induced of class
$C^r$}
iff it contains a connected subbundle $\twQ\subseteq Q$ of class $C^r$ with
structure group $\twH$ such that
\begin{equation} \label{Gdefholind}
Q=\twQ\cdot\rmC_G^2\left(\twH\right)\,.
\end{equation}
\end{Definition}
We remark that the notion of Howe subgroup is common in the literature,
cf.~\cite{Moeglin}. The notions of Howe subbundle and holonomy-induced subbundle 
are, to our knowledge, new. Moreover, we note that $H\subseteq G$ is a Howe 
subgroup iff $\rmC_G^2(H)=H$. In particular, the subset $K\subseteq G$ in the
definition can be chosen to be $\rmC_G(H)$.

The set of holonomy-induced Howe subbundles of $P$ of class $C^0$ is acted upon in 
a natural way by the group of continuous
gauge transformations $\gau^{C^0}$ and by the structure group $G$. Let
$\Howe_\ast(P)$ denote the corresponding set of conjugacy classes. We note that the actions
of $\gau^{C^0}$ and $G$ commute. Moreover, two subbundles of class $C^0$ of $P$ are 
conjugate under the action of $\gau^{C^0}$ iff they are isomorphic. 
$\Howe_\ast(P)$ carries a natural partial ordering, 
which is defined similarly to that of orbit types:
$\eta\geq\eta'$ iff there are representatives $Q$ of $\eta$ and $Q'$ of
$\eta'$ such that $Q\subseteq Q'$. 
\begin{Theorem} \label{Totgau}
Let $M$ be compact, $\dim M\geq 2$. Then the assignment $\Theta$ induces,
by passing to quotients, an order-preserving bijection from
$\OT\left(\con^k,\gau^{k+1}\right)$ onto $\Howe_\ast(P)$.
\end{Theorem}
{\it Proof:} Let $\tau\in\OT\left(\con^k,\gau^{k+1}\right)$ and choose a
representative $S\subseteq\gau^{k+1}$. There exists $A\in\con^k$ such that
$S=\gau^{k+1}_A$. According to Lemma \rref{LThS}(a), $\Theta(S)$
can be obtained by extending the subbundle $P_A\subseteq P$ to the structure
group $\rmC_G^2(H_A)$. Since $P_A$ is of class $C^0$, so is
$\Theta(S)$. Since $P_A$ is connected, $\Theta(S)$ is holonomy-induced of
class $C^0$. Moreover, it is obviously Howe. According to Lemma
\rref{LThS}(c), when passing to quotients the class of $\Theta(S)$ does not
depend on the chosen representative $S$ of $\tau$. Thus, indeed,
$\Theta$ projects to a map from $\OT\left(\con^k,\gau^{k+1}\right)$ to
$\Howe_\ast(P)$.

To check that this map is surjective, let $Q\subseteq P$ be a
holonomy-induced Howe subbundle of class $C^0$. Let $\twQ\subseteq Q$ be a
connected subbundle of class $C^0$, with structure group $\twH$, such that
\gref{Gdefholind} holds. Due to well-known smoothing theorems
\cite[Ch.~I, \S 4]{Hirzebruch}, $\twQ$ 
and $Q$ are $C^0$-isomorphic to $C^\infty$-subbundles of $P$. Hence, up to
the action of $\gau^{C^0}$, we may assume that $\twQ$ and $Q$ are of
class $C^\infty$ themselves. Moreover, up to the action of $G$, $p_0\in\twQ$.
Since $M$ is compact and $\dim M\geq 2$, $\twQ$ carries a 
$C^\infty$-connection with holonomy group $\twH$ \cite[Ch.~II, Thm.~8.2]{KoNo}.
This connection prolongs to a unique (smooth) $A\in\con^k$ obeying
$P_A=\twQ$ and $H_A=\twH$. Then Lemma \rref{LThS}(a) and \gref{Gdefholind}
imply
$$
\Theta\left(\gau^{k+1}_A\right) = \twQ\cdot\rmC_G^2\left(\twH\right) = Q\,.
$$
This proves surjectivity.

To show that the projected map is injective, let
$\tau,\tau'\in\OT\left(\con^k,\gau^{k+1}\right)$. Choose representatives
$S,S'$ and assume that there exist $g\in\gau^{C^0}$, $a\in G$ such that
\begin{equation} \label{GThconj}
\Theta\left(S'\right)
=
\vartheta_g\left(\Theta\left(S\right)\right)\cdot a\,.
\end{equation}
Consider the following lemma.
\begin{Lemma} \label{Lsmoothing}
Let $A\in\con^k$ and let $Q\subseteq P$ be a subbundle of class $C^\infty$.
If there exists $h\in\gau^{C^0}$ such that
\begin{equation} \label{GQthh}
\Theta\left(\gau^{k+1}_A\right) = \vartheta_h(Q)
\end{equation}
then $h$ may be chosen from $\gau^{k+1}$.
\end{Lemma}
Before proving the lemma, let us assume that it holds and finish the
arguments. Again, due to smoothing theorems, $\Theta(S)$ is
$C^0$-isomorphic to some $C^\infty$-subbundle $Q\subseteq P$. Then there
exists $h\in\gau^{C^0}$ such that $\Theta(S)=\vartheta_h(Q)$. Due to Lemma
\rref{Lsmoothing}, we can choose $h\in\gau^{k+1}$. Moreover, due to 
\gref{GThconj},
$\Theta(S') = \vartheta_{gh}\left(Q\cdot a\right)$. By application of Lemma
\rref{Lsmoothing} again, we can achieve $gh\in\gau^{k+1}$. This shows
that we may assume, from the beginning, $g\in\gau^{k+1}$.

Now consider \gref{GThconj}. Since $p_0\in\Theta(S)$, $\vartheta_g(p_0)\cdot a=
p_0\cdot\left(g(p_0)a\right)\in\Theta(S')$. Since also $p_0\in\Theta(S')$,
$g(p_0)a$ is an element of the structure group of $\Theta(S')$. Then
$\Theta(S')\cdot\left(a^{-1}g(p_0)^{-1}\right) = \Theta(S')$, so that
\gref{GThconj} and Lemma \rref{LThS}(c) yield
$$
\Theta\left(S'\right)
= \vartheta_g\left(\Theta(S)\right)\cdot g(p_0)^{-1}
= \Theta\left(gSg^{-1}\right)\,.
$$
Due to Lemma \rref{LThS}(b), then $S'=gSg^{-1}$. This proves injectivity.
\\

{\it Proof of Lemma \rref{Lsmoothing}:} Let $A$ and $Q$ be given. Under the
assumption that \gref{GQthh} holds, $\Theta\left(\gau^{k+1}_A\right)$ and $Q$
have the same structure group $H$. There exist an open covering
$\{ U_i\}$ and local trivializations
$$
\xi_i:U_i\times H\rightarrow\Theta\left(\gau^{k+1}_A\right)|_{U_i}\,,
~~~
\eta_i:U_i\times H\rightarrow Q|_{U_i}
$$
of $\Theta\left(\gau^{k+1}_A\right)$ and $Q$, respectively. These define
local trivializations
$$
\twxi_i,\tweta_i:U_i\times G\rightarrow P|_{U_i}
$$
of $P$ over $\{ U_i \}$. Here $\eta_i$, $\tweta_i$ can be chosen from the
class $C^\infty$.
As for $\xi_i$ and $\twxi_i$, we note that $\Theta\left(\gau^{k+1}_A\right)$
contains the holonomy bundle $P_A$. A standard way of constructing local
cross sections in $P_A$
goes as follows (cf. the proof of Lemma 1 in \cite[Ch.~II,\S 7.1]{KoNo}): 
Choose a local chart about some $x\in M$ and a point $p$ in 
the fibre of $P_A$ over $x$. Take the pre-images, under
the local chart map, of straight lines running through $x$ and  lift
them to $A$-horizontal paths running through $p$. Since $A$ is of class $W^k$, the lifts are
of class $W^{k+1}$. Hence, so is the local cross section and, therefore, the
local trivialization defined by that cross section.
This shows that $\xi_i$ and $\twxi_i$ may be chosen from the class $W^{k+1}$.

Due to \gref{GQthh}, the family $\{ \vartheta_h\circ\eta_i \}$
defines a local trivialization of class $C^0$ of
$\Theta\left(\gau^{k+1}_A\right)$ over $\{ U_i \}$. Hence, there
exists a vertical automorphism $\vartheta'$ of class $C^0$ of
$\Theta\left(\gau^{k+1}_A\right)$ such that
$\xi_i = \vartheta'\circ\vartheta_h\circ\eta_i$ $\forall i$. By equivariant
prolongation, $\vartheta'$ defines a unique element
$h'\in\gau^{C^0}$. Since $\vartheta_{h'}$ leaves
$\Theta\left(\gau^{k+1}_A\right)$ invariant,
$\Theta\left(\gau^{k+1}_A\right)=\vartheta_{h'h}(Q)$. Thus, by
possibly redefining $h$ we may assume that $h'=\II$, i.e., that $\vartheta'$
is trivial. Then
\begin{equation} \label{Getai}
\twxi_i = \vartheta_h\circ\tweta_i~~~~\forall~i\,.
\end{equation}
As we shall argue now, \gref{Getai} implies $h\in\gau^{k+1}$. To see this,
note that $h\in\gau^{k+1}$ iff the local representatives
$h_i=h\circ\tweta_i\circ\iota$ are of class $W^{k+1}$, see \cite{MitterViallet}.
Here $\iota$ denotes the embedding $U_i\rightarrow U_i\times
G$, $x\mapsto(x,\II)$. Using $\tweta_i(x,h_i(x))=\vartheta_h\circ
\tweta_i(x,\II)$, $\forall~x\in U_i$,
we find
\begin{equation} \label{Ghi}
h_i=\pr_2\circ\tweta_i^{-1}\circ\vartheta_h\circ\tweta_i\circ \iota\,,
\end{equation}
where $\pr_2$ is the canonical projection $U_i\times G\rightarrow G$.
Inserting \gref{Getai} into \gref{Ghi} yields
$$
h_i=\pr_2\circ\tweta_i^{-1}\circ\twxi_i\circ \iota~~~~\forall~i\,.
$$
Here $\twxi_i$ is of class $W^{k+1}$ and the other maps are of class
$C^\infty$. Thus, according to the composition rules of Sobolev mappings,
$h_i$ is of class $W^{k+1}$. It follows $h\in\gau^{k+1}$. This proves
Lemma \rref{Lsmoothing} and, therefore, Theorem \rref{Totgau}.
\qed
\\

{\it Remarks:}
\\
1. As one important consequence of Theorem \rref{Totgau},
$\OT\left(\con^k,\gau^{k+1}\right)$ does not depend on $k$. 
\\
2. General arguments show that $\Howe_\ast(P)$ is countable, see 
\cite[\S 4.2]{KoRo}. Hence, so is $\OT\left(\con^k,\gau^{k+1}\right)$.
Countability of $\OT\left(\con^k,\gau^{k+1}\right)$ is a necessary condition 
for this set to define a stratification in the sense of \cite{KoRo:Strat}.
It was first stated in Theorem 4.2.1 in \cite{KoRo}. In fact, the
proof of this theorem already contains most of the arguments needed to
prove Theorem \rref{Totgau}. Unfortunately, although in the proof of Theorem
4.2.1 the authors used that isomorphy of evolution subbundles
implies conjugacy under the action of $\gau^{k+1}$, they did not give an
argument for that. Such an argument is provided by our Lemma 
\rref{Lsmoothing}.
\\
3. Theorem \rref{Totgau} also shows that the notion of holonomy-induced Howe subbundle
may be viewed as an abstract version of the notion of evolution subbundle
generated by a connection.
\\
4. The geometric ideas behind the proof of Theorem \rref{Totgau} are
also contained in \cite[\S 2]{Heil:OrbSpa}. However, a rigorous proof was
not given there.
\\

In view of Theorem \rref{Totgau}, we are left with the problem of determining
the set $\Howe_\ast(P)$ together with its partial ordering. This leads us to 
the following
\\

\noindent
{\bf Programme}
\\[0.5ex]
{\bf Step 1}\hspace{0.5cm}Determination of the Howe subgroups of $G$. Since
$G$-action on subbundles conjugates the structure group, classification
up to conjugacy is sufficient.

{\bf Step 2}\hspace{0.5cm}Determination of Howe subbundles of $P$. Since subbundles are
conjugate by $\gau^{C^0}$ iff they are isomorphic, classification up to
isomorphy is sufficient.

{\bf Step 3}\hspace{0.5cm}Specification of the Howe subbundles which are
holonomy-induced.

{\bf Step 4}\hspace{0.5cm}Factorization by $G$-action

{\bf Step 5}\hspace{0.5cm}Determination of the natural partial ordering.
\\

In the present paper, we perform steps 1--4 for the group $G=\rmSU n$. 
As already noted, the
determination of the natural partial ordering, which includes the study of
the natural partial ordering of Howe subgroups, will be published in a
subsequent paper.
\section{The Howe Subgroups of $\rmSU n$}
\label{SHSG}
Let $\Howe(\rmSU n)$ denote the set of conjugacy classes of Howe subgroups
of $\rmSU n$. In order to derive $\Howe(\rmSU n)$, we consider $\rmSU n$ as a 
subset of the general linear algebra $\rmgl(n,\CC)$, viewed as an associative
algebra.

In the literature it is customary to consider, instead of Howe subgroups,
{\it Howe dual pairs}. A Howe dual pair is an ordered pair of subgroups
$(H_1,H_2)$ of $G$ such that
$$
H_1=\rmC_G(H_2)\,, ~~~~~ H_2=\rmC_G(H_1)\,.
$$
The assignment $H\mapsto(H,\rmC_G(H))$ defines a $1:1$-relation between
Howe subgroups and Howe dual pairs. One also defines Howe
subalgebras and Howe dual pairs in an associative algebra. We remark that, usually, one
restricts attention to {\it reductive} Howe dual pairs. This means, one
requires that any finite dimensional representation of the members be
completely reducible. In our case this condition is automatically satisfied,
because $\rmSU n$ is compact and Howe subgroups are always closed. Reductive
Howe dual pairs play an important role in the representation theory
of Lie groups. This was first observed by R.~Howe \cite{Howe:Theta}. Although
for $\rmSU n$ it is not necessary to go into the details of the
classification theory of reductive Howe dual pairs, we note that there exist,
essentially, two methods. One applies to the isometry groups of Hermitian
spaces and uses the theory of Hermitian forms
\cite{Moeglin,Przebinda,meine}. The other method applies to complex 
semisimple Lie algebras and uses root space techniques \cite{Rubenthaler}.
\\

Let $\rmK(n)$ denote the collection of pairs of sequences (of equal
length) of positive integers 
$$ 
J=\left(\bfk,\bfm\right) =
\left((k_1,\dots, k_r),(m_1,\dots, m_r)
\right)\,,~~r=1,2,3,\dots,n,
$$
 which obey
\begin{equation} \label{Gsumcond}
\bfk \cdot\bfm = \sum_{i=1}^r k_im_i = n\,.
\end{equation}
For a given element $J=(\bfk,\bfm)$ of $\rmK(n)$, let $g$ denote the greatest 
common divisor of the integers $m_1,\dots, m_r$. Define 
a sequence $\twbfm=\left(\twm_1,\dots,\twm_r\right)$ by $g\twm_i=m_i$
$\forall~i$. Moreover, for any permutation $\sigma$ of $r$ elements, define 
$\sigma J=(\sigma\bfk,\sigma\bfm)$.

Any $J\in\rmK(n)$ generates a canonical decomposition
\begin{equation} \label{GdefdecompJ}
\CC^n=\left(\CC^{k_1}\otimes\CC^{m_1}\right)
\oplus\cdots\oplus
\left(\CC^{k_r}\otimes\CC^{m_r}\right)\,.
\end{equation}
This decomposition, in turn, induces an injective homomorphism
\begin{eqnarray} \nonumber
\rmgl(k_1,\CC)\times\cdots\times\rmgl(k_r,\CC)
& \rightarrow &
\rmgl(n,\CC)
\\ \label{Gstarep}
(D_1,\dots, D_r)
& \mapsto &
\left(D_1\otimes\II_{m_1}\right)
\oplus\cdots\oplus
\left(D_r\otimes\II_{m_r}\right)\,.
\end{eqnarray}
Written as a matrix w.r.t.~the canonical basis, the elements of the image look
like 
$$
\ddeckmatrix{
\vveckmatrix{D_1}{0}{0}{0}{D_1}{0}{0}{0}{D_1}
}{0}{0}{
\vveckmatrix{D_r}{0}{0}{0}{D_r}{0}{0}{0}{D_r}
}
$$
($m_1$ times $D_1$, $m_2$ times $D_2$, etc.).
We denote the image of this homomorphism by $\rmgl(J,\CC)$, its intersection
with $\rmU n$ by $\UJ$ and its intersection with $\rmSU n$ by $\SUJ$.
Note that $\UJ$ is the image of the restriction of
\gref{Gstarep} to $\rmU k_1\times\cdots\times\rmU k_r$.

\begin{Lemma} \label{LHSG}
~\\
{\rm (a)} A $\ast$-subalgebra of $\rmgl(n,\CC)$ is Howe if and only if it is
conjugate, under the action of $\rmSU n$ by inner automorphisms, to
$\rmgl(J,\CC)$ for some $J\in\rmK(n)$.

{\rm (b)} A subgroup of $\rmU n$ (resp. $\rmSU n$) is Howe if and only if it
is conjugate, under the action of $\rmSU n$ by inner automorphisms, to $\UJ$
(resp. $\SUJ$) for some $J\in\rmK(n)$.
\end{Lemma}
{\it Remark:} Assertion (a) is a version of the structure theorem for finite-dimensional 
von Neumann algebras.
\\

{\it Proof:}
(a) For any $J\in\rmK(n)$, $\rmgl(J,\CC)$ is a $\ast$-subalgebra of $\rmgl(n,\CC)$
containing $\II_n$. Thus, $\rmgl(J,\CC)$ is a von Neumann algebra and the
Double Commutant Theorem says that $\rmgl(J,\CC)$ is
Howe. Conversely, let $L$ be a Howe $\ast$-subalgebra of $\rmgl(n,\CC)$. Let
$L'$ denote the centralizer of $L$ in $\rmgl(n,\CC)$ and let $\twL$ denote the 
subalgebra generated by $L$ and $L'$. Decompose
\begin{equation} \label{GdecompCn}
\CC^n=V_1\oplus\cdots\oplus V_r
\end{equation}
into mutually orthogonal, $\twL$-irreducible subspaces. Due to Schur's
Lemma, for each $i$,
\begin{equation} \label{GdecompVi}
V_i=W_i\otimes\CC^{m_i}\,,
\end{equation}
where $W_i$ is $L$-irreducible and $m_i$ is a positive integer. Moreover,
\begin{eqnarray}\label{GLpVi}
L'|_{V_i} 
& \subseteq &
\left\{\id_{W_i}\otimes D_i'~|~D_i'\in\rmgl(m_i,\CC)\right\}\,,
\\ 
\label{GLVi}
L|_{V_i} 
& \subseteq &
\left\{D_i\otimes\II_{m_i}~|~D_i\in\End(W_i)\right\}\,.
\end{eqnarray}
Since $L'$ is the full centralizer of $L$, one has equality in \gref{GLpVi}.
As a consequence, since $L$ is the full centralizer of $L'$, there holds
equality in \gref{GLVi}. Now let $k_i=\dim W_i$ and $J=(\bfk,\bfm)$. By 
identifying $W_i\cong\CC^{k_i}$, we arrive at a decomposition of $\CC^n$ of
the form \gref{GdefdecompJ}. There exists $T\in\rmSU n$ transforming this
decomposition into the canonical one induced by $J$. Then the inner
automorphism of $\rmgl(n,\CC)$ defined by $T$ transforms $L$ into
$\rmgl(J,\CC)$. 

(b) We give only the proof for $\rmSU n$. Let $H$ be a Howe subgroup of 
$\rmSU n$. Then
$$
H=\rmC_{\rmSU n}(K)=\rmC_{\rmgl(n,\CC)}(K)\cap\rmSU n
$$
for some subset $K\subseteq\rmSU n$. By construction, $\rmC_{\rmgl(n,\CC)}(K)$
is a Howe $\ast$-subalgebra of $\rmgl(n,\CC)$. Hence, by virtue of (a), it is 
conjugate, under $\rmSU n$-action, to $\rmgl(J,\CC)$ for some $J$. Then $H$ is 
conjugate, in $\rmSU n$, to $\rmgl(J,\CC)\cap\rmSU n=\SUJ$.
Conversely, let $J\in\rmK(n)$. Due to (a), there exists $J'\in\rmK(n)$ and 
$T\in\rmSU
n$ such that $\rmgl(J,\CC)=\rmC_{\rmgl(n,\CC)}\left(T\rmgl(J',\CC)T^{-1}\right)$. Taking
into account that $\left(T\rmgl(J',\CC)T^{-1}\right)\cap\rmSU n$ spans
$\left(T\rmgl(J',\CC)T^{-1}\right)$, we obtain
$$
\begin{array}{rcl}
\SUJ
& = &
\rmgl(J,\CC)\cap\rmSU n
\\
& = &
\rmC_{\rmgl(n,\CC)}\left(T\rmgl(J',\CC)T^{-1}\cap\rmSU n\right)\cap\rmSU n
\\
& = &
\rmC_{\rmSU n}\left(T\rmgl(J',\CC)T^{-1}\cap\rmSU n\right)\,.
\end{array}
$$
This shows that $\SUJ$ is Howe.
\qed

\begin{Lemma}\label{Lcjg}
Let $J,J'\in\rmK(n)$. Then $\rmgl(J,\CC)$ and $\rmgl(J',\CC)$ are conjugate under
the action of $\rmSU n$ by inner automorphisms if and
only if there exists a permutation $\sigma$ of $1,\dots, r$ such that
$J'=\sigma J$.
\end{Lemma}
{\it Proof:} Let $T\in\rmSU n$ such that
$\rmgl(J',\CC)=T^{-1}\rmgl(J,\CC) T$. Then
$\rmgl(J,\CC)$ and $\rmgl(J',\CC)$ are isomorphic. Hence, there
exists a permutation $\sigma$ of $1,\dots, r$ such that
$\bfk'=\sigma\bfk$. Moreover, $T$ is an isomorphism of the
representations $$
\begin{array}{l}
\rmgl(k_1,\CC)\times\cdots\times\rmgl(k_r,\CC)
\stackrel{J}{\longrightarrow}
\rmgl(n,\CC)\,,
\\
\rmgl(k_1,\CC)\times\cdots\times\rmgl(k_r,\CC)
\stackrel{\sigma}{\longrightarrow}
\rmgl(k_{\sigma(1)},\CC)\times\cdots\times\rmgl(k_{\sigma(r)},\CC)
\stackrel{J'}{\longrightarrow}\rmgl(n,\CC)\,,
\end{array}
$$
where $J$, $J'$ indicate the embeddings \gref{Gstarep} defined by $J$ and
$J'$, respectively. In particular, it does not change the multiplicities
of the irreducible factors. This implies $\bfm'=\sigma\bfm$. It follows
$J'=\sigma J$.

Conversely, assume that $J'=\sigma J$ for some
permutation $\sigma$ of $1,\dots, r$. Consider the canonical decompositions
\gref{GdefdecompJ} defined by $J$ and $J'$, respectively. There exists
$T\in\rmSU n$ which maps the factors
$\CC^{k'_i}\otimes\CC^{l'_i}$ of the second decomposition identically onto 
the factors $\CC^{k_{\sigma(i)}}\otimes\CC^{m_{\sigma(i)}}$ of the
first one, for $i=1,\dots, r$. It is not difficult to see that
$\rmgl(J',\CC)=T^{-1}\rmgl(J,\CC)T$.
\qed
\\

As a consequence of Lemma \rref{Lcjg}, we introduce an equivalence relation
on the set $\rmK(n)$ by writing $J\sim J'$ iff there exists a permutation of
$1,\dots, r$ such that $J'=\sigma J$. Let $\hat{\rmK}(n)$ denote the set of
equivalence classes.
\begin{Theorem} \label{THSG}
The assignment $J\mapsto\SUJ$ induces a bijection from $\hat{\rmK}(n)$ onto
$\Howe(\rmSU n)$.
\end{Theorem}
{\it Proof:} According to Lemma \rref{LHSG}, the assignment $J\mapsto\SUJ$
induces a surjective map $\rmK(n)\rightarrow\Howe(\rmSU n)$. Due to Lemma
\rref{Lcjg}, this map projects to $\hat{\rmK}(n)$ and the projected map is injective.
\qed
\\

This concludes the classification of Howe subgroups of $\rmSU n$, i.e., Step
$1$ of our programme.
\section{The Howe Subbundles of $\rmSU n$-Bundles}
\label{SHSB}
In this chapter, let $P$ be a principal $\rmSU n$-bundle over $M$,
$\dim M\leq 4$.
We are going to derive the Howe subbundles of $P$ up to
isomorphy. Since the action of $\rmSU n$ on subbundles conjugates the
structure group, it suffices to consider reductions of $P$ to the subgroups
$\SUJ$, $J\in\rmK(n)$.
Thus, let $J\in\rmK(n)$ be fixed. Let $\Bun{M}{\SUJ}$ denote the set
of isomorphism classes of principal $\SUJ$-bundles over $M$ (where
principal bundle isomorphisms are assumed to commute with the structure
group action and to project to the identical map on the base space). 
Moreover, let $\Red{P}{\SUJ}$ denote the set of
isomorphism classes of reductions of $P$ to the subgroup $\SUJ\subseteq\rmSU
n$.

We shall first derive a description of $\Bun{M}{\SUJ}$ in terms of suitable
characteristic classes and then give a characterization of the subset
$\Red{P}{\SUJ}$.
The classification of $\Bun{M}{\SUJ}$ will be performed by
constructing the Postnikov tower of the classifying space $\BSUJ$ up to level
$5$. For the convenience of the reader, the basics of this method will be
briefly explained below.

Note that in the sequel maps of topological spaces are always assumed to be 
continuous, without explicitly stating this.
\subsection{Preliminaries}
\label{ShsbSSprelims}
\paragraph{Universal Bundles and Classifying Spaces}

Let $G$ be a Lie group. As a basic fact in bundle theory, there exists a
so-called {\it universal $G$-bundle}
\begin{equation} \label{Guvlbun}
G\hookrightarrow\rmE G\rightarrow\rmB G
\end{equation}
with the following property: For any \CW-complex (hence, in particular, any
manifold) $X$ the assignment
\begin{equation} \label{Ghtpclf}
[X,\rmB G]\longrightarrow\Bun{X}{G}\,,~~f\mapsto f^\ast\rmE G\,,
\end{equation}
is a bijection \cite{Husemoller}. Here $[\cdot,\cdot]$ means the
set of homotopy classes of maps and $f^\ast$ denotes the pull-back
of bundles. Both $\rmE G$ and $\rmB G$ can be realized as
\CW-complexes. They are unique up to homotopy equivalence. $\rmB
G$ is called the {\it classifying space of $G$}. The homotopy
class of maps $X\rightarrow\rmB G$ associated to $P\in\Bun{X}{G}$
by virtue of \gref{Ghtpclf} is called the {\it classifying map of
$P$}. We denote it by $\ka{P}$. Note that a principal $G$-bundle
is universal iff its total space is contractible. In particular,
\begin{equation} \label{GGandBG}
\pi_i(G)\cong\pi_{i+1}(\rmB G)\,,~~i=0,1,2,\dots\,.
\end{equation}
This is an immediate consequence of the exact homotopy sequence of fibre
spaces \cite{Bredon:Top}.
\paragraph{Associated Principal Bundles Defined by Homomorphisms}

Let $\varphi:G\rightarrow G'$ be a Lie group homomorphism and let
$P\in\Bun{X}{G}$. By virtue of the action
$$
G\times G'\rightarrow G'\,,~~(a,a')\mapsto\varphi(a)a'\,,
$$
$G'$ becomes a left $G$-space and we have an associated bundle
$\ab{P}{\varphi} = P\times_G G'$. Observe that $\ab{P}{\varphi}$ can be
viewed as a principal $G'$-bundle with right $G'$-action
$$
\ab{P}{\varphi}\times G'\longrightarrow\ab{P}{\varphi}\,,~~
\left([(p,a')],b'\right)\mapsto [(p,a'b')]\,.
$$
One has the natural bundle morphism
\begin{equation} \label{Gnatbumor}
\psi:P\rightarrow\ab{P}{\varphi}\,,~~p\mapsto[(p,\II_{G'})]\,.
\end{equation}
It obeys $\psi(p\cdot a)=\psi(p)\cdot\varphi(a)$ and projects to the
identical map on $X$.

In the special case where $\varphi$ is a Lie subgroup
embedding, the natural bundle morphism \gref{Gnatbumor} is an embedding of
$P$ onto a subbundle of $\ab{P}{\varphi}$. In this case, if no confusion about 
$\varphi$ can arise, we shall often write $\ab{P}{G'}$ instead of
$\ab{P}{\varphi}$. Note that $\ab{P}{\varphi}$ is the extension of $P$ 
by $G'$ and $P$ is a reduction of $\ab{P}{\varphi}$ to the subgroup 
$(G,\varphi)$.
\paragraph{Classifying Maps Associated to Homomorphisms}

Again, let $\varphi:G\rightarrow G'$ be a homomorphism. There
exists a map $\rmB\varphi:\rmB G\rightarrow\rmB G'$, associated to
$\varphi$, which is defined as the classifying map of the
principal $G'$-bundle $\ab{(\rmE G)}{\varphi}$ associated to the
universal $G$-bundle $\rmE G$. It has the following functorial
property: For $\varphi:G\rightarrow G'$ and
$\varphi':G'\rightarrow G''$ there holds
\begin{equation} \label{GBphipsi}
\rmB(\psi\circ\varphi)=\rmB\psi\circ\rmB\varphi\,.
\end{equation}
Using $\rmB\varphi$, the classifying map of $\ab{P}{\varphi}$ can be
expressed through that of $P$:
\begin{equation} \label{Gclfmapassbun}
\ka{\ab{P}{\varphi}} = \rmB\varphi\circ\ka{P}\,.
\end{equation}
In the special case where $\varphi$ is a normal Lie subgroup 
embedding, the short exact sequence of Lie group homomorphisms
\begin{equation}\label{Gshexsqc}
\II
\longrightarrow G
\stackrel{\varphi}{\longrightarrow} G'
\stackrel{p}{\longrightarrow} G/G'
\longrightarrow \II
\end{equation}
induces a principal bundle
\begin{equation} \label{GBphibun}
G'/G\hookrightarrow\rmB G\stackrel{\rmB\varphi}{\longrightarrow}\rmB G'\,.
\end{equation}
The classifying map of this bundle is $\rmB p$ \cite{Borel}, where $p$
denotes the natural projection.
\paragraph{Characteristic Classes}

Let $G$ be a Lie group. Consider the cohomology ring $H^\ast(\rmB G,\pi)$ of
the classifying space with values in some Abelian group $\pi$. For any
$P\in\Bun{X}{G}$, the homomorphism $\left(\ka{P}\right)^\ast$, induced on cohomology, maps
$H^\ast(\rmB G,\pi)$ to $H^\ast(X,\pi)$.
Therefore, given $\gamma\in H^\ast(\rmB G,\pi)$, one can define a map
\begin{equation} \label{Gccasmap}
\chi_\gamma:\Bun{X}{G}\rightarrow
H^\ast(X,\pi)\,,~~~P\mapsto\left(\ka{P}\right)^\ast\gamma\,.
\end{equation}
This is called the {\it characteristic class} for $G$-bundles over $X$
defined by $\gamma$. By construction, one has the following universal
property of characteristic classes:
Let $f:X\rightarrow X'$ be a map and let $P'\in\Bun{X'}{G}$. Then
\begin{equation} \label{GccfastB}
\chi_\gamma\left(f^\ast P'\right) = f^\ast\chi_\gamma\left(P'\right)\,.
\end{equation}
Observe that if two bundles are
isomorphic then their images under arbitrary characteristic classes coincide,
whereas the converse, in general, does not hold. This is due to the fact that
characteristic classes
can control maps $X\rightarrow\rmB G$ only on the level of the homomorphisms
induced on cohomology. In general, the latter do not give sufficient
information on the homotopy properties of the maps. In certain cases,
however, they do. For example, such cases are obtained by specifying $G$ to
be $\rmU 1$ or discrete, or by restricting $X$ in dimension.
In these cases there exist sets of characteristic classes which classify
$\Bun{X}{G}$. In the sequel we will utilize this to determine
$\Bun{M}{\SUJ}$.
\paragraph{Eilenberg-MacLane Spaces}
Let $\pi$ be a group and $n$ a positive integer. An arcwise connected
\CW-complex $X$ is called an {\it Eilenberg-MacLane space of type
$K(\pi,n)$} iff $\pi_n(X)=\pi$ and $\pi_i(X)=0$ for $i\neq n$.
Eilenberg-MacLane spaces exist for any choice of $\pi$ and $n$, provided
$\pi$ is commutative for $n\geq 2$. They are unique up to homotopy
equivalence.

The simplest example of an Eilenberg-MacLane space is the $1$-sphere
$\sphere{1}$, which is of type $K(\ZZ,1)$. Two further examples, $K(\ZZ,2)$
and $K(\ZZ_g,1)$, are briefly discussed in Appendix \rref{AEMLS}.
Note that Eilenberg-MacLane
spaces are, apart from very special examples, infinite dimensional.
Also note that, up to homotopy equivalence one has
\begin{equation} \label{GprodEMLS}
K(\pi_1\times\pi_2,n)=K(\pi_1,n)\times K(\pi_2,n)\,.
\end{equation}
Now assume $\pi$ to be commutative also in the case $n=1$.
Due to the Universal Coefficient Theorem, 
$\Hom\left(H_n(K(\pi,n)),\pi\right)$ is isomorphic to a subgroup of 
$H^n(K(\pi,n),\pi)$. Due to the Hurewicz Theorem, $H_n(K(\pi,n))
\cong\pi_n(K(\pi,n))=\pi$. It follows that $H^n(K(\pi,n),\pi)$ contains
elements which correspond to isomorphisms $H_n(K(\pi,n))\rightarrow\pi$. Such
elements are called {\it characteristic}.
If $\gamma\in H^n(K(\pi,n),\pi)$ is characteristic then for any \CW-complex
$X$, the map
\begin{equation} \label{G[]=H}
[X,K(\pi,n)]\rightarrow H^n(X,\pi)\,, ~~f\mapsto f^\ast\gamma\,,
\end{equation}
is a bijection \cite[\S VII.12]{Bredon:Top}. In this sense,
Eilenberg-MacLane spaces provide a link between homotopy properties and
cohomology. We remark that the bijection
\gref{G[]=H} induces an Abelian group structure on the set $[X,K(\pi,n)]$.
\paragraph{Path-Loop-Fibration}
Let $X$ be an arcwise connected topological space. Consider the path-loop
fibration over $X$
\begin{equation} \label{Gpathloopfibr}
\Omega(X)\hookrightarrow P(X)\longrightarrow X\,,
\end{equation}
where $\Omega(X)$ and $P(X)$ denote the loop space and the path space of
$X$, respectively (both based at some point $x_0\in X$). Since $P(X)$ is 
contractible, the exact homotopy sequence induced by the fibration
\gref{Gpathloopfibr} implies
$\pi_i(\Omega(X))\cong \pi_{i+1}(X)$, $i=0,1,2,\dots$. Thus, for
$X=K(\pi,n)$, $$
\pi_i\left(\Omega(K(\pi,n+1))\right)\cong\pi_{i+1}\left(K(\pi,n+1)\right)=
\left\{\begin{array}{cll} \pi & | & i=n \\ 0 & | &
\mbox{otherwise}
\end{array}\right.
$$
Hence, $\Omega(K(\pi,n+1))=K(\pi,n)$ $\forall n$ and the path-loop fibration
over $K(\pi,n+1)$ reads
\begin{equation} \label{GpathloopfibrEMLS}
K(\pi,n)\hookrightarrow P(K(\pi,n+1))\longrightarrow K(\pi,n+1)\,.
\end{equation}

\paragraph{Postnikov Tower}
\label{CHSBSprelimsSSPost}
A map $f:X\rightarrow X'$ of topological spaces is called an
{\it $n$-equivalence} iff the homomorphism induced on homotopy groups
$f_\ast:\pi_i(X)\rightarrow\pi_i(X')$ is an isomorphism for $i<n$ and
surjective for $i=n$. One also defines the notion of an $\infty$-equivalence,
which is often called weak homotopy equivalence.

Let $f:X\rightarrow X'$ be an $n$-equivalence and let $Y$ be a \CW-complex.
Then the map $[Y,X]\rightarrow[Y,X']$, $g\mapsto f\circ g$, is bijective for
$\dim Y<n$ and surjective for $\dim Y=n$
\cite[Ch.~VII,~Cor.~11.13]{Bredon:Top}.

A \CW-complex $Y$ is called {\it $n$-simple} iff it is arcwise connected and the
action of $\pi_1(Y)$ on $\pi_i(Y)$ is trivial for $1\leq i\leq n$. It is
called {\it simple} iff it is $n$-simple for all $n$.

The following theorem describes how a simple \CW-complex can be approximated 
by $n$-equivalent spaces constructed from Eilenberg-MacLane spaces.
\begin{Theorem} \label{TPostnikov}
Let $Y$ be a simple \CW-complex. There exist
\\
{\rm (a)} a sequence of \CW-complexes $Y_n$ and principal fibrations
\begin{equation} \label{GPostnikovprifibr}
K(\pi_n(Y),n)\hookrightarrow Y_{n+1}\stackrel{q_n}{\longrightarrow}Y_n\,,
~~n=1,2,3,\dots\,,
\end{equation}
induced by maps $\theta_n:Y_n\rightarrow K\left(\pi_n(Y),n+1\right)$,
\\
{\rm (b)} a sequence of $n$-equivalences $y_n:Y\rightarrow Y_n$,
$n=1,2,3,\dots$, 
\\
such that $Y_1=\onepoint$ (one point space) and $q_n\circ y_{n+1}=y_n$ for all 
$n$.
\end{Theorem}
{\it Proof:}
The assumption that $Y$ be a simple \CW-complex implies that the constant map
$Y\rightarrow\onepoint$ is a simple map (see \cite[Ch.~VII,~Def.~13.4]{Bredon:Top} 
for a definition of the latter). Thus, the assertion is a consequence of a more 
general theorem about simple maps given in \cite[Ch.~VII,~Thm.~13.7]{Bredon:Top}.
\qed
\\

{\it Remarks:}

1. The sequence of spaces and maps $\left(Y_n,y_n,q_n\right)$,
$n=1,2,3,\dots$, is called a {\it Postnikov tower}, or {\it
Postnikov system}, or {\it Postnikov decomposition} of $Y$. 

2. For the principal fibrations \gref{GPostnikovprifibr} to be induced by a
map $\theta_n:Y_n\rightarrow K(\pi_n(Y),n+1)$ means that they are given as
pull-back of the path-loop fibration \gref{GpathloopfibrEMLS} over
$K(\pi_n(Y),n+1)$.
\\

The theorem allows one to successively construct $Y_n$ for given $n$,
starting from $Y_1=\onepoint$. For example, for $n=5$ such constructions have 
been
carried out for $Y=\rmB\rmU k$ in \cite[\S 4.2]{AvisIsham}, or for
$Y=\rmB\rmP\rmU k$, where $\rmP\rmU k$ denotes the projective unitary group,
in \cite{Woodward}. In the sequel, we shall construct $\left(\BSUJ\right)_5$.
For this purpose, we need information about the low-dimensional
homotopy groups of $\SUJ$.

\subsection{The Homotopy Groups of $\SUJ$}
For $a$ being a positive integer, denote the canonical embedding 
$\ZZ_a\hookrightarrow\rmU 1$ 
by $j_a$ and the endomorphism of $\rmU 1$ mapping $z\mapsto z^a$ by $p_a$. 
Moreover, let $j_J$ and $i_J$ denote the natural embeddings 
$\SUJ\hookrightarrow\UJ$ and $\UJ\hookrightarrow\rmU n$, respectively. 

Using the natural projections $\prU{J}{i}:\UJ\rightarrow\rmU k_i\,$, we 
define a homomorphism
\begin{equation} \label{GdeflU}
\lambda_J:\rmU J\rightarrow\rmU 1\,,~~D \mapsto
\prod\nolimits_{i=1}^r p_{\twm_i}\circ\det\nolimits_{\rmU
k_i}\circ\,\prU{J}{i}(D)\,.
\end{equation}
The following diagram commutes:
\begin{equation}\label{Gctvdgrdet}
\resetparms
\setsqparms[1`1`1`1;700`300]
\Square[\UJ`\rmU n`\rmU 1`\rmU 1;i_J`\lambda_J`\det\nolimits_{\rmU
n}`p_g]
\end{equation}
Accordingly, the restriction of $\lambda_J$ to the subgroup $\SUJ$ takes
values in $j_g(\ZZ_g)\subseteq\rmU 1$. Thus, we can define a homomorphism
$\lambda^\rmS_J:\SUJ\rightarrow\ZZ_g$ by the following commutative diagram:
\begin{equation}\label{Gctvdgr}
\resetparms
\setsqparms[1`1`1`1;700`300]
\Square[\SUJ`\UJ`\ZZ_g`\rmU 1;j_J`\lambda^\rmS_J`\lambda_J`j_g]
\end{equation}
Let $(\SUJ)_0$ denote the arcwise connected component of the identity. Note that 
it is also a connected component, because $\SUJ$ is a closed subgroup of 
$\rmGL(n,\CC)$.
\begin{Lemma} \label{Lcoco}
The homomorphism $\lambda^\rmS_J$ projects to an isomorphism
$\SUJ/(\SUJ)_0\rightarrow\ZZ_g$.
\end{Lemma}
{\it Proof:} Consider the homomorphism $\lambda^\rmS_J:\SUJ\rightarrow\ZZ_g$.
The target space being discrete, $\lambda^\rmS_J$ must be constant on
connected components. Hence $(\SUJ)_0\subseteq\ker\lambda^\rmS_J$, so that
$\lambda^\rmS_J$ projects to a homomorphism $\SUJ/(\SUJ)_0\rightarrow\ZZ_g$.
The latter is surjective, because $\lambda^\rmS_J$ is surjective. To prove 
injectivity, we show $\ker\lambda^\rmS_J\subseteq(\SUJ)_0$.
Let $D\in\ker\lambda^\rmS_J$ and denote $D_i=\prU{J}{i}\circ j_J (D)$.
Define a homomorphism
$$
\varphi:\rmU 1^r\rightarrow\rmU
1\,,~(z_1,\dots,z_r)\mapsto z_1^{\twm_1}\cdots
z_r^{\twm_r}\,.
$$
Then 
$$
\lambda^\rmS_J(D)=\varphi\left(\det\nolimits_{\rmU k_1}D_1,\dots,
\det\nolimits_{\rmU k_r}D_r\right)\,.
$$
By assumption, $(\det\nolimits_{\rmU k_1}D_1,\dots,\det\nolimits_{\rmU k_r}
D_r)\in\ker\varphi$.
Since the exponents defining $\varphi$ have greatest common divisor 1,
$\ker\varphi$ is connected. Thus, there exists a path
$(\gamma_1(t),\dots,\gamma_r(t))$ in $\ker\varphi$ running from
$(\det\nolimits_{\rmU
k_1} D_1,\dots,\det\nolimits_{\rmU k_r} D_r)$ to $(1,\dots, 1)$.
For each $i=1,\dots, r$, define a path $G_i(t)$ in $\rmU k_i$ as follows:
First, go from $D_i$ to $(\det\nolimits_{\rmU k_i} D_i)\oplus\II_{k_i-1}$,
keeping the determinant constant, thus using connectedness of $\rmSU k_i$.
Next, use the path \linebreak $\gamma_i(t)
\oplus\II_{k_i-1}$ to get to $\II_{k_i}$. By construction, the image of
$(G_1(t),\dots, G_r(t))$ under the embedding \gref{Gstarep} is a path in
$\SUJ$ connecting $D$ with $\II_n$. This proves 
$\ker\lambda^\rmS_J\subseteq(\SUJ)_0$.
\qed
\begin{Theorem} \label{Thtpgr}
The homotopy groups of $\SUJ$ are
$$
\pi_i(\SUJ)\cong\left\{\begin{array}{ccl}
\ZZ_g & | & i=0
\\
\ZZ^{\oplus(r-1)} & | & i=1
\\
\pi_i(\rmU k_1)\oplus\cdots\oplus\pi_i(\rmU k_r) & | & i>1\,.
\end{array}\right.
$$
In particular, $\pi_1(\SUJ)$ and $\pi_3(\SUJ)$ are torsion-free.
\end{Theorem}
{\it Proof:} For $i=0$, $\pi_0(\SUJ)=\SUJ/(\SUJ)_0$. This group is given by
Lemma \rref{Lcoco}. For $i>1$, the assertion follows immediately
from the exact homotopy sequence induced by the bundle $\SUJ\hookrightarrow
\UJ\stackrel{\det_{\rmU n}}{\longrightarrow}\rmU 1$. For $i=1$, consider the 
following portion of this sequence:
$$
\arraycolsep0.2em
\begin{array}{ccccccccccc}
\pi_2(\rmU 1) & \longrightarrow &
\pi_1(\SUJ) & \longrightarrow &
\pi_1(\UJ) & \stackrel{(\det_{\rmU n})_\ast}{\longrightarrow} &
\pi_1(\rmU 1) & \longrightarrow &
\pi_0(\SUJ) & \longrightarrow &
\pi_0(\UJ)
\\
0 & \longrightarrow &
\pi_1(\SUJ) & \longrightarrow &
\ZZ^{\oplus r} & \longrightarrow &
\ZZ & \longrightarrow &
\ZZ_g & \longrightarrow &
0\,.
\end{array}
$$
One has $\ZZ^{\oplus r}/\ker((\det_{\rmU n})_\ast)\cong\im((\det_{\rmU n})_\ast)$.
Exactness implies $\ker((\det_{\rmU n})_\ast)\cong\pi_1(\SUJ)$ and
$\im((\det_{\rmU n})_\ast)=g\ZZ\cong\ZZ$. 
It follows $\pi_1(\SUJ)\cong\ZZ^{\oplus (r-1)}$, as asserted.
\qed
\subsection{The Postnikov Tower of $\BSUJ$ up to Level 5}
\label{CHSBScc}
Let $r^\ast$ denote the number of indices $i$ for which $k_i>1$. 
\begin{Theorem} \label{TBG5}
The $5$th level of the Postnikov tower of $\BSUJ$ is given by
\begin{equation} \label{GBG5}
(\BSUJ)_5=K(\ZZ_g,1)\times \prod_{j=1}^{r-1}K(\ZZ,2)\times 
\prod_{j=1}^{r^\ast}K(\ZZ,4)\,.
\end{equation}
\end{Theorem}
{\it Proof:}
First, we check that $\BSUJ$ is a simple space. To see this, note that any 
inner automorphism of $\SUJ$ is generated by an
element of $(\SUJ)_0$, hence is homotopic to the identity automorphism.
Consequently, the natural action of $\pi_0(\SUJ)$ on $\pi_{i-1}(\SUJ)$,
$i=1,2,3,\dots$ , induced by inner automorphisms, is trivial. Since the
natural isomorphisms $\pi_{i-1}(\SUJ)\cong\pi_i(\BSUJ)$ transform this 
action into that of $\pi_1(\BSUJ)$ on $\pi_i(\BSUJ)$,
the latter is trivial, too. Thus, we can apply Theorem \rref{TPostnikov} 
to construct the Postnikov tower of $\BSUJ$ up to level $5$. According to 
Theorem \rref{Thtpgr}, the relevant homotopy groups are
\begin{equation} \label{Ghtpgr}
\pi_1(\BSUJ)=\ZZ_g\,,~~~~~
\pi_2(\BSUJ)=\ZZ^{\oplus (r-1)}\,,~~~~~
\pi_3(\BSUJ)=0\,,~~~~~
\pi_4(\BSUJ)=\ZZ^{\oplus r^\ast}\,.
\end{equation}
Moreover, we note that $H^\ast(K(\ZZ,2),\ZZ)$ is torsion-free, and that
\begin{equation}\label{GHiEMLS}
H^{2i+1}(K(\ZZ,2),\ZZ) = 0\,,~~~~H^{2i+1}(K(\ZZ_g,1),\ZZ) = 0\,,~~~~
i=0,1,2,\dots\,,
\end{equation}
see Appendix \rref{AEMLS}. We start with $(\BSUJ)_1=\onepoint$.

$(\BSUJ)_2:$~~ Being a fibration over $(\BSUJ)_1$, $(\BSUJ)_2$ must
coincide with the fibre:
\begin{equation} \label{GBG2}
(\BSUJ)_2=K(\ZZ_g,1)\,.
\end{equation}
$(\BSUJ)_3:$~~ In view of \gref{GBG2} and \gref{Ghtpgr}, $(\BSUJ)_3$ is the
total space of a fibration
\begin{equation} \label{GBSUJ3}
K(\ZZ^{\oplus (r-1)},2)\hookrightarrow(\BSUJ)_3\stackrel{q_2}{\longrightarrow}
K(\ZZ_g,1)
\end{equation}
induced from the path-loop fibration over $K(\ZZ^{\oplus (r-1)},3)$ by
some map $\theta_2:K(\ZZ_g,1)\rightarrow K(\ZZ^{\oplus (r-1)},3)$. Note that
$K(\ZZ^{\oplus (r-1)},n)=\prod_{j=1}^{r-1}K(\ZZ,n)$ $\forall n$. Then, due to 
\gref{G[]=H},
$$
[K(\ZZ_g,1),K(\ZZ^{\oplus (r-1)},3)]=\prod_{i=1}^{r-1} H^3(K(\ZZ_g,1),\ZZ)\,.
$$
Here the rhs. is trivial by \gref{GHiEMLS}. Hence, $\theta_2$ is
homotopic to a constant map, so that the fibration \gref{GBSUJ3} is trivial.
It follows
\begin{equation} \label{GBG3}
(\BSUJ)_3=K(\ZZ_g,1)\times\prod_{j=1}^{r-1}K(\ZZ,2)\,.
\end{equation}
$(\BSUJ)_4:$~~ In view of \gref{Ghtpgr}, $(\BSUJ)_4$ is given by a
fibration over $(\BSUJ)_3$ with fibre $K(0,3)=\onepoint$. Hence, it just
coincides with the base space.

$(\BSUJ)_5:$~~ According to \gref{Ghtpgr} and \gref{GBG3}, $(\BSUJ)_5$ is the
total space of a fibration
\begin{equation}\label{GfibrBG5}
K\left(\ZZ^{\oplus r^\ast},4\right)\hookrightarrow(\BSUJ)_5\stackrel{q_4}{\longrightarrow}
K\left(\ZZ_g,1\right)\times\prod_{j=1}^{r-1}K(\ZZ,2)\,,
\end{equation}
which is induced by a map $\theta_4$ from the base to $K\left(\ZZ^{\oplus
r^\ast},5\right)$.
Similarly to the case of $\theta_2$, 
\begin{equation} \label{G[BG3]}
\left[K(\ZZ_g,1)
\times
\prod_{j=1}^{r-1}K(\ZZ,2),K\left(\ZZ^{\oplus r^\ast},5\right)\right]
=
\prod_{i=1}^{r^\ast} H^5\left(
K(\ZZ_g,1)\times\prod_{j=1}^{r-1}K(\ZZ,2),\ZZ
\right)\,.
\end{equation}
Since $H^\ast(K(\ZZ,2),\ZZ)$ is torsion-free, we can apply the K\"unneth Theorem 
for cohomology \cite[Ch.~XIII, Cor.~11.3]{Massey} to obtain
\begin{eqnarray}\nonumber
&&
H^5\left(K(\ZZ_g,1)\times\prod_{j=1}^{r-1}K(\ZZ,2),\ZZ\right)
\\ \nonumber
&&
\cong
\bigoplus_{{j,j_1,\dots, j_{r\mbox{\tiny -}1}}\atop
{j+j_1+\cdots+ j_{r-1}=5}}
H^j(K(\ZZ_g,1),\ZZ)\otimes
H^{j_1}(K(\ZZ,2),\ZZ)\otimes\cdots\otimes H^{j_{r-1}}(K(\ZZ,2),
\ZZ)\,.
\end{eqnarray}
By \gref{GHiEMLS}, each summand of the rhs. is trivial, because it contains a 
tensor factor of odd degree. Hence, \gref{G[BG3]} is trivial, and so is the
fibration \gref{GfibrBG5}. This proves the assertion.
\qed
\\

The fact that $(\BSUJ)_5$ is a direct product of Eilenberg-MacLane spaces
immediately yields the following corollary.
\begin{Corollary}\label{Cccsuffice}
Let $J\in\rmK(n)$ and $\dim M\leq 4$. Let $P,
P'\in\Bun{M}{\SUJ}$. Assume that for any characteristic class $\alpha$
defined by an element of $H^1(\BSUJ,\ZZ_g)$, $H^2(\BSUJ,\ZZ)$, or
$H^4(\BSUJ,\ZZ)$ there holds $\alpha(P)=\alpha(P')$. Then $P$ and $P'$ are
isomorphic.
\end{Corollary}
{\it Proof:} Let $\pr_1$, $\pr_{21},\dots,\pr_{2r\mbox{\scriptsize -}1}$, and
$\pr_{41},\dots,\pr_{4{r^\ast}}$ denote the natural projections of the direct
product $K(\ZZ_g,1)\times
\prod_{i=1}^{r-1}K(\ZZ,2)\times\prod_{i=1}^{r^\ast}K(\ZZ,4)$ onto its
factors. Let $\gamma_1$, $\gamma_2$, and $\gamma_4$ be characteristic
elements of $H^1(K(\ZZ_g,1),\ZZ_g)$, $H^2(K(\ZZ,2),\ZZ)$, and
$H^4(K(\ZZ,4),\ZZ)$, respectively. Consider the map
\begin{equation}\label{GCTBG5}
\renewcommand{\arraystretch}{1.4}
\arraycolsep0.2em
\begin{array}[b]{rcl}
\varphi : 
[M,\BSUJ] 
& \rightarrow & 
[M,(\BSUJ)_5]
\\
& \rightarrow &
[M,K(\ZZ_g,1)]
\times
\prod_{i=1}^{r-1} [M,K(\ZZ,2)]
\times
\prod_{i=1}^{r^\ast} [M,K(\ZZ,4)]
\\
& \rightarrow &
H^1(M,\ZZ_g)
\times
\prod_{i=1}^{r-1} H^2(M,\ZZ)
\times
\prod_{i=1}^{r^\ast} H^4(M,\ZZ)\,,
%
\\
f 
& \mapsto &
\left(
f^\ast(\pr_1\circ y_5)^\ast\gamma_1,
\{
f^\ast(\pr_{2i}\circ y_5)^\ast\gamma_2
\}_{i=1}^{r-1},
\{
f^\ast(\pr_{4i}\circ y_5)^\ast\gamma_4
\}_{i=1}^{r^\ast}
\right),
\end{array}
\end{equation}
where $y_5:\BSUJ\rightarrow(\BSUJ)_5$ is the $5$-equivalence provided
by Theorem \rref{TPostnikov}. According to Theorem \rref{TBG5}, the 
second step of $\varphi$ and, therefore, the whole map, is a bijection.

Now let $P,P'\in\Bun{M}{\SUJ}$ as proposed in the assertion. Then, by assumption, 
the homomorphisms $\left(\ka{P}\right)^\ast$ and $\left(\ka{P'}\right)^\ast$,
induced on $H^1(\BSUJ,\ZZ_g)$, $H^2(\BSUJ,\ZZ)$, and $H^4(\BSUJ,\ZZ)$, coincide. 
This implies $\varphi\left(\ka{P}\right) =
\varphi\left(\ka{P'}\right)$. Hence, $\ka{P}$ and $\ka{P'}$ are homotopic. 
This proves the corollary.
\qed
\\

We remark that, of course, the cohomology elements
$\left(\pr_1\circ y_5\right)^\ast\gamma_1$,
$\left(\pr_{2i}\circ y_5\right)^\ast\gamma_2$, $i=1,\dots, r-1$,
and
$\left(\pr_{4i}\circ y_5\right)^\ast\gamma_4$, $i=1,\dots, r^\ast$ define
a set of characteristic classes which classifies $\Bun{M}{\SUJ}$. These
classes are independent and surjective. However, they are hard to handle, because
we do not know the homomorphism $y_5^\ast$ explicitly.
Therefore, we prefer to work with characteristic classes defined by some
natural generators of the cohomology groups in question. The price we have
to pay for this is that the classes so constructed are subject to
a relation and that we have to determine their image explicitly.
\subsection{Generators for $H^\ast(\BSUJ,\ZZ)$}
Instead of generators for the groups $H^2(\BUJ,\ZZ)$ and $H^4(\BUJ,\ZZ)$ only,
we can construct generators for the whole cohomology
algebra $H^\ast(\BSUJ,\ZZ)$ without any additional effort.

Consider the homomorphisms and induced homomorphisms
\begin{equation}\label{GprjJ}
\begin{array}[b]{rcccl}
\SUJ
& \stackrel{j_J}{\longrightarrow} &
\UJ
& \stackrel{\prU{J}{i}}{\longrightarrow} &
\rmU k_i\,,
\\
H^\ast(\BSUJ,\ZZ)
& \stackrel{\left(\rmB j_J\right)^\ast}{\longleftarrow} &
H^\ast(\BUJ,\ZZ)
& \stackrel{\left(\rmB\prU{J}{i}\right)^\ast}{\longleftarrow} &
H^\ast(\rmB\rmU k_i,\ZZ)\,,
\end{array}
\end{equation}
where $i=1,\dots, r$. Recall that the cohomology algebra
$H^\ast(\rmB\rmU k,\ZZ)$ is generated freely over $\ZZ$ by elements
$\gamma_{\rmU k}^{(2j)}\in H^{2j}(\rmB\rmU k,\ZZ)$, $j=1,\dots, k$, see
\cite{Borel}. We denote
\begin{equation}\label{GdefgUk}
\gamma_{\rmU k}
=
1+\gamma_{\rmU k}^{(2)}+\cdots+\gamma_{\rmU k}^{(2k)}\,.
\end{equation}
The generators $\gamma_{\rmU k_i}^{(2j)}$ define elements
\begin{eqnarray}  \label{GdeftwgJi2j}
\twgamma_{J,i}^{(2j)} 
& = &
\left(\rmB\prU{J}{i}\right)^\ast\gamma_{\rmU k_i}^{(2j)}\,,
\\ \label{GdefgJi2j}
\gamma_{J,i}^{(2j)} 
& = &
\left(\rmB j_J\right)^\ast\left(\rmB\prU{J}{i}\right)^\ast
\gamma_{\rmU k_i}^{(2j)}
\end{eqnarray}
of $H^{2j}(\BUJ,\ZZ)$ and $H^{2j}(\BSUJ,\ZZ)$, respectively. 
We denote
\begin{eqnarray} \label{GdeftwgJi}
\twgamma_{J,i} 
& = &
1+\twgamma_{J,i}^{(2)}+\cdots+\twgamma_{J,i}^{(2k_i)}\,,~~i=1,\dots, r\,,
\\ \label{GdefgJi}
\gamma_{J,i}
& = &
1+\gamma_{J,i}^{(2)}+\cdots+\gamma_{J,i}^{(2k_i)}\,,~~i=1,\dots, r\,,
\end{eqnarray}
as well as $\twgamma_J=(\twgamma_{J,1},\dots,\twgamma_{J,r})$ and
$\gamma_J=(\gamma_{J,1},\dots,\gamma_{J,r})$.
\begin{Lemma} \label{LcohomZ}
The cohomology algebra $H^\ast(\BUJ,\ZZ)$ is generated freely over $\ZZ$ 
by the elements $\twgamma_{J,i}^{(2j)}$, $j=1,\dots,k_i$, $i=1,\dots,r$.
\end{Lemma}
{\it Proof:} Consider the isomorphism and induced isomorphism
\begin{equation} \label{GPHast(BUJ,Z)1}
\begin{array}[b]{ccccc}
\UJ
& \stackrel{d_r}{\longrightarrow} &
\Pi_i\UJ
& \stackrel{\Pi_i\prU{J}{i}}{\longrightarrow} &
\Pi_i\rmU k_i\,,
\\
H^\ast\left(\BUJ,\ZZ\right)
& \stackrel{d_r^\ast}{\longleftarrow} &
H^\ast\left(\Pi_i\BUJ,\ZZ\right)
& \stackrel{\left(\Pi_i\rmB\prU{J}{i}\right)^\ast}{\longleftarrow} &
H^\ast\left(\Pi_i\rmB\rmU k_i,\ZZ\right)\,.
\end{array}
\end{equation}
where $d_r$ denotes $r$-fold diagonal embedding and $\Pi_i$ is a shorthand
notation for $\Pi_{i=1}^r$. Due to the K\"unneth Theorem for cohomology 
\cite[Ch.~XIII, Cor.~11.3]{Massey},
$H^\ast\left(\Pi_i\rmB\rmU k_i,\ZZ\right)$ is
generated freely over $\ZZ$ by the elements
$$
1_{\rmB\rmU k_1}\times\cdots\times 1_{\rmB\rmU k_{i-1}}
\times
\gamma_{\rmU k_i}^{(2j)}
\times
1_{\rmB\rmU k_{i+1}}\times\cdots\times 1_{\rmB\rmU k_r}\,,~~
j=1,\dots, k_i, i=1,\dots, r\,.
$$
Here $\times$ stands for the cohomology cross product, and $1_{\rmB\rmU k_i}$
denotes the generator of $H^0(\rmB\rmU k_i,\ZZ)$. The assertion follows by
applying \gref{GPHast(BUJ,Z)1} to these generators and using
\begin{equation} \label{Gdastcross}
d_r^\ast(\alpha_1\times\cdots\times\alpha_r)
=
\alpha_1\smile\dots\smile\alpha_r~~~~
\forall~~\alpha_i\in H^\ast(\rmB\rmU k_i,\ZZ)\,.~\qed
\end{equation}
\begin{Lemma} \label{LPcohomZ1}
$\left(\rmB j_J\right)^\ast$ is surjective.
\end{Lemma}
{\it Proof:}
According to \gref{GBphibun} and due to $\UJ/\SUJ\cong\rmU 1$, $\rmB j_J$ 
is the projection in a principal bundle
\begin{equation} \label{Geta}
\rmU 1\hookrightarrow\BSUJ\stackrel{\rmB j_J}{\longrightarrow}\BUJ\,.
\end{equation}
Denote this bundle by $\eta$. Due to $\pi_1(\BUJ)\cong\pi_0(\UJ)=0$, $\eta$
is orientable, see \cite[Def.~7.3.3]{DodsonParker}. Therefore, it induces a
Gysin sequence, see \cite[\S 7.3.1]{DodsonParker},
\begin{equation} \label{GGysinetaJ}
\begin{array}[b]{l}
H^1(\BUJ,\ZZ)\stackrel{\left(\rmB j_J\right)^\ast}{\longrightarrow}
H^1(\BSUJ,\ZZ)\stackrel{\sigma^\ast}{\longrightarrow}
H^0(\BUJ,\ZZ)\stackrel{\smile c_1(\eta)}{\longrightarrow}
H^2(\BUJ,\ZZ)
\longrightarrow\phantom{\cdots}
\\
\phantom{H^1(\BUJ,\ZZ)}
\stackrel{\left(\rmB j_J\right)^\ast}{\longrightarrow}
H^2(\BSUJ,\ZZ)\stackrel{\sigma^\ast}{\longrightarrow}
H^1(\BUJ,\ZZ)\stackrel{\smile c_1(\eta)}{\longrightarrow}
H^3(\BUJ,\ZZ)
\longrightarrow\phantom{\cdots}
\\
\phantom{H^1(\BUJ,\ZZ)}
\stackrel{\left(\rmB j_J\right)^\ast}{\longrightarrow}
H^3(\BSUJ,\ZZ)\stackrel{\sigma^\ast}{\longrightarrow}
H^2(\BUJ,\ZZ)\stackrel{\smile c_1(\eta)}{\longrightarrow}
H^4(\BUJ,\ZZ)\longrightarrow
\cdots\,.
\end{array}
\end{equation}
(On the level of differential forms, $\sigma^\ast$ is given by integration 
over the fibre.)
If $\eta$ was trivial, $\pi_1(\BSUJ)$ would coincide with
$\pi_1(\BUJ\times\rmU 1) = \ZZ$, which would contradict Theorem
\rref{Thtpgr}. Hence, $\eta$ is nontrivial, so that $c_1(\eta)\neq
0$. Since $H^\ast(\BUJ,\ZZ)$ has no zero divisors by Lemma \rref{LcohomZ}, 
it follows 
that multiplication by $c_1(\eta)$ is an injective operation on
$H^\ast(\BUJ,\ZZ)$. Then exactness of the Gysin sequence 
\gref{GGysinetaJ} implies that the homomorphism $\sigma^\ast$ is 
trivial and, therefore, $\left(\rmB j_J\right)^\ast$ is surjective. 
\qed
\\

Lemmas \rref{LcohomZ} and \rref{LPcohomZ1} imply the following corollary.
\begin{Corollary} \label{CcohomZ}
The cohomology algebra $H^\ast(\BSUJ,\ZZ)$ is generated over 
$\ZZ$ by the elements $\gamma_{J,i}^{(2j)}$, $j=1,\dots, k_i$, 
$i=1,\dots, r$. 
\qed
\end{Corollary}
{\it Remark:} The generators $\gamma_{J,i}^{(2)}$ of $H^\ast(\BSUJ,\ZZ)$ 
are subject to a relation which is, however, irrelevant for our purposes.
For the sake of completeness, we derive this relation in
Appendix \rref{AcohomZ}.
\subsection{Generator for $H^1(\BSUJ,\ZZ_g)$}
Since $\BSUJ$ is connected, $H^1(\BSUJ,\ZZ_g)$ can be computed by means of
the Hurewicz and the Universal Coefficient Theorems:
$$
\begin{array}{rcl}
H^1(\BSUJ,\ZZ_g)
& \cong &
\Hom(H_1(\BSUJ),\ZZ_g) \oplus \Ext(H_0(\BSUJ),\ZZ_g)
\\ & \cong &
\Hom(\pi_1(\BSUJ),\ZZ_g) \oplus \Ext(\ZZ,\ZZ_g)\,.
\end{array}
$$
Due to Theorem \rref{Thtpgr}, $\pi_1(\BSUJ)\cong\pi_0(\SUJ)\cong\ZZ_g$.
Moreover, $\Ext(\ZZ,\ZZ_g)=0$. Hence,
$H^1(\BSUJ,\ZZ_g)$ is isomorphic to $\ZZ_g$. It is therefore generated by
a single element. Apparently, we are free to choose {\it any} of the
generators to work with. However, there exists a relation between this
generator and the generators $\gamma_{J,i}^{(2)}$ of $H^2(\BSUJ,\ZZ)$. This
can be seen as follows. Consider the short exact sequence
\begin{equation} \label{Gexsqcmgrg}
0\longrightarrow
\ZZ\stackrel{\mu_g}{\longrightarrow}
\ZZ\stackrel{\varrho_g}{\longrightarrow}
\ZZ_g\longrightarrow 0\,,
\end{equation}
where $\mu_g$ denotes multiplication by $g$ and $\varrho_g$ reduction modulo
$g$. This induces a long exact sequence (see \cite[\S IV.5]{Bredon:Top})
\begin{equation} \label{GBock}
\cdots\stackrel{\beta_g}{\longrightarrow}
H^i(\cdot,\ZZ)\stackrel{\mu_g}{\longrightarrow}
H^i(\cdot,\ZZ)\stackrel{\varrho_g}{\longrightarrow}
H^i(\cdot,\ZZ_g)\stackrel{\beta_g}{\longrightarrow}
H^{i+1}(\cdot,\ZZ)\stackrel{\mu_g}{\longrightarrow}\cdots\,.
\end{equation}
Here we have denoted the coefficient homomorphisms induced by $\mu_g$ and
$\varrho_g$ by the same letters, i.e., $\mu_g$ maps $\alpha\mapsto
g\,\alpha$ and $\varrho_g$ maps $\alpha\mapsto\alpha\,\mod\, g$. Usually, 
the connecting
homomorphism $\beta_g$ is called {\it Bockstein homomorphism}.
Application of $\beta_g$ to an arbitrary generator of $H^1(\BSUJ,\ZZ_g)$
yields an element of $H^2(\BSUJ,\ZZ)$ which can be expressed in terms of the
$\gamma_{J,i}^{(2)}$. In order to keep track of this relation, we have to
choose a {\it specific} generator of $H^1(\BSUJ,\ZZ_g)$. This will be
constructed now.

Consider the homomorphism $\lambda^\rmS_J:\SUJ\rightarrow\ZZ_g$ and the 
induced homomorphism
\begin{equation} \label{GBlSJast}
\left(\rmB\lambda^\rmS_J\right)^\ast
:
H^1(\rmB\ZZ_g,\ZZ_g)
\longrightarrow
H^1(\BSUJ,\ZZ_g)\,.
\end{equation}
Due to Lemma \rref{Lcoco},
${\lambda^\rmS_J}_\ast:\pi_0(\SUJ)\rightarrow\pi_0(\ZZ_g)$ is an isomorphism. 
Hence, so is 
$\left(\rmB\lambda^\rmS_J\right)_\ast:\pi_1(\BSUJ)\rightarrow\pi_1(\rmB\ZZ_g)$. 
Then the Hurewicz and Universal Coefficient Theorems imply that
\gref{GBlSJast} is an isomorphism. Thus, generators of $H^1(\BSUJ,\ZZ_g)$
can be obtained as the images of generators of $H^1(\rmB\ZZ_g,\ZZ_g)$ under 
$\left(\rmB\lambda^\rmS_J\right)^\ast$.
\begin{Lemma} \label{Ldeltag}
There exists a unique element $\delta_g\in H^1(\rmB\ZZ_g,\ZZ_g)$ such that
\begin{equation} \label{Gbgdg}
\beta_g(\delta_g)=\left(\rmB j_g\right)^\ast\gamma_{\rmU 1}^{(2)}\,.
\end{equation}
This element is a generator of $H^1(\rmB\ZZ_g,\ZZ_g)$.
\end{Lemma}
{\it Proof:} First we notice that both $\beta_g(\delta_g)$ and
$\left(\rmB j_g\right)^\ast\gamma_{\rmU 1}^{(2)}$ are elements of
$H^2(\rmB\ZZ_g,\ZZ_g)$ so that Eq.~\gref{Gbgdg} makes sense.

Next, consider the following portion of the exact sequence
\gref{GBock}:
\begin{equation} \label{GportionBZg}
\cdots\longrightarrow
H^1(\rmB\ZZ_g,\ZZ)\stackrel{\varrho_g}{\longrightarrow}
H^1(\rmB\ZZ_g,\ZZ_g)\stackrel{\beta_g}{\longrightarrow}
H^2(\rmB\ZZ_g,\ZZ)\stackrel{\mu_g}{\longrightarrow}
H^2(\rmB\ZZ_g,\ZZ)\longrightarrow\cdots\,.
\end{equation}
To determine the cohomology groups, we note that a model for $\rmB\ZZ_g$ is
given by the lens space $\lens{g}{\infty}$, see Appendix \rref{AEMLS}. Thus,
from Eq.~\gref{GHiLg} in this appendix we infer 
\begin{eqnarray}\label{GH1Lg}
H^1(\rmB\ZZ_g,\ZZ) & = & 0
\\ \label{GH2Lg}
H^2(\rmB\ZZ_g,\ZZ) & = & \ZZ_g\,.
\end{eqnarray}
Due to \gref{GH1Lg}, $\beta_g$ is injective in \gref{GportionBZg}. Due to
\gref{GH2Lg}, $\mu_g$ is trivial in \gref{GportionBZg} so that $\beta_g$ is
also surjective there. Hence, we can define
$\delta_g=\beta_g^{-1}\circ\left(\rmB j_g\right)^\ast\gamma_{\rmU 1}^{(2)}$. 
In order to check that this is a generator, we consider
$J^\circ=((1),(g))\in\rmK(g)$. We observe that $\ZZ_g\cong\SUJ^\circ$, $\rmU
1\cong\UJ^\circ$, and that $j_g$ corresponds, by virtue of these
isomorphisms, to $j_{J^\circ}:\SUJ^\circ\rightarrow\UJ^\circ$. Then Lemma
\rref{LPcohomZ1} implies that $\left(\rmB j_g\right)^\ast$ is surjective.
In particular, $H^2(\rmB\ZZ_g,\ZZ)$ is generated by $\left(\rmB
j_g\right)^\ast\gamma_{\rmU 1}^{(2)}$. Hence, $H^1(\rmB\ZZ_g,\ZZ_g)$ is
generated by $\delta_g$.
\qed
\\

We define 
\begin{equation}\label{GdefdJ}
\delta_J=\left(\rmB\lambda^\rmS_J\right)^\ast\delta_g\,. 
\end{equation}
Then Lemma \rref{Ldeltag} yields the following corollary.
\begin{Corollary}\label{Cdg}
$H^1(\BSUJ,\ZZ_g)$ is generated by $\delta_J$, where
$\beta_g(\delta_J)
=
\left(\rmB\lambda^\rmS_J\right)^\ast\left(\rmB j_g\right)^\ast
\gamma_{\rmU 1}^{(2)}$.\qed
\end{Corollary}
\subsection{The Relation between Generators}
In this subsection, we are going to derive the relation between $\delta_J$
and the $\gamma_{J,i}^{(2)}$, i.e., to compute $\beta_g(\delta_J)$ in terms
of the latter.

For any topological space $X$, let $H^\rmeven_0(X,\ZZ)$ denote the
subset of $H^\rmeven(X,\ZZ)$ consisting of elements of the form
$1+\alpha^{(2)}+\alpha^{(4)}+\dots$ . For any
finite sequence of nonnegative integers $\bfa=(a_1,\dots, a_s)$,
define a polynomial function
\begin{equation} \label{GdefEm}
E_{\bfa}:\prod_{i=1}^s H^\rmeven_0(X,\ZZ)\rightarrow H^\rmeven_0(X,\ZZ)\,,~~
(\alpha_1,\dots,\alpha_s)
\mapsto
\alpha_1^{a_1}\smile\dots\smile\alpha_s^{a_s}\,,
\end{equation}
where powers are taken w.r.t.~the cup product. By construction, for any map
$f:X\rightarrow Y$,
\begin{equation} \label{GfastEm}
f^\ast E_{\bfa}(\alpha_1,\dots,\alpha_s) = E_{\bfa}(f^\ast\alpha_1,\dots,
f^\ast\alpha_s)\,.
\end{equation}
Let us derive explicit expressions for the components of $E_{\bfa}$
of $2$nd and $4$th degree.
\begin{Lemma} \label{LEJ2,EJ4}
Let $\alpha=(\alpha_1,\dots,\alpha_s)\in\prod_{i=1}^sH^\rmeven_0(X,\ZZ)$.
Then
\begin{eqnarray} \label{GEJ2}
\hspace*{-0.5cm}E_{\bfa}^{(2)}(\alpha)
& = &
\sum_{i=1}^s a_i\alpha_i^{(2)}\,,
\\ \label{GEJ4}
\hspace*{-0.5cm}E_{\bfa}^{(4)}(\alpha)
& = &
\sum_{i=1}^s a_i\alpha_i^{(4)} + \sum_{i=1}^s\frac{a_i(a_i-1)}{2}
\alpha_i^{(2)}\smile\alpha_i^{(2)} + \sum_{i<j=2}^s a_ia_j\alpha_i^{(2)}
\smile\alpha_j^{(2)}\,.
\end{eqnarray}
\end{Lemma}
{\it Proof:} This is a straightforward computation which we
only indicate here. Let $\alpha$ be given. Without loss of generality we
may assume that the components of $\alpha_i$ of degree higher than $4$
vanish. For the cup product of elements of this form one has the
following formula:
$$
\begin{array}[b]{l}
\left(1+\beta^{(2)}+\beta^{(4)}\right)\smile\left(1+\gamma^{(2)}+\gamma^{(
4)}\right)
\\[0.1cm]
\phantom{\left(1+\beta^{(2)}+\beta^{(4)}\right)\smile\left(1+
\right.}
= 1 + \left(\beta^{(2)}+\gamma^{(2)}\right) +
\left(\beta^{(4)}+\gamma^{(4)}+\beta^{(2)}\smile\gamma^{(2)}\right)\,.
\end{array}
$$
We can iterate this to obtain
$$
\alpha_i^{a_i} = 1+\left(a_i\alpha_i\right)+\left(a_i\alpha_i^{(4)}+
\frac{a_i(a_i-1)}{2}\alpha_i^{(2)}\smile\alpha_i^{(2)}\right)
$$
and then to compute the product of all the factors $\alpha_i^{a_i}$. This yields
the assertion.
\qed
\\

As an immediate consequence of \gref{GEJ2}, for any $l\in\ZZ$,
\begin{equation}\label{GgEJ2}
E_{l\,\bfa}^{(2)}=l\,E _{\bfa}^{(2)}\,.
\end{equation}
\begin{Lemma} \label{LBiJast}
The following two formulae hold:
\begin{eqnarray} \label{GBiJast}
\left(\rmB\;\!i_J\right)^\ast\gamma_{\rmU n}
& = &
E_{\bfm}\left(\twgamma_J\right)\,,
\\ \label{GBlJast}
\left(\rmB\lambda_J\right)^\ast\gamma_{\rmU 1}^{(2)}
& = &
E_{\twbfm}^{(2)}\left(\twgamma_J\right)\,.
\end{eqnarray}
\end{Lemma}
{\it Proof:} First, consider \gref{GBiJast}.
We decompose $i_J$ as follows:
\begin{equation} \label{GiJ}
i_J:\UJ\stackrel{d_r}{\longrightarrow}
\Pi_i\UJ\stackrel{\Pi_i\prU{J}{i}}{\longrightarrow}
\Pi_i\rmU k_i\stackrel{\Pi_i d_{m_i}}{\longrightarrow}
\Pi_i(\rmU k_i\stackrel{m_i}{\times\cdots\times}\rmU k_i)
\stackrel{j}{\longrightarrow}\rmU n\,.
\end{equation}
Here $d_r$, $d_{m_i}$ denote $r$-fold and $m_i$-fold diagonal embedding,
respectively, and $j$ stands for the natural (blockwise) embedding. 
According to \gref{GiJ}, $\left(\rmB\!\;i_J\right)^\ast$ decomposes as
\begin{equation}\label{GdecompBiJ}
\begin{array}[b]{l}
\left(\rmB\!\;i_J\right)^\ast:
H^\ast\left(\rmB\rmU n,\ZZ\right)
\stackrel{\left(\rmB j\right)^\ast}{\longrightarrow}
H^\ast\left(\Pi_i(\rmB\rmU k_i
\stackrel{m_i}{\times\cdots\times}
\rmB\rmU k_i),\ZZ\right)
\\
\phantom{\left(\rmB\!\;i_J\right)^\ast:}
\stackrel{\left(\Pi_i d_{m_i}\right)^\ast}{\longrightarrow}
H^\ast\left(\Pi_i\rmB\rmU k_i,\ZZ\right)
\stackrel{\left(\Pi_i\rmB\prU{J}{i}\right)^\ast}{\longrightarrow}
H^\ast\left(\Pi_i\BUJ,\ZZ\right) \stackrel{d_r^\ast}{\longrightarrow}
H^\ast\left(\BUJ,\ZZ\right)\,.
\end{array}
\end{equation}
Under the assumption that we have chosen the generators
$\gamma_{\rmU k}^{(2j)}$ for different $k$ in a consistent way 
(namely, such that the universal properties of Chern classes 
hold for the characteristic classes
defined by these elements), there holds the relation
\begin{equation} \label{GBjastgammaUn}
\left(\rmB j\right)^\ast\gamma_{\rmU n} =
(\gamma_{\rmU k_1}\stackrel{m_1}{\times\cdots\times}\gamma_{\rmU k_1})
\times\cdots\times
(\gamma_{\rmU k_r}\stackrel{m_r}{\times\cdots\times}\gamma_{\rmU k_r})\,.
\end{equation}
Using this, as well as \gref{Gdastcross}, we obtain
$$
\arraycolsep0.1em
\renewcommand{\arraystretch}{1.3}
\begin{array}[b]{rcl}
\left(\rmB\!\;i_J\right)^\ast\gamma_{\rmU n}
& = &
d_r^\ast\circ
\left(\Pi_i\rmB\prU{J}{i}\right)^\ast\circ
\left(\Pi_i d_{m_i}\right)^\ast\circ
\left(\rmB j\right)^\ast \gamma_{\rmU n}
\\
&=&
d_r^\ast\circ
\left(\Pi_i\rmB\prU{J}{i}\right)^\ast\circ
\left(\Pi_i d_{m_i}\right)^\ast
\left(
(\gamma_{\rmU k_1}\stackrel{m_1}{\times\cdots\times}\gamma_{\rmU k_1})
\times\cdots\times
(\gamma_{\rmU k_r}\stackrel{m_r}{\times\cdots\times}\gamma_{\rmU k_r})
\right)
\\
&=&
d_r^\ast\circ
\left(\Pi_i\rmB\prU{J}{i}\right)^\ast
\left(\gamma_{\rmU k_1}^{m_1}
\times\cdots\times
\gamma_{\rmU k_r}^{m_r}\right)
\\
&=&
d_r^\ast
\left(\twgamma_{J,1}^{m_1}\times\cdots\times\twgamma_{J,r}^{m_r}\right)
\\
&=&
\twgamma_{J,1}^{m_1} \smile\dots\smile \twgamma_{J,r}^{m_r}\,.
\end{array}
\renewcommand{\arraystretch}{1}
$$
This yields \gref{GBiJast}. Now consider \gref{GBlJast}.
The commutative diagram \gref{Gctvdgrdet} implies
\begin{equation} \label{GLBiJast1}
\left(\rmB\lambda_J\right)^\ast\left(\rmB p_g\right)^\ast
\gamma_{\rmU 1}^{(2)}
=
\left(\rmB\!\;i_J\right)^\ast\left(\rmB \det\nolimits_{\rmU n}\right)^\ast
\gamma_{\rmU 1}^{(2)}\,.
\end{equation}
We have
\begin{eqnarray}\label{GLBiJast2}
\left(\rmB\det\nolimits_{\rmU n}\right)^\ast\gamma_{\rmU 1}^{(2)}
& = &
\gamma_{\rmU n}^{(2)}\,,
\\ \label{GLBiJast3}
\left(\rmB p_g\right)^\ast\gamma_{\rmU 1}^{(2)}
& = &
g\,\gamma_{\rmU 1}^{(2)}\,.
\end{eqnarray}
Formula \gref{GLBiJast3} follows, by virtue of the Hurewicz and the Universal
Coefficient Theorems, from the fact that the homomorphism
$\left(\rmB p_g\right)_\ast:\pi_2(\rmB\rmU 1)\rightarrow\pi_2(\rmB\rmU 1)$
is given by multiplication by $g$.
Inserting Eqs. \gref{GLBiJast2} and \gref{GLBiJast3} into \gref{GLBiJast1} 
we obtain
$$
\begin{array}{rclcl}
g\left(\rmB\lambda_J\right)^\ast\gamma_{\rmU 1}^{(2)}
& = &
E_{\bfm}^{(2)}\left(\twgamma_J\right)
& | & 
\mbox{by \gref{GBiJast}}
\\
& = &
g\,E_{\twbfm}^{(2)}\left(\twgamma_J\right)
& | &
\mbox{by \gref{GgEJ2}.}
\end{array}
$$
Since this relation holds in $H^2(\BUJ,\ZZ)$ which is free Abelian, it
implies \gref{GBlJast}.
\qed
\begin{Theorem} \label{TbgdJ}
There holds the relation
$\beta_{g}(\delta_J)=E_{\twbfm}^{(2)}(\gamma_J)$.
\end{Theorem}
{\it Proof:} We compute
$$
\begin{array}{rclcl}
\beta_g(\delta_J)
& = &
\left(\rmB\lambda^\rmS_J\right)^\ast\left(\rmB j_g\right)^\ast
\gamma_{\rmU 1}^{(2)}
& | & \mbox{by Corollary \rref{Cdg}}
\\
& = &
\left(\rmB j_J\right)^\ast\left(\rmB\lambda_J\right)^\ast
\gamma_{\rmU 1}^{(2)}
& | & \mbox{by \gref{Gctvdgr}}
\\
& = &
\left(\rmB j_J\right)^\ast E^{(2)}_{\twbfm}\left(\twgamma_J\right)
& | & \mbox{by \gref{GBlJast}}
\\
& = &
E^{(2)}_{\twbfm}\left(\gamma_J\right)\,.
&&
\end{array}
$$
\vspace{-1cm}

\qed
\subsection{Characteristic Classes for $\SUJ$-Bundles}
Using the cohomology elements $\gamma_{J,i}^{(2j)}$ and $\delta_J$
constructed above, we define the following characteristic classes for
$\SUJ$-bundles over a manifold $M$:
\begin{eqnarray} \label{GdefaJi}
\alpha_{J,i} & : & \Bun{M}{\SUJ}\rightarrow H^\rmeven_0(M,\ZZ)\,,~~
Q\mapsto\left(\ka{Q}\right)^\ast\gamma_{J,i}\,,~~
i=1,\dots, r
\\
\xi_J & : & \Bun{M}{\SUJ}\rightarrow H^1(M,\ZZ_g)\,,~~
Q\mapsto\left(\ka{Q}\right)^\ast\delta_J\,.
\end{eqnarray}
Sorted by degree,
$\alpha_{J,i}(Q)=1+\alpha_{J,i}^{(2)}(Q) +\cdots+ \alpha_{J,i}^{(2k_i)}(Q)$,
where $\alpha_{J,i}^{(2j)}(Q)=\left(\ka{Q}\right)^\ast\gamma_{J,i}^{(2j)}$. 
Moreover, we introduce the notation
$\alpha_J(Q) = (\alpha_{J,1}(Q),\dots,\alpha_{J,r}(Q))$.
Then
\begin{equation} \label{GdefaJ}
\alpha_J(Q) = \left(\ka{Q}\right)^\ast\gamma_J
\end{equation}
and $\alpha_J$ can be viewed as a map from $\Bun{M}{\SUJ}$ to the set
\begin{equation} \label{GdefHJM}
H^{(J)}(M,\ZZ)=\prod_{i=1}^r \prod_{j=1}^{k_i} H^{2j}(M,\ZZ)\,.
\end{equation}
By construction, the relation which holds for $\gamma_J$ and $\delta_J$
carries over to the characteristic classes $\alpha_J$ and $\xi_J$.
By virtue of \gref{GfastEm}, from Theorem \rref{TbgdJ} we infer
\begin{equation} \label{Grelcc}
E_{\twbfm}^{(2)}\left(\alpha_J(Q)\right)
=
\beta_{g}\left(\xi_J(Q)\right)
~~~~\forall~
Q\in\Bun{M}{\SUJ}\,.
\end{equation}
In order to derive expressions for $\alpha_J$ and $\xi_J$ in terms of the 
ordinary characteristic classes for $\rmU k_i$-bundles and $\ZZ_g$-bundles,
let $Q\in\Bun{M}{SUJ}$. There are two kinds of principal bundles 
associated in a natural way to $Q$: The $\rmU k_i$-bundles
$\ab{Q}{\prU{J}{i}\circ j_J}$, $i=1,\dots,r$, and the $\ZZ_g$-bundle
$\ab{Q}{\lambda^\rmS_J}$. For the first ones, using
\gref{Gclfmapassbun} and \gref{GdefgJi2j} we compute
$$
c\left(\ab{Q}{\prU{J}{i}\circ j_J }\right)
=
\left(\ka{\ab{Q}{\prU{J}{i}\circ j_J }}\right)^\ast\gamma_{\rmU k_i}
=
\left(\ka{Q}\right)^\ast\circ\left(\rmB j_J
\right)^\ast\circ\left(\rmB\prU{J}{i}\right)^\ast
\gamma_{\rmU k_i}
=
\left(\ka{Q}\right)^\ast\gamma_{J,i}\,,
$$
so that
\begin{equation} \label{GaJc}
\alpha_{J,i}\left(Q\right) = c\left(\ab{Q}{\prU{J}{i}\circ j_J }\right)\,,~~
i=1,\dots, r\,.
\end{equation}
As for the second one, let $\chi_g$ denote the characteristic class for
$\ZZ_g$-bundles over $M$ defined by the generator $\delta_g\in
H^1(\rmB\ZZ_g,\ZZ_g)$, i.e.,
\begin{equation}\label{Gdefchig}
\chi_g(R)=\left(\ka{R}\right)^\ast\delta_g\,,~~R\in\Bun{M}{\ZZ_g}\,.
\end{equation}
Then \gref{Gclfmapassbun} and \gref{GdefdJ} yield
$
\chi_{g}\left(\ab{Q}{\lambda^\rmS_J}\right)
=
\left(\ka{\ab{Q}{\lambda^\rmS_J}}\right)^\ast\delta_{g}
=
\left(\ka{Q}\right)^\ast\circ\left(\rmB \lambda^\rmS_J\right)^\ast\delta_{g}
=
\left(\ka{Q}\right)^\ast\delta_J
$. Consequently,
\begin{equation} \label{GxJc}
\xi_J\left(Q\right) = \chi_{g}\left(\ab{Q}{\lambda^\rmS_J}\right)\,.
\end{equation}
\subsection{Classification of $\SUJ$-Bundles}
We denote 
\begin{equation} \label{GdefKMJ}
\rmK(M)_J = \left\{(\alpha,\xi)\in H^{(J)}(M,\ZZ)\times H^1(M,\ZZ_g)
~\left|~
E_{\twbfm}^{(2)}(\alpha) = \beta_{g}(\xi)\right.\right\}\,.
\end{equation}
\begin{Theorem} \label{TBunMSUJ}
Let $M$ be a manifold, $\dim M\leq 4$, and let $J\in\rmK(n)$. Then the
characteristic classes $\alpha_J$ and $\xi_J$ define a bijection from
$\Bun{M}{\SUJ}$ onto $\rmK(M)_J$.
\end{Theorem}
{\it Proof:} The map is injective by Corollary \rref{Cccsuffice}. In the
following lemma we prove that it is also surjective.
\begin{Lemma} \label{Lrangecc}
Let $M$ be a manifold, $\dim M\leq 4$, and let $J\in\rmK(n)$. Let
$(\alpha,\xi)\in\rmK(M)_J$. Then there exists
$Q\in\Bun{M}{\SUJ}$ such that $\alpha_J(Q)=\alpha$ and $\xi_J(Q)=\xi$.
\end{Lemma}
{\it Proof:} We give a construction of $Q$ in terms of $\rmU k_i$ and
$\ZZ_g$-bundles. There exists $R\in\Bun{M}{\ZZ_g}$ such that $\chi_g(R)=\xi$.
Due
to $\dim M\leq 4$, there exist also $Q_i\in\Bun{M}{\rmU k_i}$ such that
$c(Q_i)=\alpha_i$, $i=1,\dots, r$. Define $\twQ=Q_1\times_M\cdots\times_M Q_r$ 
(Whitney, or fibre, product). By identifying $\rmU k_1\times\cdots\times\rmU k_r$ with
$\UJ$, $\twQ$ becomes a $\UJ$-bundle. Then
\begin{equation} \label{GtwQpr}
\ab{\twQ}{\prU{J}{i}}\cong Q_i\,,~~i=1,\dots, r\,.
\end{equation}
Consider the $\rmU 1$-bundles $\ab{\twQ}{\lambda_J}$ and $\ab{R}{j_g}$
associated to $\twQ$ and $R$, respectively. Assume, for a moment, that
they are isomorphic. Then $R$ is a subbundle of $\ab{\twQ}{\lambda_J}$
with structure group $\ZZ_g$. Let $Q$ denote the pre-image of $R$ under the
natural bundle morphism $\twQ\rightarrow\ab{\twQ}{\lambda_J}$, see
\gref{Gnatbumor}. This is a
subbundle of $\twQ$ with structure group being the pre-image of $\ZZ_g$
under $\lambda_J$, i.e., with structure group $\SUJ$.
Using \gref{GaJc}, $\ab{Q}{j_J} = \twQ$, and \gref{GtwQpr}, for
$i=1,\dots,r$,
$$
\alpha_{J,i}(Q) 
=
c\left(\ab{Q}{\prU{J}{i}\circ j_J}\right)
=
c\left(\ab{\left(\ab{Q}{j_J}\right)}{\prU{J}{i}}\right)
=
c\left(\ab{\twQ}{\prU{J}{i}}\right)
=
c\left(Q_i\right)
=
\alpha_i\,.
$$
Moreover, by construction of $Q$, $\ab{Q}{\lambda^\rmS_J}\cong R$. 
Thus, \gref{GxJc} yields $\xi_J\left(Q\right) = \chi_J\left(R\right) = \xi$.

It remains to prove $\ab{\twQ}{\lambda_J}\cong\ab{R}{j_g}$. We compute
$$
\renewcommand{\arraystretch}{1.4}
\begin{array}[b]{rclcl}
c_1\left(\ab{\twQ}{\lambda_J}\right)
& = &
\left(\ka{\ab{\twQ}{\lambda_J}}\right)^\ast\gamma_{\rmU 1}^{(2)}
&&
\\
& = &
\left(\ka{\twQ}\right)^\ast\circ\left(\rmB\lambda_J\right)^\ast
\gamma_{\rmU 1}^{(2)}
& | &
\mbox{by \gref{Gclfmapassbun}}
\\
& = &
\left(\ka{\twQ}\right)^\ast
E_{\twbfm}^{(2)}\left(\twgamma_J\right)
& | &
\mbox{by \gref{GBlJast}}
\\
& = &
E_{\twbfm}^{(2)}\left(\left(\ka{\twQ}\right)^\ast\twgamma_J\right)
& | &
\mbox{by \gref{GfastEm}}
\\
& = &
E_{\twbfm}^{(2)}\left(
\left(\ka{\twQ}\right)^\ast\circ\left(\rmB\prU{J}{1}\right)^\ast
\gamma_{\rmU k_1}
,\dots,
\left(\ka{\twQ}\right)^\ast\circ\left(\rmB\prU{J}{r}\right)^\ast
\gamma_{\rmU k_r}
\right)
& | &
\mbox{by \gref{GdeftwgJi2j}}
\\
& = &
E_{\twbfm}^{(2)}\left(
\left(\ka{\ab{\twQ}{\prU{J}{1}}}\right)^\ast\gamma_{\rmU k_1}
,\dots,
\left(\ka{\ab{\twQ}{\prU{J}{r}}}\right)^\ast\gamma_{\rmU k_r}
\right)
& | &
\mbox{by \gref{Gclfmapassbun}}
\\
& = &
E_{\twbfm}^{(2)}\left(
\left(\ka{Q_1}\right)^\ast\gamma_{\rmU k_1}
,\dots,
\left(\ka{Q_r}\right)^\ast\gamma_{\rmU k_r}
\right)
& | &
\mbox{by \gref{GtwQpr}}
\\
& = &
E_{\twbfm}^{(2)}\left(c(Q_1),\dots, c(Q_r)\right)
&&
\\
& = &
E_{\twbfm}^{(2)}(\alpha)\,,
\end{array}
$$
as well as
$$
\begin{array}{rclcl}
c_1\left(\ab{R}{j_g}\right)
& = &
\left(\ka{\ab{R}{j_g}}\right)^\ast\gamma^{(2)}_{\rmU 1}
&&
\\
& = &
\left(\ka{R}\right)^\ast\circ\left(\rmB j_{g}\right)^\ast
\gamma_{\rmU 1}^{(2)}
& | &
\mbox{by \gref{Gclfmapassbun}}
\\
& = & \left(\ka{R}\right)^\ast\circ\beta_{g}(\delta_{g})
& | &
\mbox{by \gref{Gbgdg}}
\\
& = & \beta_{g}\circ\left(\ka{R}\right)^\ast\delta_g
& | &
\mbox{by \gref{GfastEm}}
\\
& = & \beta_{g}(\chi_g(R))
& | &
\mbox{by \gref{Gdefchig}}
\\
& = & \beta_{g}(\xi)\,.
&&
\end{array}
$$
Thus, due to $(\alpha,\xi)\in\rmK(M)_J$, $c_1(\ab{\twQ}{\lambda_J}) = 
c_1(\ab{R}{j_g})$. It follows that, indeed,
$\ab{\twQ}{\lambda_J}\cong\ab{R}{j_g}$.
This proves the lemma and, therefore, the theorem.
\qed
\subsection{Classification of $\SUJ$-Subbundles of $\rmSU n$-Bundles}
Let $P$ be a principal $\rmSU n$-bundle over a manifold $M$ and let
$J\in\rmK(n)$. We are going to characterize the subset
$\Red{P}{\SUJ}\subseteq\Bun{M}{\SUJ}$ in terms of the characteristic classes
$\alpha_J$ and $\xi_J$. Recall that for $Q\in\Bun{M}{\SUJ}$, $\ab{Q}{\rmSU n}$
denotes the extension of $Q$ by $\rmSU n$.
\begin{Lemma} \label{Lc(QSUn)}
For any $Q\in\Bun{M}{\SUJ}$,
$c\left(\ab{Q}{\rmSU n}\right)=E_{\bfm}\left(\alpha_J(Q)\right)$.
\end{Lemma}
{\it Proof:} Note that
$c\left(\ab{Q}{\rmSU n}\right)
=
c\left(\ab{Q}{\rmU n}\right)
=
c\left(\ab{Q}{i_J\circ j_J}\right)$.
Hence,
$$
\renewcommand{\arraystretch}{1.2}
\begin{array}[b]{rclcl}
c\left(\ab{Q}{\rmSU n}\right)
& = &
\left(\ka{Q}\right)^\ast
\circ\left(\rmB j_J\right)^\ast
\circ\left(\rmB\!\;i_J\right)^\ast
\gamma_{\rmU n}
& | &
\mbox{by \gref{Gclfmapassbun}}
\\
& = &
\left(\ka{Q}\right)^\ast
\circ\left(\rmB j_J\right)^\ast\circ
E_{\bfm}\left(\twgamma_J\right)
& | &
\mbox{by \gref{GBiJast}}
\\
& = &
E_{\bfm}\Big(\left(\ka{Q}\right)^\ast
\circ\left(\rmB j_J\right)^\ast\twgamma_J\Big)
& | &
\mbox{by \gref{GfastEm}}
\\
& = &
E_{\bfm}\left(\left(\ka{Q}\right)^\ast\gamma_J\right)
& | &
\mbox{by \gref{GdefgJi2j}}
\\
& = &
E_{\bfm}\left(\alpha_J(Q)\right)
& | &
\mbox{by \gref{GdefaJ}.}
\end{array}
$$
\qed
\\

We define
\begin{equation} \label{GdefKPJ}
\rmK(P)_J = \left\{\left.(\alpha,\xi)\in\rmK(M)_J \right| E_{\bfm}(\alpha)=c(P)
\right\}\,.
\end{equation}
\begin{Theorem} \label{TRedPSUJ}
Let $P$ be a principal $\rmSU n$-bundle over a manifold $M$, $\dim M\leq 4$, and let
$J\in\rmK(n)$. Then the characteristic classes $\alpha_J$, $\xi_J$ define a
bijection from $\Red{P}{\SUJ}$ onto $\rmK(P)_J$.
\end{Theorem}
{\it Proof:}
Let $Q\in\Bun{M}{\SUJ}$. Then $\left(\alpha_J(Q),\xi_J(Q)\right)\in\rmK(M)_J$.
Lemma \rref{Lc(QSUn)} implies that $\left(\alpha_J(Q),\xi_J(Q)\right)\in
\rmK(P)_J$ if and only if $c\left(\ab{Q}{\rmSU n}\right)=c(P)$. Due to
$\dim M\leq 4$, the latter is equivalent to $\ab{Q}{\rmSU n}\cong P$, i.e.,
to $Q\in\Red{P}{\SUJ}$.
\qed
\\

The equation $E_{\bfm}(\alpha)=c(P)$ actually contains the two equations
$E_{\bfm}^{(2)}(\alpha) = 0$ and $E_{\bfm}^{(4)}(\alpha)=c_2(P)$. However, under 
the assumption that $(\alpha,\xi)\in\rmK(M)_J$, the first one is redundant,
because then, due to \gref{GgEJ2},
$E_{\bfm}^{(2)}(\alpha)=g\,E_{\twbfm}^{(2)}(\alpha)=
g\,\beta_g(\xi)=0$.
Thus, the relevant equations are
\begin{eqnarray} \label{GKMJ}
E_{\twbfm}^{(2)}(\alpha)
& = &
\beta_g(\xi)\,,
\\ \label{GKPJ}
E_{\bfm}^{(4)}(\alpha)
& = &
c_2(P)\,,
\end{eqnarray}
where $\alpha\in H^{(J)}(M,\ZZ)$, $\xi\in H^1(M,\ZZ_g)$. The set of solutions of
Eq.~\gref{GKMJ} yields $\rmK(M)_J$,
hence $\Bun{M}{\SUJ}$. The set of solutions of both Eqs.
\gref{GKMJ} and \gref{GKPJ} yields $\rmK(P)_J$ and, therefore,
$\Red{P}{\SUJ}$.

This concludes the classification of Howe subbundles of $P$, i.e., Step 2 of
our programme.
\subsection{Examples}
\label{SSex}
We are going to determine $\rmK(P)_J$, i.e., to solve the system of equations
\gref{GKMJ} and \gref{GKPJ}, for several choices of $J$ and for base
manifolds $M=\sphere{4},\sphere{2}\times\sphere{2},\torus{4}$, and
$\lens{p}{3}\times\sphere{1}$. Here $\lens{p}{3}$ denotes the
$3$-dimensional lens space which is defined to be
the quotient of the restriction of the natural action of $\rmU 1$ on the
sphere $\sphere{3}\subset\CC^2$ to the subgroup $\ZZ_p$. Note that
$\lens{p}{3}$ is orientable.
We remark that there are more general lens spaces, even in $3$ dimensions.
\paragraph{Preliminary Remarks}
Due to compactness and orientability, $H^4(M,\ZZ)\cong\ZZ$.
Let us derive the Bockstein homomorphism $\beta_g:H^1(M,\ZZ_g)\rightarrow
H^2(M,\ZZ)$. Since for products of spheres the integer-valued cohomology is
torsion-free, $\beta_g$ is trivial here. On the other hand, consider
$M=\lens{p}{3}\times\sphere{1}$. From the exact homotopy sequence induced 
by the fibration $\ZZ_p\hookrightarrow\sphere{3}\rightarrow\lens{p}{3}$ we 
infer
\begin{equation} \label{Gpi1Lp3}
\pi_1\left(\lens{p}{3}\right)\cong\ZZ_p\,.
\end{equation}
According to the Hurewicz and Universal Coefficient Theorems, then
\begin{equation}\label{GH1Ldp}
H^1\left(\lens{p}{3},\ZZ_g\right)
\cong \Hom\left(\pi_1\left(\lens{p}{3}\right),\ZZ_g\right)
\cong\Hom\left(\ZZ_p,\ZZ_g\right)
\cong\ZZ_{\langle p,g\rangle}
\,,
\end{equation}
where $\langle p,g \rangle$ denotes the greatest common divisor of $p$ and
$g$. Let $\gamma_{\lens{p}{3};\ZZ_g}^{(1)}$ and $\gamma_{\sphere{1}}^{(1)}$ 
be generators of $H^1(\lens{p}{3},\ZZ_g)$ and $H^1(\sphere{1},\ZZ)$,
respectively. Due to the K\"unneth Theorem for cohomology 
\cite[Ch. XIII, Thm. 11.2]{Massey} and \gref{GH1Ldp},
\begin{equation}\label{GH1LdpS}
\begin{array}[b]{rcl}
H^1(\lens{p}{3}\times\sphere{1},\ZZ_g)
& \cong &
H^1(\lens{p}{3},\ZZ_g)\otimes H^0(\sphere{1},\ZZ) \oplus
H^0(\lens{p}{3},\ZZ_g)\otimes H^1(\sphere{1},\ZZ)
\\
& \cong & 
\ZZ_{\langle p,g\rangle}\oplus\ZZ_g\,,
\end{array}
\end{equation}
where the first factor is generated by $\gamma_{\lens{p}{3};\ZZ_g}^{(1)}
\!\times\! 1_{\sphere{1}}$ and  the second one by
$1_{\lens{p}{3};\ZZ_g}\!\times\!\gamma_{\sphere{1}}^{(1)}$. 
Consider the following portion of the exact sequence \gref{GBock}:
\begin{equation}\label{Gsqcbg}
H^1(\lens{p}{3},\ZZ)
\stackrel{\varrho_g}{\longrightarrow}
H^1(\lens{p}{3},\ZZ_g)
\stackrel{\beta_g}{\longrightarrow}
H^2(\lens{p}{3},\ZZ)\,.
\end{equation}
One has, see \cite[\S 10, p.~363]{Bredon:Top},
\begin{equation} \label{GH1lens}
H^1(\lens{p}{3},\ZZ) =  0\,,~~~~H^2(\lens{p}{3},\ZZ)\cong\ZZ_p
\end{equation}
(the first equality follows also from \gref{Gpi1Lp3}). In view of 
\gref{GH1lens} and \gref{GH1Ldp}, \gref{Gsqcbg} implies
\begin{equation}\label{GbgLdp}
\beta_g\left(\gamma_{\lens{p}{3};\ZZ_g}^{(1)}\right) 
= 
\frac{p}{\langle p,g \rangle}\gamma_{\lens{p}{3};\ZZ}^{(2)}\,,
\end{equation}
where $\gamma_{\lens{p}{3};\ZZ}^{(2)}$ is an appropriately chosen generator
of $H^2(\lens{p}{3},\ZZ)$. Moreover, due to \gref{GH1lens},
$\beta_g(1_{\lens{p}{3};\ZZ_g})=0$. Thus, we obtain
\begin{equation} \label{GbgLdpS}
\beta_g\left(\gamma_{\lens{p}{3};\ZZ_g}^{(1)}\!\times\!1_{\sphere{1}}\right) 
= 
\frac{p}{\langle p,g \rangle}\gamma_{\lens{p}{3};\ZZ}^{(2)}\!\times\!
1_{\sphere{1}}
\,,~~~~
\beta_g\left(1_{\lens{p}{3};\ZZ_g}\!\times\!\gamma_{\sphere{1}}^{(1)}\right)
= 0\,.
\end{equation}
Finally, note that via the K\"unneth Theorem, \gref{GH1lens} implies
$H^2\left(\lens{p}{3}\times\sphere{1},\ZZ\right)\cong
\ZZ_p$. Since $H^4(\lens{p}{3}\times\sphere{1},\ZZ)$ is torsion-free, then the cup 
product $\alpha_1^{(2)}\smile\alpha_2^{(2)}$ is trivial.

Now let us discuss some special choices for $J$. For brevity, we write
$J$ in the form $J=(k_1,\dots,k_r|m_1,\dots,m_r)$.
\paragraph{\boldmath $J=(1|n)\in\rmK(n)$}
Here $\SUJ=\ZZ_n$, the center of $\rmSU n$. Moreover,
$g=n$. Variables are $\xi\in H^1(M,\ZZ_n)$ and $\alpha=1+\alpha^{(2)}$,
$\alpha^{(2)}\in H^2(M,\ZZ)$. The system of equations \gref{GKMJ} and
\gref{GKPJ} reads
\begin{eqnarray}\label{G(1|n)KMJ}
\alpha^{(2)} & = & \beta_n(\xi)
\\ \label{G(1|n)KPJ}
\frac{n(n-1)}{2}\alpha^{(2)}\smile\alpha^{(2)} & = & c_2(P)\,.
\end{eqnarray}
Note that we have used Lemma \rref{LEJ2,EJ4}. Eq.~\gref{G(1|n)KMJ} merely
expresses $\alpha^{(2)}$ in terms of $\xi$. In particular, it yields
$n\,\alpha^{(2)}=0$, so that Eq.~\gref{G(1|n)KPJ}
requires $c_2(P)=0$. As a result, $\rmK(P)_J$ is nonempty iff $P$ is trivial and
is then parametrized by $\xi$. This coincides with what is known about
$\ZZ_n$-subbundles of $\rmSU n$-bundles.
\paragraph{\boldmath $J=(n|1)\in\rmK(n)$}
Here $\SUJ=\rmSU n$, the whole group. Due to $g=1$, the only
variable is $\alpha=1+\alpha^{(2)}+\alpha^{(4)}$, where
$\alpha^{(2j)}\in H^{2j}(M,\ZZ)$, $j=1,2$. The system of equations
\gref{GKMJ} and \gref{GKPJ} is $$ \alpha^{(2)} =
0\,,~~~~\alpha^{(4)} = c_2(P)\,. $$ Thus, $\rmK(P)_J$ consists of $P$
itself.
\paragraph{\boldmath $J=(1,1|2,2)\in\rmK(4)$}
One can check that $\SUJ$ has connected components
$$
\{\diag(z,z,z^{-1},z^{-1})~|~z\in\rmU 1\}\,,~~~~
\{\diag(z,z,-z^{-1},-z^{-1})|z\in\rmU 1\}\,.
$$
It is, therefore, isomorphic to $\rmU 1\times\ZZ_2$. Variables are $\xi\in
H^1(M,\ZZ_2)$ and $\alpha_i=1+\alpha_i^{(2)}$, $\alpha_i^{(2)}\in
H^2(M,\ZZ)$, $i=1,2$. The system of equations under consideration is
\begin{eqnarray}\label{G(1,1|2,2)KMJ}
\alpha_1^{(2)}+\alpha_2^{(2)} & = & \beta_2(\xi)
\\ \label{G(1,1|2,2)KPJ}
\alpha_1^{(2)}\smile\alpha_1^{(2)} + \alpha_2^{(2)}\smile\alpha_2^{(2)}
+ 4 \alpha_1^{(2)}\smile\alpha_2^{(2)} & = & c_2(P)\,.
\end{eqnarray}
We solve Eq.~\gref{G(1,1|2,2)KMJ}  w.r.t.~$\alpha_2^{(2)}$ and
insert it into Eq.~\gref{G(1,1|2,2)KPJ}. Since $H^4(M,\ZZ)$ is torsion-free, 
products including $\beta_2(\xi)$ vanish. Thus, we obtain
\begin{equation} \label{G(1,1|2,2)}
-2\alpha_1^{(2)}\smile\alpha_1^{(2)} = c_2(P)\,.
\end{equation}
For base manifold $M=\sphere{4}$, $H^2(M,\ZZ)=0$. Hence, $\rmK(P)_J$ is 
nonempty iff $c_2(P)=0$,
in which case it contains the (necessarily trivial) $\rmU 1\times\ZZ_2$-bundle 
over $\sphere{4}$.

For $M=\lens{p}{3}\times\sphere{1}$, in case $c_2(P)=0$, $\rmK(P)_J$ is
parametrized by $\xi\in H^1(M,\ZZ_g)
\cong\ZZ_{\langle 2,p \rangle}\oplus\ZZ_2$ and $\alpha_1^{(2)}\in
H^2(M,\ZZ)\cong\ZZ_p$. Otherwise it is empty.

For $M=\sphere{2}\times\sphere{2}$, $H^1(M,\ZZ_g)=0$. Let 
$\gamma_{\sphere{2}}^{(2)}$ be a generator of $H^2(\sphere{2},\ZZ)$. Due to the
K\"unneth Theorem, $H^2(M,\ZZ)\cong\ZZ\oplus\ZZ$, where it is generated by
$\gamma_{\sphere{2}}^{(2)}\!\times\! 1_{\sphere{2}}$ and 
$1_{\sphere{2}}\!\times\!\gamma_{\sphere{2}}^{(2)}$. Moreover, $H^4(M,\ZZ)$ is
generated by $\gamma_{\sphere{2}}^{(2)}\!\times\!\gamma_{\sphere{2}}^{(2)}$. 
Writing 
\begin{equation}\label{Ga1S2S2}
\alpha_1^{(2)}=a~\gamma_{\sphere{2}}^{(2)}\!\times\! 1_{\sphere{2}}
+b~1_{\sphere{2}}\!\times\!\gamma_{\sphere{2}}^{(2)}
\end{equation}
with $a,b\in\ZZ$, Eq.~\gref{G(1,1|2,2)} becomes
$$
-4ab~\gamma_{\sphere{2}}^{(2)}\!\times\!\gamma_{\sphere{2}}^{(2)}
= 
c_2(P)\,.
$$
If $c_2(P)=0$, there are two series of solutions: $a=0$ and $b\in\ZZ$ as well
as $a\in\ZZ$ and $b=0$. Here $\rmK(P)_J$ is infinite. If
$c_2(P)=4l~\gamma_{\sphere{2}}^{(2)}\!\times\!\gamma_{\sphere{2}}^{(2)}$,
$l\neq 0$, then $a=q$ and $b=-l/q$, where $q$ runs through the (positive and
negative) divisors of $l$. Hence, in this case, the cardinality of $\rmK(P)_J$
is twice the number of divisors of $l$. If $c_2(P)$ is not
divisible by $4$ then $\rmK(P)_J$ is empty.

Finally, for $M=\torus{4}$ one has $H^1(M,\ZZ)\cong\ZZ_2^{\oplus 4}$ and 
$H^2(M,\ZZ)\cong\ZZ^{\oplus 6}$. Moreover, $H^2(M,\ZZ)$ is 
generated by elements 
$\gamma^{(2)}_{\torus{4};ij}$, $1\leq i<j\geq 4$, where 
$$
\gamma_{\torus{4};12}^{(2)} 
= 
\gamma_{\sphere{1}}^{(1)}\!\times\!
\gamma_{\sphere{1}}^{(1)}\!\times\!
1_{\sphere{1}}\!\times\! 
1_{\sphere{1}}
\,,~~~~
\gamma_{\torus{4};13}^{(2)} 
= 
\gamma_{\sphere{1}}^{(1)}\!\times\! 
1_{\sphere{1}}\!\times\!
\gamma_{\sphere{1}}^{(1)}\!\times\!
1_{\sphere{1}}
\,,~~~~\mbox{etc.,} 
$$
whereas $H^4(M,\ZZ)$ is generated by
$\gamma_{\torus{4}}^{(4)} = \gamma_{\sphere{1}}^{(1)}\!\times\! 
\gamma_{\sphere{1}}^{(1)}\!\times\! \gamma_{\sphere{1}}^{(1)}\!\times\! 
\gamma_{\sphere{1}}^{(1)}$. One can check 
\begin{equation}\label{GcupT4}
\gamma^{(2)}_{\torus{4};ij}\smile\gamma^{(2)}_{\torus{4};kl}
=
\epsilon_{ijkl}~\gamma^{(4)}_{\torus{4}}\,,
\end{equation}
where $\epsilon_{ijkl}$ denotes the totally antisymmetric tensor in $4$
dimensions. Writing 
\begin{equation}\label{Ga1T4}
\alpha_1^{(2)}=\sum_{1\leq i<j\leq 4} a_{ij}\gamma^{(2)}_{\torus{4};ij}
\end{equation}
and using \gref{GcupT4}, Eq.~\gref{G(1,1|2,2)} yields 
$$
-4\left(a_{12}a_{34}-a_{13}a_{24}+a_{14}a_{23}\right)\gamma_{\torus{4}}^{(4)}
=
c_2(P)\,.
$$
Hence, we find that $\rmK(P)_J$ is again nonempty iff $c_2(P)$ is divisible by $4$, 
in which case it now has always infinitely many elements.

To conclude, let us point out that, when passing to $\hat{\rmK}(P)$, one has to
identify the symbol containing $(\alpha_1,\alpha_2)$ with that containing
$(\alpha_2,\alpha_1)$.
\paragraph{\boldmath $J=(1,1|2,3)\in\rmK(5)$}
The subgroup $\SUJ$ of $\rmSU 5$ consists of matrices of the form
$\diag(z_1,z_1,z_2,z_2,z_2)$, where $z_1,z_2\in\rmU 1$ such that $z_1^2z_2^3=1$. 
We can
parametrize $z_1=z^3$, $z_2=z^{-2}$, $z\in\rmU 1$.
Hence, $\SUJ$ is isomorphic to $\rmU 1$. Variables are
$\alpha_i=1+\alpha_i^{(2)}$, $i=1,2$. The equations to be solved read
\begin{eqnarray}\label{G(1,1|2,3)KMJ}
2\alpha_1^{(2)}+3\alpha_2^{(2)} & = & 0
\\ \label{G(1,1|2,3)KPJ}
\alpha_1^{(2)}\smile\alpha_1^{(2)}
+ 3 \alpha_2^{(2)}\smile\alpha_2^{(2)}
+ 6 \alpha_1^{(2)}\smile\alpha_2^{(2)}
& = & c_2(P)
\end{eqnarray}
Eq.~\gref{G(1,1|2,3)KMJ} can be parametrized by $\alpha_1^{(2)} = 3\eta$,
$\alpha_2^{(2)}=-2\eta$, where $\eta\in H^2(M,\ZZ)$. Then Eq.
\gref{G(1,1|2,3)KPJ} becomes
$$
-15 \eta\smile\eta = c_2(P)\,.
$$
The discussion of this equation is analogous to that of Eq.
\gref{G(1,1|2,2)} above. 
\paragraph{\boldmath $J=(2,3|1,1)\in\rmK(5)$}
Here $\SUJ \cong \rmS(\rmU 2\times\rmU 3)$, the symmetry
group of the standard model. In the grand unified $\rmSU 5$-model
this is the subgroup to which $\rmSU 5$ is broken by the heavy
Higgs field. Moreover, the subgroup $\SUJ$ is the centralizer of the subgroup
$\rmSU(1,1|2,3)$ discussed above. Variables are
$\alpha_i=1+\alpha_i^{(2)}+\alpha_i^{(4)}$, where
$\alpha_i^{(2j)}\in H^{2j}(M,\ZZ)$, $i,j=1,2$. Eqs. \gref{GKMJ}
and \gref{GKPJ} read
\begin{eqnarray}\label{G(2,3|1,1)KMJ}
\alpha_1^{(2)}+\alpha_2^{(2)}
& = &
0\,,
\\ \label{G(2,3|1,1)KPJ}
\alpha_1^{(4)}+\alpha_2^{(4)}+\alpha_1^{(2)}\smile\alpha_2^{(2)}
& = &
c_2(P)\,.
\end{eqnarray}
Using \gref{G(2,3|1,1)KMJ} to replace $\alpha_2^{(2)}$ in
\gref{G(2,3|1,1)KPJ} we obtain for the latter
$$
\alpha_2^{(4)}=c_2(P)-\alpha_1^{(4)}+\alpha_1^{(2)}\smile\alpha_1^{(2)}\,.
$$
Thus, $\rmK(P)_J$ can be parametrized by $\alpha_1$ (or $\alpha_2$), i.e., by the
Chern class of one of the factors $\rmU 2$ or $\rmU 3$. This has been known for a 
long time \cite{Isham}.
\paragraph{\boldmath $J=(2|2)$}
The subgroup $\SUJ$ of $\rmSU 4$ consists of matrices $D\oplus D$, where $D\in\rmU
2$ such that $(\det D)^2=1$. Hence, it has connected components 
$\{D\oplus D | D\in\rmSU 2\}$ and $\{(iD)\oplus(iD) | D\in\rmSU 2\}$.
One can check that the map $\rmSU 2\times\ZZ_4\rightarrow\SUJ$, 
$(D,a)\mapsto e^{2\pi ia/4}D$, 
induces an isomorphism from $(\rmSU 2\times\ZZ_4)/\ZZ_2$ onto $\SUJ$.

Now consider $\rmK(P)_J$. Variables are $\xi\in
H^1(M,\ZZ_2)$ and $\alpha=1+\alpha^{(2)}+\alpha^{(4)}$. We have
\begin{eqnarray} \label{G(2|2)KMJ}
\alpha^{(2)} & = & \beta_2(\xi)
\\ \label{G(2|2)KPJ}
\alpha^{(2)}\smile\alpha^{(2)} + 2\alpha^{(4)} & = & c_2(P)
\end{eqnarray}
Eq.~\gref{G(2|2)KMJ} fixes $\alpha^{(2)}$ in terms of $\xi$. For
example, in case $M=\lens{p}{3}\times\sphere{1}$, by expanding 
$
\xi 
= 
\xi_\rmL~\gamma_{\lens{p}{3};\ZZ_g}^{(1)}\!\times\!1_{\sphere{1}} 
+
\xi_\rmS~1_{\lens{p}{3}}\!\times\!\gamma_{\sphere{1}}^{(1)}
$,
Eqs. \gref{GbgLdpS} and \gref{G(2|2)KMJ} imply
$$
\alpha^{(2)} = \left\{\begin{array}{ccl}
q\xi_L~\gamma_{\lens{p}{3};\ZZ}^{(2)}\!\times\!1_{\sphere{1}}
& | & p=2q \\
0 & | & p=2q+1.
\end{array}\right.
$$
For general $M$, due to \gref{G(2|2)KMJ}, Eq.~\gref{G(2|2)KPJ} becomes 
$2 \alpha^{(4)} =
c_2(P)$. Thus, $\rmK(P)_J$ is nonempty iff $c_2(P)$ is even and is then 
parametrized by $\xi\in H^1(M,\ZZ_2)$. 

Let us point out the following. Consider Eq.~\gref{G(2|2)KMJ} which
defines the set $\rmK(M)_J$ classifying $\SUJ$-bundles
over $M$. In case $M$ is a product of spheres (or, more generally, has 
trivial $\beta_2$ in degree $1$), this classification coincides with that of
$(\rmSU 2\times\ZZ_2)$-bundles, although this group is only locally
isomorphic to $\SUJ$. In case $M=\lens{p}{3}\times\sphere{1}$, there is a
difference though. Here the $\SUJ$-bundles which are
nontrivial over the factor $\lens{p}{3}$ have a magnetic 
charge, whereas an $(\rmSU 2\times\ZZ_2)$-bundle can never have one.
\section{Holonomy-Induced Howe Subbundles}
\label{Sholind}
In the next step of our programme we have to specify the Howe subbundles
which are holonomy-induced. However, in fact, this turns out not to be 
necessary here because for Howe
subbundles of principal $\rmSU n$-bundles the condition of being
holonomy-induced is redundant. This will be proved now.
\begin{Lemma} \label{Ldimarg}
Let $H\subseteq H'\subseteq\rmSU n$ be Howe subgroups. If $\dim H=\dim H'$
then $H=H'$.
\end{Lemma}
{\it Proof:}
There exist $J,J'\in\rmK(n)$ such that $H$ and $H'$ are conjugate to
$\SUJ$ and $\SUJ'$, respectively.
Consider $\UJ$ and $\UJ'$. Due to $H\subseteq H'$, there exists $D\in\rmSU n$
such that $D^{-1}\UJ D\subseteq\UJ'$. Moreover,
\begin{equation} \label{GLdimarg2}
\dim\UJ'=\dim\SUJ'+1=\dim H'+1 = \dim H+1 = \dim\SUJ+1=\dim\UJ\,.
\end{equation}
$\UJ'$ being connected and $D^{-1}\UJ D$ being closed in $\UJ'$,
\gref{GLdimarg2} implies $D^{-1}\UJ D=\UJ'$. Then
$D^{-1}\SUJ D = D^{-1}\left(\UJ\cap\rmSU n\right)D =
\left(D^{-1}\UJ D\right)\cap\rmSU n = \UJ'\cap\rmSU n=\SUJ'$. It follows $H=H'$.
\qed
\begin{Theorem}\label{Tcond(A)}
Any Howe subbundle of a principal $\rmSU n$-bundle is holonomy-induced.
\end{Theorem}
{\it Proof:} Let $P$ be a principal $\rmSU n$-bundle and let $Q$ be a Howe
subbundle of $P$ with structure group $H$. Denote the structure group of a
connected component of $Q$ by $\tilde{H}$. It is easily seen that $Q$ is
holonomy-induced provided
\begin{equation} \label{GTcond(A)proof}
\rmC_{\rmSU n}^2(\tilde{H})=H\,.
\end{equation}
Since $H$ is a Howe subgroup, we have $\tilde{H} \subseteq \rmC_{\rmSU n}^2(\tilde{H})
\subseteq H$. Since $\tilde{H}$ and $H$ are of the same dimension, so
are $\rmC_{\rmSU n}^2(\tilde{H})$ and $H$. By virtue of Lemma
\rref{Ldimarg}, this implies \gref{GTcond(A)proof}. 
\qed
\\

For the reader who wonders whether there exist Howe subbundles
which are not holonomy-induced we give an example here. Consider the Lie
group $\rmSO 3$. One checks that the subgroup $H=\{\II_3,\diag(-1,-1,1)\}$
is Howe. Thus, the subbundle $Q=M\times H$ of the bundle $M\times\rmSO 3$ is
Howe. Any connected subbundle $\twQ$ of $Q$ has the center $\{\II_3\}$ as its 
structure group. Since the center is Howe itself, $\twQ\cdot \rmC_G^2(\{\II_3\}) 
= \twQ\neq Q$.
\section{Factorization by $\rmSU n$-Action}
\label{SP}
In Step 4 of our programme to determine $\Howe_\ast(P)$, we actually have to
take the disjoint
union of $\Red{P}{H}$ over all Howe subgroups $H$ of $\rmSU n$ and to
factorize by the action of $\rmSU n$. Since $\rmSU n$-action on subbundles
conjugates their structure groups, however, it suffices to take the union
only over $\SUJ$, $J\in\rmK(n)$:
\begin{equation} \label{GsHSB}
\bigsqcup_{J\in\rmK(n)}\Red{P}{\SUJ}\,.
\end{equation}
According to this, define
\begin{equation} \label{GdefKP}
\rmK(P)=\bigsqcup_{J\in\rmK(n)}\rmK(P)_J\,.
\end{equation}
We shall denote the elements of $\rmK(P)$ by $\calJ$ and write them in the form
$\calJ=(J;\alpha,\xi)$, where $J\in\rmK(n)$ and $(\alpha,\xi)\in\rmK(P)_J$.
Due to Theorem \rref{TRedPSUJ}, the collection of characteristic
classes $\{(\alpha_J,\xi_J)|J\in\rmK(n)\}$ defines a bijection from \gref{GsHSB}
onto $\rmK(P)$. Now we reverse this bijection: Let $\calJ\in\rmK(P)$,
$\calJ=(J;\alpha,\xi)$.
Then define $ Q_\calJ$ to be the isomorphism class of $\SUJ$-subbundles of $P$
which obey
\begin{equation} \label{GdefQJ}
\alpha_J( Q_\calJ)=\alpha\,,~~~~\xi_J( Q_\calJ)=\xi\,.
\end{equation}
\begin{Lemma} \label{LcjgP}
Let $\calJ,\calJ'\in\rmK(P)$, where $\calJ=(J;\alpha,\xi)$ and
$\calJ'=(J';\alpha',\xi')$. There exists $D\in\rmSU n$ such that
$ Q_{\calJ'}= Q_\calJ\cdot D$ (up to isomorphy) 
if and only if $\xi'=\xi$, as well as $J'=\sigma J$ and $\alpha'=\sigma\alpha$
for some permutation $\sigma$ of $1,\dots,r$.
\end{Lemma}
{\it Proof:}
For $J_1,J_2\in\rmK(n)$ and $D\in\rmSU n$ such that
$D^{-1}\,\SUJ_1\,D\subseteq\SUJ_2$, define embeddings
\begin{eqnarray}\nonumber
h_{J_1J_2}^D & : & \UJ_1\rightarrow\UJ_2\,,~~C\mapsto D^{-1}CD
\\ \nonumber
h_{J_1J_2}^{D;\rmS} & : &\SUJ_1\rightarrow\SUJ_2\,,~~C\mapsto D^{-1}CD\,.
\end{eqnarray}
First, assume that we are given $D\in\rmSU n$ such that $ Q_{\calJ'}\cong Q_\calJ\cdot D$. 
Then $D^{-1}\,\SUJ\,D=\SUJ'$, so that we can consider the isomorphisms
$h_{JJ'}^{D;\rmS}:\SUJ\rightarrow\SUJ'$ and
$h_{JJ'}^D:\UJ\rightarrow\UJ'$. One can
check that
\begin{equation} \label{GQJA=QJhAJ}
 Q_\calJ\cdot D\cong\ab{ Q_\calJ}{h_{JJ'}^{D;\rmS}}\,.
\end{equation}
Moreover, there exists a permutation $\sigma$ of $1,\dots, r$ such that
$h_{JJ'}^D$ maps the $\sigma(i)$-th factor of $\UJ$ isomorphically onto the
$i$-th factor of $\UJ'$. Then, in particular, $J'=\sigma J$. Moreover, there 
exists $C\in\rmSU n$ such that
$\prU{J'}{i}\circ h_{JJ'}^C = \prU{J}{\sigma(i)}$ $\forall~i$
(in fact,
$C$ has been constructed in the proof of Lemma \rref{Lcjg}). Then
\begin{equation} \label{GphAJ}
\prU{J'}{i}\circ j_J \circ h_{JJ'}^{C;\rmS}=\prU{J}{\sigma(i)}\circ j_J \,.
\end{equation}
Next, define $B=DC^{-1}\in\rmSU n$. The corresponding homomorphism
$h_{JJ}^B$ is an automorphism of $\UJ$
which leaves each factor invariant separately. One can check that it is an
{\it inner} automorphism of $\UJ$. Then $h_{JJ}^{B;\rmS}$ is an inner automorphism
of $\SUJ$ and can, therefore, be generated by an element of the connected
component of the identity. Then $\rmB h_{JJ}^{B;\rmS}$ is homotopic to 
$\rmB\id_{\SUJ}\equiv\id_{\BSUJ}$. As a consequence, the classifying map of the 
subbundle $Q_\calJ\cdot B$, which is, due to \gref{GQJA=QJhAJ} and
\gref{Gclfmapassbun}, given by 
$
\ka{ Q_\calJ\cdot B} = \ka{\ab{ Q_\calJ}{h_{JJ}^{B;\rmS}}} = \rmB h_{JJ}^{B;\rmS}\circ\ka{ 
Q_\calJ}$,
is homotopic to that of $Q_\calJ$.
It follows $Q_{\calJ'}\cong Q_\calJ\cdot
C$. Then, due to \gref{GQJA=QJhAJ},
\begin{equation}\label{GQLpQL}
Q_{\calJ'} \cong \ab{ Q_\calJ}{h_{JJ'}^{C;\rmS}}\,.
\end{equation}
We compute the characteristic classes of $\ab{Q_\calJ}{h_{JJ'}^{C;\rmS}}$ in
terms of those of $Q_\calJ$:
\begin{equation} \label{GaJpi}
\begin{array}[b]{rclcl}
\alpha_{J',i}\left(\ab{ Q_\calJ}{h_{JJ'}^{C;\rmS}}\right)
& = &
c\left(\ab{\left(\ab{ Q_\calJ}{h_{JJ'}^{C;\rmS}}\right)}{\prU{J'}{i}\circ j_J }\right)
& | &
\mbox{by \gref{GaJc}}
\\
& = &
c\left(\ab{ Q_\calJ}{\prU{J'}{i}\circ j_J\circ h_{JJ'}^{C;\rmS}}\right)
&&
\\
& = &
c\left(\ab{ Q_\calJ}{\prU{J}{\sigma(i)}\circ j_J }\right)
& | &
\mbox{by \gref{GphAJ}}
\\
& = &
\alpha_{J,\sigma(i)}\left( Q_\calJ\right)\,,
&&
\end{array}
\end{equation}
as well as, using $\lambda^\rmS_{J'}\circ h_{JJ'}^{C;\rmS} =
\lambda^\rmS_J$ and \gref{GxJc}, 
\begin{equation} \label{GxJp}
\xi_{J'}\left(\ab{ Q_\calJ}{h_{JJ'}^{C;\rmS}}\right)
=
\chi_g\left(\ab{\left(\ab{ Q_\calJ}{h_{JJ'}^{C;\rmS}}
\right)}{\lambda^\rmS_{J'}}\right)
=
\chi_g\left(\ab{ Q_\calJ}{\lambda^\rmS_{J}}\right)
=
\xi_J\left( Q_\calJ\right)\,.
\end{equation}
Thus,
$$
\begin{array}{rclcl}
\alpha'_i
& = &
\alpha_{J',i}\left( Q_{\calJ'}\right)
& | & 
\mbox{by \gref{GdefQJ}}
\\
& = &
\alpha_{J',i}\left(\ab{ Q_\calJ}{h_{JJ'}^{C;\rmS}}\right)
& | &
\mbox{by \gref{GQLpQL}}
\\
& = &
\alpha_{J,\sigma(i)}\left( Q_\calJ\right)
& | &
\mbox{by \gref{GaJpi}}
\\
& = & 
\alpha_{\sigma(i)}
& | & 
\mbox{by \gref{GdefQJ},}
\end{array}
$$
i.e., $\alpha'=\sigma\alpha$. In an analogous way, using \gref{GxJp}, we
obtain $\xi'=\xi$. 

To prove the converse implication, assume $\xi'=\xi$ and let $\sigma$ be a permutation of
$1,\dots, r$ such that $J'=\sigma J$ and $\alpha'=\sigma\alpha$. Due to
Lemma \rref{Lcjg}, there exists $D\in\rmSU n$
such that $\SUJ'=D^{-1}\,\SUJ\,D$. As shown in the proof of this lemma, $D$
can be chosen in such a way that \gref{GphAJ} holds. Then \gref{GaJpi}
holds, too. We compute the
characteristic classes of $Q_\calJ\cdot D$:
$$
\begin{array}{rclcl}
\alpha_{J',i}\left( Q_\calJ\cdot D\right)
& = &
\alpha_{J^\prime,i}\left(\ab{ Q_\calJ}{h_{JJ'}^{D;\rmS}}\right)
& | & 
\mbox{by \gref{GQJA=QJhAJ}}
\\
& = &
\alpha_{J,\sigma(i)}\left( Q_\calJ\right)
& | & 
\mbox{by \gref{GaJpi}}
\\
& = &
\alpha_{\sigma(i)}
& | & 
\mbox{by \gref{GdefQJ}}
\\
& = &
\alpha'_i
& | &
\mbox{by assumption}
\\
& = &
\alpha_{J',i}\left( Q_{\calJ'}\right)
& | & 
\mbox{by \gref{GdefQJ}.}
\end{array}
$$
Analogously, using \gref{GxJp}, we find $\xi_{J'}\left( Q_\calJ\cdot D\right)
=
\xi_{J'}\left( Q_{\calJ'}\right)$. It follows $Q_\calJ\cdot D \cong Q_{\calJ'}$.
\qed
\\

As suggested by Lemma \rref{LcjgP}, we introduce an equivalence relation
on the set $\rmK(P)$: Let $\calJ,\calJ'\in\rmK(P)$, where $\calJ=(J;\alpha,\xi)$ and
$\calJ'=(J';\alpha',\xi')$. Write $\calJ\sim\calJ'$ iff $\xi'=\xi$ and there
exists a permutation $\sigma$ of $1,\dots, r$ such that $J'=\sigma J$ and
$\alpha'=\sigma\alpha$.
The set of equivalence classes will be denoted by $\hat{\rmK}(P)$.
\begin{Theorem} \label{THSB}
The assignment $\calJ\mapsto Q_\calJ$ induces a bijection from
$\hat{\rmK}(P)$ onto $\Howe_\ast(P)$.
\end{Theorem}
{\it Proof:}
The assignment $\calJ\mapsto Q_\calJ$ induces a map
$\rmK(P)\rightarrow\Howe_\ast(P)$.
This map is surjective by construction. Due to Lemma \rref{LcjgP}, it
projects to $\hat{\rmK}(P)$ and the projected map is injective.
\qed
\\

With Theorem \rref{THSB} we have accomplished the determination of
$\Howe_\ast(P)$ and, therefore, of the set of orbit types
$\OT\left(\con^k,\gau^{k+1}\right)$. Calculations for the latter can now
be performed entirely on the level of the classifying set $\hat{\rmK}(P)$.
\section{Example: Gauge Orbit Types for $\rmSU 2$}
\label{SSU2}
In this section, we are going to determine $\OT\left(\con^k,\gau^{k+1}\right)$ for
an $\rmSU 2$-gauge theory over the base manifolds discussed in Subsection
\rref{SSex}. The set $\rmK(2)$ contains the elements
$$
J_a=(1|2)\,,~~~~
J_b=(1,1|1,1)\,,~~~~
J_c=(2|1)\,.
$$
Here $\SUJ_a=\{\pm\II\}\cong\ZZ_2$ is the center, 
$\SUJ_b=\left\{\left.\diag\left(z,z^{-1}\right)~\right|~z\in\rmU
1\right\}\cong\rmU 1$ is the toral subgroup
and $\SUJ_c=\rmSU 2$. The strata corresponding to the
elements of $\rmK(P)_{J_a}$, $\rmK(P)_{J_b}$, $\rmK(P)_{J_c}$ are, in the
respective order, those with
stabilizers isomorphic to $\rmSU 2$, $\rmU 1$, and the generic stratum.
Accordingly, we shall refer to the first class as $\rmSU 2$-strata and to
the second class as $\rmU 1$-strata. We have
$$
\rmK(P) = \rmK(P)_{J_a}\cup\rmK(P)_{J_b}\cup\rmK(P)_{J_c}
$$
(disjoint union). As we already know, $\rmK(P)_{J_a}$ is parametrized
by $\xi\in H^1(M,\ZZ_2)$ in case $P$ is trivial and is empty
otherwise. Moreover, $\rmK(P)_{J_c}$ consists of $P$ itself. For
$\rmK(P)_{J_b}$,
variables are $\alpha_i=1+\alpha_i^{(2)}$,
$\alpha_i^{(2)}\in H^2(M,\ZZ)$, $i=1,2$. The system of equations \gref{GKMJ},
\gref{GKPJ} reads
\begin{eqnarray}\label{GSU21}
\alpha_1^{(2)}+\alpha_2^{(2)} & = & 0
\\ \label{GSU22}
\alpha_1^{(2)}\smile\alpha_2^{(2)} & = & c_2(P)
\end{eqnarray}
Using \gref{GSU21} to replace $\alpha_2^{(2)}$ in \gref{GSU22} we obtain
\begin{equation} \label{GSU23}
-\alpha_1^{(2)}\smile\alpha_1^{(2)} = c_2(P)\,.
\end{equation}
Note that here $\alpha_1^{(2)}$ is just the first Chern class of a reduction
of $P$ to the subgroup $\rmU 1$. According to this, Eq.~\gref{GSU23} has
been derived in the discussion of spontaneous symmetry breaking of $\rmSU 2$
to $\rmU 1$, see \cite{Isham}. Note also that when passing from $\rmK(P)$ to
$\hat{\rmK}(P)$, the pairs $(\alpha_1,\alpha_2)$ and $(\alpha_2,\alpha_1)$ label the
same class of subbundles. Hence, solutions $\alpha_1^{(2)}$ of \gref{GSU23}
have to be identified with their negative.

Let us now discuss Eq.~\gref{GSU23} as well as the set $\hat{\rmK}(P)$ in
dependence of $P$ for some specific base manifolds $M$.
\paragraph{\boldmath$M=\sphere{4}$}
Since $H^2(M,\ZZ)=0$, Eq.~\gref{GSU23} requires $c_2(P)=0$. Thus,
in case $P$
is trivial, $\hat{\rmK}(P)$ contains the $\ZZ_2$-bundle, the $\rmU 1$-bundle (both
necessarily trivial) and $P$ itself. Accordingly, in the gauge orbit space
there exist, besides the generic stratum, an $\rmSU 2$-stratum and a
$\rmU 1$-stratum.

We remark that, due to the base manifold being a
sphere, this result can be obtained much easier by homotopy arguments. It
is, therefore, well known. The structure of the gauge orbit space
in the present situation has been studied in great
detail in \cite{FSS}. It has been shown that the two
nongeneric strata of the gauge orbit space can be parametrized by means of
an affine subspace of $\con^k$ which is acted upon by the Weyl group
of $\rmSU 2$ (the latter being just $\ZZ_2$).

In case $P$ is nontrivial, both $\rmK(P)_{J_b}$ and $\rmK(P)_{J_a}$ are empty,
so that $\hat{\rmK}(P)$ contains only the bundle $P$ itself. Accordingly, there do
not exist nongeneric strata in the gauge orbit space.
\paragraph{\boldmath$M=\sphere{2}\times\sphere{2}$}
We write $\alpha_1^{(2)}$ in the form \gref{Ga1S2S2}. Then Eq.~\gref{GSU23}
becomes
$$
-2ab~\gamma_{\sphere{2}}^{(2)}\!\times\!\gamma_{\sphere{2}}^{(2)}
=
c_2(P)\,.
$$
Thus, if $P$ is trivial then $\hat{\rmK}(P)$ contains, in addition to $P$ itself, 
the $\ZZ_2$-bundle, which
is trivial, and the $\rmU 1$-bundles labelled by $a\geq 0$, $b=0$ and $a=0$,
$b\geq 0$, i.e., which are trivial over one of the $2$-spheres. 
The corresponding nongeneric
strata in the gauge orbit space are one $\rmSU 2$-stratum and infinitely
many $\rmU 1$-strata.

In case $c_2(P)=2l~\gamma_{\sphere{2}}^{(2)}\!\times\!\gamma_{\sphere{2}}^{(2)}$, 
$l\neq 0$, $\hat{\rmK}(P)$ contains the $\rmU 1$-bundles with $a=q$ and
$b=-l/q$, where $q$ is a (positive) divisor of $m$. Hence, here the nongeneric 
part of the gauge orbit space consists of finitely many $\rmU 1$-strata.

Finally, in case 
$c_2(P)=(2m+1)~\gamma_{\sphere{2}}^{(2)}\!\times\!\gamma_{\sphere{2}}^{(2)}$, 
$\hat{\rmK}(P)$ contains only $P$ itself and the gauge
orbit space consists only of the generic stratum.
\paragraph{\boldmath$M=\torus{4}$}
We use the notation introduced in Subsection \rref{SSex}. Writing $\alpha_1^{(2)}$ 
in the form \gref{Ga1T4} and using \gref{GcupT4}, Eq.~\gref{GSU23} reads
\begin{equation} \label{GSU24}
-2\left(a_{12}a_{34}-a_{13}a_{24}+a_{14}a_{23}\right)
\gamma_{\torus{4}}^{(4)} = c_2(P)\,.
\end{equation}
Hence, the result is similar to that for the case
$M=\sphere{2}\times\sphere{2}$. The
only difference is that, due to $H^1(M,\ZZ_2)\cong\ZZ_2^{\oplus 4}$,
there exist $16$ different $\ZZ_2$-bundles which are all contained in
$\hat{\rmK}(P)$
for $P$ being trivial. Accordingly, in this case the gauge orbit space
contains $16$ $\rmSU 2$-strata. Moreover, in case
$c_2(P)=2m\gamma_{\torus{4}}^{(4)}$, $m\neq 0$,
the number of solutions of \gref{GSU24} is infinite. Therefore, in this case
there exist infinitely many $\rmU 1$-strata in the gauge orbit space.

We remark that pure Yang-Mills theory on $\torus{4}$ has been discussed in
\cite{Tok1,Tok2}, where the authors have studied the
maximal Abelian gauge for gauge group $\rmSU n$. They have found that in
case $P$ is nontrivial this gauge fixing is necessarily singular on Dirac
strings joining magnetically charged defects in the base manifold. It is
an interesting question whether there is a relation between such defects
and the nongeneric strata of the gauge orbit space.
\paragraph{\boldmath$M=\lens{p}{3}\times\sphere{1}$}
Here \gref{GSU23} requires $c_2(P)=0$. Thus, in case $P$ is trivial $\hat{\rmK}(P)$
contains, in addition to $P$ itself, the $\ZZ_2$-bundles over
$\lens{p}{3}\times\sphere{1}$, which are labelled by the elements of
$$
H^1(\lens{p}{3}\times\sphere{1},\ZZ_2)\cong\left\{\begin{array}{ccl}
\ZZ_2\oplus\ZZ_2 & | & p\neq \mbox{$0$, even}
\\
\ZZ_2 & | & \mbox{otherwise}
\end{array}\right.
$$
as well as the $\rmU 1$-bundles over $\lens{p}{3}\times\sphere{1}$, which
are labelled by
$\alpha_1^{(2)}\in H^2(\lens{p}{3}\times\sphere{1},\ZZ)\cong\ZZ_p$,
modulo the identification $\alpha_1^{(2)}\sim-\alpha_1^{(2)}$. According to
this, if $p$ is even, the
nongeneric part of the gauge orbit space contains $4$ $\rmSU 2$-strata and
$(p/2+1)$ $\rmU 1$-strata. If $p$ is odd, it contains $2$ $\rmSU 2$-strata and
$((p+1)/2)$ $\rmU 1$-strata.

In case $P$ is nontrivial there do not exist nongeneric strata in the gauge
orbit space.
\section[Application: \\ Kinematical Nodes in $2+1$ Dimensional Chern-Simons 
Theory]{Application: Kinematical Nodes in \\ $2\!+\!1$ Dimensional
Chern-Simons Theory}
\label{SNodes}
Following \cite{Asorey:Nodes}, we consider Chern-Simons theory with gauge
group $\rmSU n$ in the Hamiltonian approach. The Hamiltonian in Schr\"odinger 
representation is given by 
$$
H
=
-\frac{\Lambda}{2}\int_{\Sigma_s} \frac{\diff^2x}{\sqrt{h}}
\Tr\left|\frac{\delta}{\delta A_\mu}+\frac{\rmi
\ell}{4\pi}\epsilon^{\mu\nu}A_\nu \right|^2 +
\frac{1}{4\Lambda}\int_{\Sigma_s}\diff^2 x\sqrt{h}
\Tr\left(F_{\mu\nu}F^{\mu\nu}\right)\,. 
$$ 
Geometrically, $A_\mu$
and $F_{\mu\nu}$ are the local representatives of a connection $A$ in a
given $\rmSU n$-bundle $P$ over the $2$ dimensional space
$\Sigma_s$ and its curvature $F$, respectively. We assume
$\Sigma_s$ to be a Riemann surface of genus $s$. Note
that, due to the dimension of $\Sigma_s$, $c_2(P)=0$, i.e., $P$ is
trivial. The configuration space is the
gauge orbit space $\orb^k$ associated to $P$. Physical states are
given by cross sections of a complex line bundle $\eta$ over 
(the topological space) $\orb^k$ with
first Chern class $c_1(\eta)=\ell$. Correspondingly, they can be thought
of as functionals $\psi$ on the space $\con^k$ of connections
in $P$ which are subject to the Gauss law condition
\begin{equation} \label{GGauss}
\nabla_A^\mu\frac{\delta}{\delta A^\mu(x)}\psi(A)
=
\frac{\rmi\ell}{4\pi}\epsilon^{\mu\nu}\partial_\mu A_\nu(x)\psi(A)\,.
\end{equation}
Here $\nabla_A$ denotes the covariant derivative w.r.t.~$A$. 
In \cite{Asorey:Nodes} it has been shown that if $A$ carries a 
nontrivial magnetic charge, i.e., if it can be reduced to some 
subbundle of $P$ with nontrivial
first Chern class, all physical states obey $\psi(A)=0$. Such a connection
is called a {\it kinematical} node. (Note that,
due to $\eta$ being nontrivial, there exist also dynamical 
nodes, which differ from state to state.) The authors of 
\cite{Asorey:Nodes} argue that
nodal gauge field configurations are relevant for the confinement
mechanism. In the following we shall show that being a node is a
property of strata. For that purpose, we reformulate the result of
\cite{Asorey:Nodes} in our language.
\begin{Theorem} \label{TNodes}
Let $A\in\con^k$ have orbit type $[\calJ]\in\hat{\rmK}(P)$, where
$\calJ=(J;\alpha,\xi)$. If $\alpha_i^{(2)}\neq 0$ for some $i$ then $A$
is a kinematical node, i.e., $\psi(A)=0$ for all physical states $\psi$.
\end{Theorem}
{\it Proof:} The proof follows the lines of
\cite{Asorey:Nodes}. By assumption, $A$ can be 
reduced to a connection on an $\SUJ$-subbundle $Q\subseteq P$ of 
class $(\alpha_J(Q),\xi_J(Q))=(\alpha,\xi)$.
$\Sigma_s$ being a compact orientable $2$-manifold, $H^2(\Sigma_s,\ZZ)\cong\ZZ$. 
Hence, $\alpha_i^{(2)}=c_i\gamma^{(2)}$, where $c_i\in\ZZ$ 
and $\gamma^{(2)}$ is a generator of $H^2(\Sigma_s,\ZZ)$. We define a $1$-parameter 
subgroup $\{ \tilde{\Phi}_t~|~t\in\RR \}$ of $\rmSU n$ by
$$
\twPhi_t
=
\left(\exp\left\{i\frac{c_1}{k_1}t\right\}\II_{k_1}
\otimes\II_{m_1}\right)
\oplus\cdots\oplus
\left(\exp\left\{i\frac{c_r}{k_r}t\right\}\II_{k_r}
\otimes\II_{m_r}\right)\,.
$$
Using
$$
\renewcommand{\arraystretch}{1.2}
\begin{array}{rclcl}
(m_1c_1+\cdots+m_rc_r)\gamma^{(2)}
& = &
g\left(\twm_1\alpha_1^{(2)}+\cdots+\twm_r\alpha_r^{(2)}\right)
&&
\\
& = &
g\,E_{\twbfm}^{(2)}(\alpha)
& | & 
\mbox{by \gref{GEJ2}}
\\
& = &
g\,\beta_g(\xi)
& | & 
\mbox{by \gref{GKMJ}}
\\
& = &
0
& | & 
\mbox{by $g\,\beta_g=0$,}
\end{array}
$$
one checks $\det\twPhi_t=\exp\{\rmi t(m_1c_1+\cdots+m_rc_r)\}=1$.
Commuting with $\SUJ$ for all $t$, $\twPhi_t$ defines a $1$-parameter 
subgroup $\{\Phi_t~|~t\in\RR\}$ of $\gau^{k+1}$ by
$$
\Phi_t(q)=\twPhi_t~~~\forall~q\in Q,t\in\RR\,.
$$
Each element of this subgroup is constant on $Q$. Hence, so is the 
generator $\phi = \dot{\Phi}(0)$:
\begin{equation}\label{GgenonQ}
\phi(q)
=
\rmi\left[\left(\frac{c_1}{k_1}~\II_{k_1}\otimes\II_{m_1}\right)
\oplus\cdots\oplus
\left(\frac{c_r}{k_r}~\II_{k_r}\otimes\II_{m_r}\right)\right]
~~~~\forall~q\in Q\,.
\end{equation}
In particular,
\begin{equation} \label{GAso1}
\nabla_A\phi=0\,.
\end{equation}
According to this, for any state $\psi$, 
$$
\int_{\Sigma_s}\Tr\left(
\phi\nabla_A^\mu\frac{\delta}{\delta A^\mu}
\right)
\psi(A)
=
-\int_{\Sigma_s}\Tr\left(
(\nabla_A^\mu\phi)\frac{\delta}{\delta A^\mu}
\right)
\psi(A)
=
0\,.
$$
For physical states, the Gauss law implies
\begin{equation} \label{GAso2}
\int_{\Sigma_s}\Tr\left(
\phi \diff A
\right)
\psi(A) = 0\,.
\end{equation}
Using \gref{GAso1}, as well as the structure equation
$F = \diff A+\frac{1}{2}[A,A]$, we obtain
\begin{equation} \label{GAso3}
\begin{array}{rcl}
\int_{\Sigma_s}\Tr\left(\phi \diff A\right)
& = &
\int_{\Sigma_s}\Tr\left(\phi \diff A-\nabla_A\phi\wedge A\right)
\\
& = &
\int_{\Sigma_s}\Tr\left(
\phi\diff A
- \diff\phi\wedge A
- [A,\phi]\wedge A
\right)
\\
& = &
\int_{\Sigma_s}\Tr\left(
2\phi\diff A + \phi[A,A]
\right)
\\
& = &
2 \int_{\Sigma_s}\Tr\left(\phi F\right)\,.
\end{array}
\end{equation}
Since $A$ is reducible to $Q$, $F$ has block
structure $\left(F_1\otimes\II_{m_1}\right)
\oplus\cdots\oplus
\left(F_r\otimes\II_{m_r}\right)$
with $F_j$ being $(k_j\times k_j)$-matrices. Thus, using \gref{GgenonQ},
\begin{equation} \label{GAso4}
\int_{\Sigma_s}\Tr\left(\phi F\right)
=
\rmi\sum_{j=1}^r \frac{m_j}{k_j}c_j
\int_{\Sigma_s} \Tr F_j\,.
\end{equation}
Now $c_j$ being just the first Chern classes of the elementary
factors of the $\rmU k_1\times\cdots\times\rmU k_r$-bundle $\ab{Q}{\UJ}$,
we have
$$
\int_{\Sigma_s} \Tr F_j = -2\pi \rmi c_j\,, j=1,\dots, r\,.
$$
Inserting this into \gref{GAso4} and the latter into \gref{GAso3} we obtain
$$
\int_{\Sigma_s}\Tr\left(\phi\diff A\right)
=
4\pi \sum_{j=1}^r \frac{m_j}{k_j}\left(c_j\right)^2\,.
$$
Thus, in view of \gref{GAso2}, 
\begin{equation}\label{GAsoend}
4\pi \sum_{j=1}^r \frac{m_j}{k_j}\left(c_j\right)^2\psi(A) 
= 0\,.
\end{equation}
It follows that if one of the $c_j$ is nonzero then $\psi(A)=0$
for all physical states $\psi$, i.e., $A$ is a kinematical node. This proves 
the theorem.
\qed
\\

{\it Remark:} Let us compare \gref{GAsoend} with Formula $(6)$ in 
\cite{Asorey:Nodes}. Define $k'_i=k_im_i$ and $m'_i=1$.
Then $J'=(\bfk',\bfm')\in\rmK(n)$ and $\UJ\subseteq\UJ'$. Let
$\varphi:\UJ\rightarrow\UJ'$ denote the canonical embedding. Consider the
subbundle $\ab{Q}{\varphi}\subseteq P$. One can check that the elementary
factors of this subbundle have first Chern classes $c'_i=m_ic_i$. Inserting
$k'_i$, $l'_i$, and $c'_i$ into \gref {GAsoend} one obtains Formula
$(6)$ in \cite{Asorey:Nodes}. In fact, the authors of
\cite{Asorey:Nodes} use that $A$ is reducible to $Q^\prime$, rather than that
it is even reducible to $Q$. This argument being ''coarser'' than ours, it
still suffices to prove that any connection which is reducible to a
subbundle with nontrivial magnetic charge is a kinematical node. 
\\ 

As a consequence of Theorem \rref{TNodes}, one can speak of nodal and nonnodal
strata. This information
can be read off directly from the labels of the strata.

Let us discuss this in some more detail. Let $J\in\rmK(n)$ be given and 
consider Eqs. \gref{GKMJ} and \gref{GKPJ}. Variables are
$\xi\in H^1(\Sigma_s,\ZZ_g)$ and $\alpha_i^{(2)}\in H^2(\Sigma_s,\ZZ)$, 
$i=1,\dots, r$.
Since $H^2(\Sigma_s,\ZZ)$ is torsion-free, $\beta_g$ is trivial. Moreover, due to
$H^4(\Sigma_s,\ZZ)=0$, $a_i\smile a_j$ vanishes.
Thus, the system of equations \gref{GKMJ}, \gref{GKPJ} reduces to
\begin{equation}\label{GNodes1}
E_{\twbfm}^{(2)}(\alpha)=0\,.
\end{equation}
Writing $\alpha_i^{(2)}=c_i\gamma^{(2)}$ again, \gref{GNodes1} becomes
$$
\sum_{i=1}^r \twm_i c_i = 0\,.
$$
The set of solutions of this equation is a subgroup 
$G_{\twbfm}\subseteq\ZZ^{\oplus r}$. According to Theorem \rref{TNodes}, the
nonnodal strata are parametrized by $\xi$ and the neutral element of
$G_{\twbfm}$, whereas the nodal strata are labelled by $\xi$ and all the other
elements of $G_{\twbfm}$. For example, in the case of $\rmSU 2$ we obtain the
following.
\\
$J=(1|2)$: Here $G_{\twbfm}=\{0\}\subseteq\ZZ$, hence all
strata are nonnodal.
\\
$J=(1,1|1,1)$: We have $G_{\twbfm}=\{(c,-c)|c\in\ZZ\}\subseteq
\ZZ^{\oplus 2}$. Since also $\xi=0$, each value of $c$ labels one stratum.
That corresponding to $c=0$ is nonnodal, the others are nodal.
\\
$J=(2|1)$: Here we have the generic stratum, which is
nonnodal.
\section{Summary}
Starting from a principal $\rmSU n$-bundle $P$ over a compact connected
orientable Riemannian $4$-manifold $M$, we have derived a classification of
the orbit types of the action of the group of gauge transformations of $P$
on the space of connections in $P$. Orbit types are known to label the
elements of the natural stratification, given by Kondracki and
Rogulski \cite{KoRo}, of the gauge orbit space associated to $P$.
The interest in this stratification is due to the fact that the role of
nongeneric strata in gauge physics is not clarified yet.

In order to accomplish the classification, we have utilized that
orbit types are 1:1 with a certain class of subbundles of $P$
(called holonomy-induced Howe subbundles), factorized by the
natural actions of vertical automorphisms of $P$ and of the
structure group. We have shown that such classes of subbundles are
labelled by symbols $[(J;\alpha,\xi)]$, where 
$$
J=((k_1,\dots,k_r),(m_1,\dots, m_r))
$$
is a pair of sequences of positive integers obeying 
$$
\sum_{i=1}^r k_im_i=n\,,
$$
$\alpha=(\alpha_1,\dots,\alpha_r)$ where $\alpha_i\in H^\ast(M,\ZZ)$ are 
admissible values of the Chern class of $\rmU k_i$-bundles over $M$, and
$\xi\in H^1(M,\ZZ_g)$ with $g$ being the greatest common divisor
of $(m_1,\dots, m_r)$. The cohomology elements $\alpha_i$ and $\xi$ are 
subject to the relations
\begin{eqnarray}\nonumber
\sum_{i=1}^r \frac{m_i}{g} \alpha_i^{(2)} & = & \beta_g(\xi)\,,
\\ \nonumber
\alpha_1^{m_1}\smile\dots\smile\alpha_r^{m_r} & = & c(P)\,,
\end{eqnarray}
where $\beta_g: H^1(M,\ZZ_g)\rightarrow H^2(M,\ZZ)$ is the connecting
homomorphism associated to the short exact sequence of coefficient groups in
cohomology
$$
0\rightarrow\ZZ\rightarrow\ZZ\rightarrow\ZZ_g\rightarrow 0\,.
$$
Finally, for any permutation $\sigma$ of $\{1,\dots, r\}$, the symbols
$$
\begin{array}{c}
[((k_1,\dots, k_r),(m_1,\dots, m_r);(\alpha_1,\dots,\alpha_r),\xi)]\,,
\\{}
[((k_{\sigma(1)},\dots, k_{\sigma(r)}),(m_{\sigma(1)},\dots,
m_{\sigma(r)});(\alpha_{\sigma(1)},\dots, \alpha_{\sigma(r)}),\xi)]
\end{array}
$$
have to be identified.

The result obtained enables one to determine which strata are present in the
gauge orbit space,
depending on the topology of the base manifold and the topological sector,
i.e., the isomorphism class of $P$. For some examples we have discussed this
dependence in detail. We have also shown that our result can be used to
reformulate a sufficient condition on a connection to be a node for all
physical states (when the latter are viewed as functionals of connections
subject to the Gauss law). This condition has been derived in
\cite{Asorey:Nodes} and applies to Chern-Simons Theory in $2+1$ dimensions.

In this way, our result may be viewed as one more step
towards a systematic investigation of the physical effects related to 
nongeneric strata of the gauge orbit space.

We remark that orbit types still carry more information about
the stratification structure. Namely, their partial ordering encodes how the
strata are patched together in order to build up the gauge orbit space (cf.
\cite[Thm. (4.3.5)]{KoRo}). A derivation of this partial
ordering will be published separately.
\section*{Acknowledgements}
\addcontentsline{toc}{section}{Acknowledgements}
The authors would like to thank C.~Fleischhack, S.~Kolb, A.~Strohmaier, 
and S.~Boller for interesting discussions. 
They are also grateful to T.~Friedrich, J.~Hilgert, and H.-B.~Rademacher 
for useful suggestions, as well as to L.M.~Woodward, C.~Isham, and 
M.~\v{C}adek who on request have been very helpful with providing specific 
information.
\begin{appendix}
\section{The Eilenberg-MacLane Spaces $K\;\!\!(\;\!\!\ZZ,\;\!\!2\;\!\!)$ 
and $K\;\!\!(\;\!\!\ZZ_g,\;\!\!1\;\!\!)$}
\label{AEMLS}
In this appendix, we construct a model for each of the Eilenberg-MacLane
spaces $K(\ZZ,2)$ and $K(\ZZ_g,1)$ and derive the integer-valued
cohomology of these spaces. Consider the natural free action of $\rmU 1$ on the 
sphere $\sphere{\infty}$ which is induced from the natural action of $\rmU 1$ 
on $\sphere{2n-1}\subset\CC^n$. The orbit space of this action is the complex
projective space $\CC\rmP^\infty$. Moreover, by viewing $\ZZ_g$ as a 
subgroup of
$\rmU 1$, this action gives rise to a natural free action of $\ZZ_g$ on
$\sphere{\infty}$. The orbit space of the latter is the lens space
$\lens{g}{\infty}$. By construction, one has principal bundles
\begin{eqnarray}\label{GpribunCP}
\rmU 1 & \hookrightarrow & \sphere{\infty}~\longrightarrow~\CP^\infty\,,
\\ \label{GpribunLg}
\ZZ_g & \hookrightarrow & \sphere{\infty}~\longrightarrow~\lens{g}{\infty}\,.
\end{eqnarray}
Due to $\pi_i(\sphere{\infty}) = 0$ $\forall i$, the exact homotopy 
sequences induced by \gref{GpribunCP}, \gref{GpribunLg} 
yield
$$
\begin{array}{ccccl}
\pi_i(\CP^\infty)
& = & 
\pi_{i-1}(\rmU 1)
& = &
\left\{
\begin{array}{ccl}
\ZZ & | & i=2\\
0 & | & i\neq 2\,,
\end{array}
\right.
\\[0.3cm]
\pi_i(\lens{g}{\infty})
& = &
\pi_{i-1}(\ZZ_g)
& = &
\left\{
\begin{array}{ccl}
\ZZ_g & | & i=1\\
0 & | & i=2,3,\dots\,,
\end{array}
\right.
\end{array}
$$
respectively. As a consequence, $\CP^\infty$ is a model of $K(\ZZ,2)$ and 
$\lens{g}{\infty}$ is a model of $K(\ZZ_g,1)$. In particular,
\begin{equation} \label{GHiCP}
H^i(K(\ZZ,2),\ZZ) 
= H^i(\CP^\infty,\ZZ)
= \left\{\begin{array}{ccl}
\ZZ & | & i\mbox{ even}\\
0 & | & i\mbox{ odd}\,,
\end{array}\right.
\end{equation}
see \cite[Ch.~VI, Prop.~10.2]{Bredon:Top}, and 
\begin{equation}\label{GHiLg}
H^i(K(\ZZ_g,1),\ZZ)
= H^i(\lens{g}{\infty},\ZZ)
= \left\{\begin{array}{ccl}
\ZZ & | & i=0\\
\ZZ_g & | & i\neq 0\mbox{, even}\\
0 & | & i\neq 0\mbox{, odd}\,,
\end{array}\right.
\end{equation}
see \cite[\S 24, p.~176]{FomenkoFuchs}.

We notice that the vanishing of all homotopy groups of $\sphere{\infty}$ 
also implies that the bundles \gref{GpribunCP} and \gref{GpribunLg} are 
universal for $\rmU 1$ and $\ZZ_g$, respectively. Hence, $\CP^\infty$ and 
$\lens{g}{\infty}$ are models of $\rmB\rmU 1$ and $\rmB\ZZ_g$,
respectively. For $\rmB\ZZ_g$, this has been used in the proof of Lemma
\rref{Ldeltag}.
\section{The Cohomology Algebra $H^\ast(\BSUJ,\ZZ)$} 
\label{AcohomZ}
Let $J\in\rmK(n)$, $J=(\bfk,\bfm)$.
Recall from Corollary \rref{CcohomZ} that the cohomology algebra 
$H^\ast(\BSUJ,\ZZ)$ is generated by the elements
$\gamma_{J,i}^{(2j)}$, $j=1,\dots,k_i$, $i=1,\dots,r$, defined in
\gref{GdefgJi2j}. Here we are going to
derive the relations these generators are subject to.
\begin{Proposition}\label{PrelZ}
The generators $\gamma_{J,i}^{(2j)}$ of $H^\ast(\BSUJ,\ZZ)$ are subject to
the relation
\begin{equation} \label{GrelZ}
E_{\bfm}^{(2)}\left(\gamma_J\right) = 0\,.
\end{equation}
\end{Proposition}
{\it Proof:} Consider the homomorphism $\left(\rmB
j_J\right)^\ast:H^\ast(\BUJ,\ZZ)\rightarrow H^\ast(\BSUJ,\ZZ)$ induced by
the embedding $j_J:\SUJ\rightarrow\UJ$.
By construction, $\gamma_{J,i}^{(2j)} = 
\left(\rmB j_J\right)^\ast\twgamma_{J,i}^{(2j)}$.
Since there are no relations between the $\twgamma_{J,i}^{(2j)}$,
see Lemma \rref{LcohomZ}, all the
relations between the $\gamma_{J,i}^{(2j)}$ are derived from
elements of the kernel of $\left(\rmB j_J\right)^\ast$. We claim that 
\begin{equation}\label{GkerBjJast}
\ker\left(\rmB j_J\right)^\ast = H^\ast(\BUJ,\ZZ)\smile
E_{\bfm}^{(2)}\left(\twgamma_J\right)\,,
\end{equation}
i.e., that $\ker\left(\rmB j_J\right)^\ast$ is generated as an ideal by the
single element $E_{\bfm}^{(2)}\left(\twgamma_J\right)$. This implies the 
assertion.

In order to prove \gref{GkerBjJast}, consider the $\rmU 1$-bundle $\eta$ given in
\gref{Geta}. The Gysin sequence \gref{GGysinetaJ} induced by $\eta$ being
exact, it yields
$$
\ker\left(\rmB j_J\right)^\ast = H^\ast(\BUJ,\ZZ)\smile c_1(\eta)\,.
$$
Let us compute $c_1(\eta)$. We note that $\eta$ is induced as a principal
bundle of the form \gref{Gshexsqc} by the short exact sequence of Lie group
homomorphisms
$$
\II
\longrightarrow\SUJ
\stackrel{j_J}{\longrightarrow}\UJ
\stackrel{\det_{\rmU n}\circ i_J}{\longrightarrow}\rmU 1
\longrightarrow\II\,,
$$
cf.~\gref{Gshexsqc}. Hence, it has classifying map 
$\rmB\det_{\rmU n}\circ\rmB\!\;i_J$. Accordingly, 
$$
\begin{array}{rclcl}
c_1(\eta)
& = &
\left(\rmB\;\!i_J\right)^\ast\circ\left(\rmB\det_{\rmU n}\right)^\ast
\gamma_{\rmU 1}^{(2)}
&&
\\
& = &
\left(\rmB\;\!i_J\right)^\ast \gamma_{\rmU n}^{(2)}
& | &
\mbox{by \gref{GLBiJast2}}
\\
& = &
E_{\bfm}^{(2)}\left(\twgamma_{J,1},\dots,\twgamma_{J,r}\right)
& | &
\mbox{by \gref{GBiJast}}
\end{array}
$$
This proves the proposition.
\qed
\\

As a consequence of Proposition \rref{PrelZ}, $H^\ast(\BSUJ,\ZZ)$ is
isomorphic to the polynomial ring 
$$
\ZZ
[x_{11},\dots,x_{1k_1},\dots,x_{r1},\dots,x_{rk_r}]
/(m_1x_{11}+\cdots+m_rx_{r1})\,,
$$
where $\deg(x_{ij})=2j$.

We remark that the relation \gref{GrelZ} is a consequence of the relation
between $\delta_J$ and $\gamma_J$ given in Theorem \rref{TbgdJ}. Namely,
using \gref{GgEJ2}, Theorem \rref{TbgdJ}, and $g\,\beta_g=0$ we find
$$
E_{\bfm}^{(2)}\left(\gamma_J\right)
=
g\,E_{\twbfm}^{(2)}\left(\gamma_J\right)
=
g\,\beta_g(\delta_J)
=
0\,.
$$
This ensures, in particular, that \gref{GrelZ} does not generate a 
relation independent of \gref{Grelcc} on the level of the characteristic 
classes $\alpha_J$. (Nonetheless, we have already
proved directly in Lemma \rref{Lrangecc} that there are no relations
independent of \gref{Grelcc}.)
\section{The Cohomology Algebra $H^\ast(\BSUJ,\ZZ_g)$}
Let $J\in\rmK(n)$, $J=(\bfk,\bfm)$. In this appendix, we derive 
$H^\ast(\BSUJ,\ZZ_g)$. Recall that $g$ denotes the greatest common
divisor of $\bfm$. Moreover, recall that 
$\varrho_g:H^\ast(\BSUJ,\ZZ)\rightarrow H^\ast(\BSUJ,\ZZ_g)$ denotes 
reduction modulo $g$. We start with some technical lemmas.
\begin{Lemma}\label{LPcohomZg1}
~
\\
{\rm (a)} Let $\alpha\in H^\ast(\BSUJ,\ZZ)$ such that
$\varrho_g\left(\alpha\smile\beta_g(\delta_J)\right)=0$. Then
$\varrho_g(\alpha)=0$ as well as $\alpha\smile\beta_g(\delta_J)=0$.
\\
{\rm (b)} One has 
$\im\beta_g\subseteq H^\ast(\BSUJ,\ZZ)\smile\beta_g(\delta_J)$.
\\
{\rm (c)} The homomorphism 
$\varrho_g\circ\beta_g:H^{2j+1}(\BSUJ,\ZZ_g)\rightarrow H^{2j+2}(\BSUJ,\ZZ_g)$ 
is injective for $j=0,1,2,\dots$ .
\end{Lemma}
{\it Proof:} (a) Let $\alpha$ be given as proposed. Then there exists $\alpha'\in
H^\ast(\BSUJ,\ZZ)$ such that
\begin{equation} \label{GabgdJ}
\alpha\smile\beta_g(\delta_J)=g\alpha'\,.
\end{equation}
Moreover, there exist
$\tilde{\alpha},\tilde{\alpha}'\in H^\ast(\BUJ,\ZZ)$ such that
$\alpha=\left(\rmB j_J\right)^\ast\tilde{\alpha}$ and
$\alpha'=\left(\rmB j_J\right)^\ast\tilde{\alpha}'$. Due to Theorem 
\rref{TbgdJ}
 and \gref{GdefgJi2j},
\begin{equation} \label{GbgdJBjJast}
\beta_g(\delta_J)=\left(\rmB
j_J\right)^\ast E^{(2)}_{\twbfm}\left(\twgamma_J\right)
\end{equation}
Using this, as well as \gref{GkerBjJast}, \gref{GabgdJ} implies that
there exists $\tilde{\alpha}''\in H^\ast(\BUJ,\ZZ)$ such that
\begin{eqnarray}\nonumber
\tilde{\alpha}\smile E_{\twbfm}\left(\twgamma_J\right)
& = &
g\tilde{\alpha}'+\tilde{\alpha}''\smile E^{(2)}_{\bfm}\left(\twgamma_J\right)
\\ \label{Gtwadivble}
& = &
g\left(
\tilde{\alpha}'+\tilde{\alpha}''\smile 
E^{(2)}_{\twbfm}\left(\twgamma_J\right)
\right)\,.
\end{eqnarray}
Taking into account that $H^\ast(\BUJ,\ZZ)$ is torsion-free, that the
elements $\twgamma_{J,i}^{(2j)}$ are generators, and that the greatest common
divisor of the integers $\twm_i$ is $1$, from \gref{Gtwadivble} we infer
that there exists $\tilde{\alpha}'''\in H^\ast(\BUJ,\ZZ)$ such that
$\tilde{\alpha}=g\tilde{\alpha}'''$. Then
$\alpha
= \left(\rmB j_J\right)^\ast\tilde{\alpha}
= g\left(\rmB j_J\right)^\ast\tilde{\alpha}'''$.
It follows $\varrho_g(\alpha)=0$. Due to $g\beta_g(\delta_J)=0$, also
$\alpha\smile\beta_g(\delta_J)=0$, as asserted.

(b) Let $\alpha\in\im\beta_g$. By exactness of
\gref{GBockportion}, $g\alpha=0$. Let $\tilde{\alpha}\in H^\ast(\BUJ,\ZZ)$
such that $\alpha=\left(\rmB j_J\right)^\ast\tilde{\alpha}$. Then
$g\tilde{\alpha}\in\ker\left(\rmB j_J\right)^\ast$. According to 
\gref{GkerBjJast}, there exists $\tilde{\alpha}'\in H^\ast(\BUJ,\ZZ)$ such
that
$$
g\tilde{\alpha}
=
\tilde{\alpha}'\smile 
E^{(2)}_{\bfm}\left(\twgamma_J\right)
=
g\left(\tilde{\alpha}'\smile
E^{(2)}_{\twbfm}\left(\twgamma_J\right)
\right)\,.
$$
Since $H^\ast(\BUJ,\ZZ)$ is free Abelian, this implies
$\tilde{\alpha}=\tilde{\alpha}'\smile
E^{(2)}_{\twbfm}\left(\twgamma_J\right)$.
Using \gref{GbgdJBjJast} we obtain
$\alpha=\left(\rmB j_J\right)^\ast\tilde{\alpha}'\smile\beta_g(\delta_J)$.
This shows (b).

(c) We note that (a) and (b) immediately imply that
$\im\beta_g\cap\ker \varrho_g=\{0\}$. Moreover, by exactness of
\gref{GBockportion}, $\beta_g$ is injective in odd degree. This proves (c).
\qed
\begin{Lemma} \label{LPcohomZg2}
The homomorphism $H^{2j}(\BSUJ,\ZZ_g)\rightarrow H^{2j+1}(\BSUJ,\ZZ_g)$, 
$\alpha\mapsto \alpha\smile\delta_J$, is an isomorphism for $j=0,1,2,\dots$ .
\end{Lemma}
{\it Proof:} First, we check injectivity. Let $\alpha\in
H^{2j}(\BSUJ,\ZZ_g)$ such that $\alpha\smile\delta_J=0$. Consider the
homomorphism $\varrho_g\circ\beta_g$. As an immediate consequence of the definition
of $\beta_g$ (namely, as the connecting homomorphism in the long exact
sequence \gref{GBock}), this is a skew-derivation, i.e., for any
$\alpha_i\in H^{j_i}(\BSUJ,\ZZ_g)$, $i=1,2$, one has
\begin{equation} \label{Gskewder}
\varrho_g\circ\beta_g(\alpha_1\smile\alpha_2)
=
(\varrho_g\circ\beta_g(\alpha_1))\smile\alpha_2
+
(-1)^{j_1}\alpha_1\smile(\varrho_g\circ\beta_g(\alpha_2))\,.
\end{equation}
Hence, due to $\beta_g$ being trivial in even degree,
\begin{equation} \label{GrgbgadJ}
\varrho_g\circ\beta_g(\alpha\smile\delta_J)
= \alpha\smile \varrho_g\circ\beta_g(\delta_J)\,.
\end{equation}
Since $\varrho_g$ is surjective in even degree, there exists $\alpha'\in
H^{2j}(\BSUJ,\ZZ)$ such that $\alpha=\varrho_g\left(\alpha'\right)$.
Inserting this on the rhs.~of \gref{GrgbgadJ} we obtain
\begin{equation}\label{GrgbgadJ2}
\varrho_g\circ\beta_g(\alpha\smile\delta_J)
=
\varrho_g\left(\alpha'\smile\beta_g(\delta_J)\right)\,.
\end{equation}
By assumption, the lhs.~of \gref{GrgbgadJ2} vanishes. Then Lemma
\rref{LPcohomZg1}(a) implies
$\alpha=\varrho_g\left(\alpha'\right)=0$.
This proves injectivity.
To show surjectivity, let $\alpha\in H^{2j+1}(\BSUJ,\ZZ_g)$.
Due to Lemma \rref{LPcohomZg1}(b), there exists $\alpha'\in H^\ast(\BSUJ,\ZZ)$ such that
$\beta_g(\alpha) = \alpha'\smile\beta_g(\delta_J)$. Then
$$
\begin{array}{rcl}
\varrho_g\circ\beta_g(\alpha)
& = &
\varrho_g\left(\alpha'\smile\beta_g(\delta_J)\right)
\\
& = &
\varrho_g\left(\alpha'\right)\smile \varrho_g\circ\beta_g(\delta_J)
\\
& = &
\varrho_g\circ\beta_g(\varrho_g\left(\alpha'\right)\smile\delta_J)\,,
\end{array}
$$
where the last equality is due to \gref{Gskewder} and the fact that $\beta_g$
is trivial in even degree. As a consequence, Lemma \rref{LPcohomZg1}(c) implies
$\alpha=\varrho_g\left(\alpha'\right)\smile\delta_J$.
This shows surjectivity and, therefore, concludes the proof of the lemma.
\qed
\begin{Lemma} \label{LdJ2}
There holds
$\delta_J\smile\delta_J=\left\{\begin{array}{ccl}
0 & | & g=2l+1
\\
l\,\varrho_g\circ\beta_g(\delta_J ) & | & g=2l
\end{array}\right.$.
\end{Lemma}
{\it Proof:} We notice that both the collections of maps
\begin{eqnarray} \label{appGcohomopn1}
\theta_1:H^1(\cdot,\ZZ_g) & \rightarrow & H^2(\cdot,\ZZ_g)\,,~~
\alpha\mapsto\alpha\smile\alpha\,,
\\ \label{appGcohomopn2}
\theta_2:H^1(\cdot,\ZZ_g) & \rightarrow & H^2(\cdot,\ZZ_g)\,,~~
\alpha\mapsto \varrho_g\circ\beta_g(\alpha)\,,
\end{eqnarray}
define a natural transformation of cohomology functors for $CW$-complexes.
Such transformations are called {\it cohomology operations of type
$(1,\ZZ_g;2,\ZZ_g)$} or, more generally, of type $(i_1,\pi_1;i_2,\pi_2)$
if they map $H^{i_1}(\cdot,\pi_1)\rightarrow H^{i_2}(\cdot,\pi_2)$
\cite[Ch.~VII,~Def.~12.2]{Bredon:Top}. Here $\pi_1,\pi_2$ are Abelian groups. One
should
note that cohomology operations need not consist of group homomorphisms.
Nevertheless, the Abelian group
structure of $H^{i_2}(\cdot,\pi_2)$ induces an according structure on
the set of cohomology operations of type $(i_1,\pi_1;i_2,\pi_2)$. Due to
a theorem of Serre \cite[Ch.~VII,~Thm.~12.3]{Bredon:Top}, there exists a group
isomorphism from the group so defined onto $H^{i_2}\left(K(\pi_1,i_1),
\pi_2\right)$. This is given by evaluating the cohomology operations at some
fixed characteristic element of $H^{i_1}\left(K(\pi_1,i_1),\pi_1\right)$.

To apply this theorem, we choose $\lens{g}{\infty}$ as a model of
$K(\ZZ_g,1)$ and $\delta_g$ as a characteristic element of
$H^1(K(\ZZ_g,1),\ZZ_g)$. Assume, for a moment, that there holds
\begin{equation} \label{Gt1dg}
\theta_1(\delta_g) = \left\{\begin{array}{ccl}
0 & | & g=2l+1
\\
l\,\theta_2(\delta_g) & | & g=2l\,.
\end{array}\right.
\end{equation}
Then the cohomology operations are related by
$$
\theta_1 = \left\{\begin{array}{ccl}
0 & | & g=2l+1
\\
l\,\theta_2 & | & g=2l\,.
\end{array}\right.
$$
Thus, in order to prove the lemma, we have to show \gref{Gt1dg}.
Although the cohomology of the spaces $\lens{g}{\infty}$ is well known,
this particular relation can rarely be found in textbooks. An exception is 
the case $g=2$, where $\lens{2}{\infty}=\RR\rmP^\infty$. For the sake of
completeness, in the following we derive \gref{Gt1dg} for
general $g$ from the case $g=2$.

One can show that $H^2(\lens{g}{\infty},\ZZ_g)\cong\ZZ_g$ and that it is generated by
$\varrho_g\circ\beta_g(\delta_g)$, where $\delta_g$ is a
generator of $H^1(\lens{g}{\infty},\ZZ_g)$ \cite{FomenkoFuchs}. Thus, there exists
$a\in\ZZ_g$ such that
$$
\theta_1(\delta_g) = \delta_g\smile\delta_g = a\,\varrho_g\circ\beta_g(\delta_g)\,.
$$
Since $2\delta_g\smile\delta_g=0$, $2a=0$. Thus, if $g$ is odd then
$a=0$. If $g$ is even then either $a=l$, where $g=2l$, or $a=0$. To rule
out the second case, it suffices to find $\delta\in H^1(\lens{g}{\infty},\ZZ_g)$ such
that $\delta\smile\delta\neq 0$. Consider the composite homomorphism
\begin{equation} \label{Gr2past}
H^1(\lens{g}{\infty},\ZZ_g)\stackrel{\varrho_2}{\longrightarrow}
H^1(\lens{g}{\infty},\ZZ_2)\stackrel{p^\ast}{\longrightarrow}
H^1(\RR\rmP^\infty,\ZZ_2)\,.
\end{equation}
Here $\varrho_2$ is reduction modulo $2$ and $p$ is the projection in the
principal bundle
$$
\ZZ_l\hookrightarrow\RR\rmP^\infty\stackrel{p}{\longrightarrow}\lens{g}{\infty}
$$
which arises by factorizing $\RR\rmP^\infty=\sphere{\infty}/\ZZ_2$ by the
residual action of $\ZZ_g/\ZZ_2\cong\ZZ_l$. We check that \gref{Gr2past} is
surjective. For the part of $\varrho_2$ this is obvious, due to $g$ being even.
For the part of $p^\ast$, we note that, by virtue of the Hurewicz and the
Universal Coefficient Theorem, $p^\ast$ is the $\ZZ_2$-dual of the
homomorphism $p_\ast:\pi_1\RR\rmP^\infty\rightarrow\lens{g}{\infty}$. The latter
is easily seen to be injective. Therefore, $p^\ast$ is surjective, as
asserted.
As a consequence, there exists $\delta\in H^1(\lens{g}{\infty},\ZZ_g)$ such
that
$p^\ast\circ \varrho_2(\delta)=\delta_2$, where $\delta_2$ denotes the generator of
$H^1(\RR\rmP^\infty,\ZZ_2)$. We compute
$$
p^\ast\circ \varrho_2(\delta\smile\delta)=(p^\ast\circ
\varrho_2(\delta))\smile(p^\ast\circ
\varrho_2(\delta))=\delta_2\smile\delta_2\,.
$$
Here the rhs.~is known to be nontrivial
\cite[Ch.~VI, Prop.~10.2]{Bredon:Top}. Hence, $\delta\smile\delta$ must have
been nontrivial. This proves the Lemma.
\qed
\begin{Proposition}\label{PcohomZg}
$H^\ast(\BSUJ,\ZZ_g)$ is generated over $\ZZ_g$ by the elements $\delta_J$
and $\varrho_g\left(\gamma_{J,i}^{(2j)}\right)$, $j=1,\dots,k_i$, $i=1,\dots,r$. The generators
are subject to the relation
\begin{equation} \label{GrelZg}
\delta_J\smile\delta_J=\left\{\begin{array}{ccl}
0 & | & g=2l+1 \\
l\,E^{(2)}_{\twbfm}\left(\gamma_J\right) & | & g=2l
\end{array}
\right.
\end{equation}
\end{Proposition}
{\it Proof:}
Consider the long exact sequence \gref{GBock} for the space $\BSUJ$.
Since the integer-valued cohomology of $\BSUJ$ is trivial in odd degree,
this sequence splits into exact portions
\begin{equation} \label{GBockportion}
\begin{array}[b]{l}
0\stackrel{\varrho_g}{\longrightarrow}
H^{2j+1}(\BSUJ,\ZZ_g)\stackrel{\beta_g}{\longrightarrow}
H^{2j+2}(\BSUJ,\ZZ)
\\
\phantom{H^{2j+1}(\BSUJ UJ}
\stackrel{\mu_g}{\longrightarrow}
H^{2j+2}(\BSUJ,\ZZ)\stackrel{\varrho_g}{\longrightarrow}
H^{2j+2}(\BSUJ,\ZZ_g)\stackrel{\beta_g}{\longrightarrow}
0\,,
\end{array}
\end{equation}
where $j=0,1,2,\dots$ .
As an immediate consequence, $\varrho_g$ is surjective in even degree. Thus,
$H^\rmeven(\BSUJ,\ZZ_g)$ is generated by
$\varrho_g\left(\gamma_{J,i}^{(2j)}\right)$, $j=1,\dots,k_i$, $i=1,\dots,r$. Then
Lemma \rref{LPcohomZg2} implies that the whole of $H^\ast(\BSUJ,\ZZ_g)$ is generated 
by these elements together with $\delta_J$.

It remains to determine the
relations the generators are subject to. First, consider relations among the
generators $\varrho_g\left(\gamma_{J,i}^{(2j)}\right)$. These arise from
the relation \gref{GrelZ} and from the elements of $\ker \varrho_g$.
Due to $E^{(2)}_{\bfm}=gE^{(2)}_{\twbfm}$, the
$\mod\,g$-reduction of
\gref{GrelZ} is trivially satisfied. Moreover, relations generated by
elements of $\ker \varrho_g$ are already taken into account by taking $\ZZ_g$ as
the base ring. Thus, there is no relation among the generators
$\varrho_g\left(\gamma_J,i^{(2j)}\right)$.

Next, consider relations involving $\delta_J$. One such relation is provided
by Lemma \rref{LdJ2}. It covers all relations in even degree, because
the latter must contain an even power of $\delta_J$, hence can be written
without $\delta_J$. Relations in odd degree, on the other hand, are of the form 
$\alpha\smile\delta_J=0$, where $\alpha$ is of even degree. Due to Lemma 
\rref{LPcohomZg2}, then $\alpha=0$. Consequently, there are no further relations.
\qed
\\

Due to Proposition \rref{PcohomZg}, if $g$ is odd then $H^\ast(\BSUJ,\ZZ_g)$ is 
isomorphic to the polynomial ring 
$$
\ZZ_g
[x,x_{11},\dots,x_{1k_1},\dots,x_{r1},\dots,x_{rk_r}]\,,
$$
whereas if $g$ is even then it is isomorphic to 
$$
\ZZ_g
[x,x_{11},\dots,x_{1k_1},\dots,x_{r1},\dots,x_{rk_r}]
/\left(x^2-\left(\twm_1x_{11}+\cdots+\twm_rx_{r1}\right)\right)\,,
$$
where $g=2l$. Here $\deg(x)=1$ and $\deg(x_{ij})=2j$.
\end{appendix}

\end{document}